\pdfoutput=1
\newcommand*{\ATLASLATEXPATH}{}
\documentclass[cernpreprint, texlive=2016, UKenglish,LANGEDIT=true]{\ATLASLATEXPATH atlasdoc}
 
\usepackage{\ATLASLATEXPATH atlaspackage}
\usepackage{\ATLASLATEXPATH atlasbiblatex}
 
\usepackage{\ATLASLATEXPATH atlasphysics}
 
\usepackage{cleveref}
\crefrangelabelformat{figure}{#3#1#4--#5\crefstripprefix{#1}{#2}#6}
\crefrangelabelformat{table}{#3#1#4--#5\crefstripprefix{#1}{#2}#6}
\usepackage{dcolumn}
\newcolumntype{d}[1]{D{.}{.}{#1} }
\usepackage{pdflscape}
\usepackage{multirow}
 
\graphicspath{{logos/}{figures/}}
 
\usepackage{ANA-STDM-2016-11-PAPER-defs}

\AtlasTitle{Measurement of the inclusive cross-section for the production of jets in association with a $Z$ boson in proton--proton collisions at 8 TeV using the ATLAS detector}
 
\PreprintIdNumber{CERN-EP-2019-133}
\AtlasVersion{3.0}
\AtlasAbstract{
The inclusive cross-section for jet production in association with a $Z$ boson decaying into an electron--positron
pair is measured as a function of the transverse momentum and the absolute rapidity of jets using 19.9 fb$^{-1}$ of
$\sqrt{s}=8$ TeV proton--proton collision data collected with the ATLAS detector at the Large Hadron Collider.
The measured $Z\text{\,+\,jets}$ cross-section is unfolded to the particle level.
The cross-section is compared with state-of-the-art Standard Model calculations, including the next-to-leading-order and  next-to-next-to-leading-order perturbative QCD calculations,
corrected for non-perturbative and QED radiation effects.
The results of the measurements cover final-state jets with transverse momenta up to 1 TeV,
and show good agreement with  fixed-order calculations.
}
 
\author{The ATLAS Collaboration}
 
\AtlasRefCode{STDM-2016-11}
 
\AtlasDate{\today}

\AtlasJournal{EPJC}

\AtlasCoverSupportingNote{Measurements of the inclusive jet production in association with a \Zboson-boson cross-section: Supporting documentation}{https://cds.cern.ch/record/2158496}
 
\AtlasCoverCommentsDeadline{23rd June 2019}
 
\AtlasCoverAnalysisTeam{Alexandre Glazov, Aliaksei Hrynevich, Nataliia Kondrashova, Pavel Starovoitov}
 
\AtlasCoverEdBoardMember{Amanda~Cooper-Sarkar, Andrew~Parker~(chair), Aranzazu~Ruiz-Martinez}

\AtlasCoverEgroupEditors{atlas-STDM-2016-11-editors@cern.ch}
 
\AtlasCoverEgroupEdBoard{atlas-STDM-2016-11-editorial-board@cern.ch}

\hypersetup{pdftitle={ATLAS document},pdfauthor={The ATLAS Collaboration}}
 
\begin{document}
 
\maketitle
 
\tableofcontents
 
\section{Introduction\label{sec:intro}}
The measurement of the  production \xs\ of jets, a collimated spray  of hadrons, in association with a \ZBoson{} (\Zjets), is an important process for testing the predictions
of perturbative quantum chromodynamics (pQCD).
It provides a benchmark for fixed-order calculations and predictions from \MC{} (MC) simulations,
which are often used to estimate the \Zjets{} background in the measurements of Standard Model processes, such as Higgs boson production,
and in searches for new physics beyond the Standard Model.
 
Various properties of \Zjets{} production have been measured in proton--antiproton collisions at $\sqs=1.96$~\TeV{} at the Tevatron~\cite{CDF_2008,D0_2009, D0_2010,CDF_2015}.
The differential \Zjets{} \xs{} is measured as functions of the \ZBoson{} transverse momentum and the jets' transverse momenta and rapidities,
and as a function of the angular separation between the \ZBoson{} and jets in final states with different jet multiplicities.
The experiments at the Large Hadron Collider (LHC)~\cite{LHC} have an increased phase space compared to previous measurements by using proton--proton collision data
at $\sqs=7$, 8 and 13~\TeV~\cite{STDM-2011-27, STDM-2012-04,ATLAS_Zjet_2015,CMS-SMP-12-004,CMS-SMP-12-004,CMS-SMP-12-017,CMS-SMP-14-013,CMS-SMP-16-015,CMS-SMP-17-005,LHCb_2011,LHCb_2012}.
The measurements at the LHC allow state-of-the-art theoretical \Zjets{} predictions to be tested.
These have recently been calculated to next-to-next-to-leading-order~(NNLO) accuracy in pQCD~\cite{zjet_nnlo,zjet_nnlo2}.
 
This paper studies the double-differential \xs{} of inclusive jet production in association with a \ZBoson{} which decays into an electron--positron pair.
The \xs{} is measured as a function of absolute jet rapidity, \absyj, and jet transverse momentum, \ptj, using the proton--proton (pp) collision data at \COME{} collected by the ATLAS experiment.
The measured \xs{} is unfolded to the particle level.
 
The  \xs{} calculated at fixed order  for  \Zjets{} production in \pp{} collisions at $\sqs=8$~\TeV{} is dominated by  quark--gluon scattering.
The \Zjets{} \xs{} is sensitive  to the gluon and sea-quark parton distribution functions (PDFs) in the proton. 
In the central \absyj{} region the \Zjets{} \xs{} probes the PDFs in the low $x$ range,
where $x$ is the proton momentum fraction,
while in the forward \absyj{}  region it examines  the intermediate and high $x$ values. The scale of the probe is set by \ptj.
 
The measured \xs{} is compared with the next-to-leading-order~(NLO) and NNLO \Zjets{} fixed-order calculations, corrected for hadronisation and the underlying event.
In addition, the data are compared with the predictions from multi-leg matrix element (ME) MC event generators supplemented with parton shower simulations.
 
The structure of the paper is as follows.
The ATLAS detector is briefly described in \Cref{sec:detector}.
This is followed by a description of the data in \Cref{sec:data} and the simulated samples in \Cref{sec:MC}.
The definition of the object reconstruction, calibration and identification procedures and a summary of the selection criteria are given in \Cref{sec:objects}.
The \Zjets{} backgrounds are discussed in \Cref{sec:backgrounds}.
The correction of the measured spectrum to the particle level is described in \Cref{sec:unfolding}.
The experimental uncertainties are discussed in \Cref{sec:uncert}.
The fixed-order calculations together with parton-to-particle-level corrections are presented in \Cref{sec:theory}.
Finally, the measured cross-section is presented and compared with the theory predictions in \Cref{sec:results}.
The quantitative comparisons with the fixed-order pQCD predictions are summarised in \Cref{sec:chi2}.
 
\section{The ATLAS detector\label{sec:detector}}
\newcommand{\AtlasCoordFootnote}{
ATLAS uses a right-handed coordinate system with its origin at the nominal interaction point (IP)
in the centre of the detector and the $z$-axis along the beam pipe.
The $x$-axis points from the IP to the centre of the LHC ring,
and the $y$-axis points upwards.
Cylindrical coordinates $(r,\phi)$ are used in the transverse plane,
$\phi$ being the azimuthal angle around the $z$-axis.
The pseudorapidity is defined in terms of the polar angle $\theta$ as $\eta = -\ln \tan(\theta/2)$.
Angular distance is measured in units of $\Delta R \equiv \sqrt{(\Delta\eta)^{2} + (\Delta\phi)^{2}}$.}

The ATLAS experiment~\cite{PERF-2007-01} at the LHC is a multipurpose particle detector
with a forward--backward symmetric cylindrical geometry and  nearly $4\pi$ coverage in
solid angle.\footnote{\AtlasCoordFootnote}
It consists of an inner tracking detector surrounded by a thin superconducting solenoid, electromagnetic and hadronic calorimeters,
and a muon spectrometer incorporating three large superconducting toroidal magnets.
 
The inner-detector system (ID) is immersed in a \SI{2}{\tesla} axial magnetic field
and provides charged-particle tracking in the range $|\eta| < 2.5$.
A high-granularity silicon pixel detector covers the \pp{} interaction region and typically provides three measurements per track.
It is followed by a silicon microstrip tracker (SCT), which usually provides four two-dimensional measurement points per track.
These silicon detectors are complemented by a transition radiation tracker (TRT),
which provides electron identification information.
 
The calorimeter system covers the pseudorapidity range $|\eta| < 4.9$.
In the region  $|\eta|< 3.2$, electromagnetic calorimetry is provided by barrel and
endcap high-granularity lead/liquid-argon (LAr) electromagnetic calorimeters,
with an additional thin LAr presampler covering $|\eta| < 1.8$ to correct for energy loss in material upstream of the calorimeters.
Hadronic calorimetry is provided by a steel/scintillator-tile calorimeter,
segmented into three barrel structures for $|\eta| < 1.7$, and two copper/LAr hadronic endcap calorimeters in the range $1.5<|\eta|< 3.2$.
The calorimetry in the forward pseudorapidity region, $3.1<|\eta|< 4.9$, is provided by the copper-tungsten/LAr calorimeters.
 
The muon spectrometer surrounds the calorimeters and contains
three large air-core toroidal superconducting magnets with eight coils each.
The field integral of the toroids ranges between \num{2.0} and \SI{6.0}{\tesla\metre} across most of the detector.
The muon spectrometer includes a system of precision tracking chambers and fast detectors for triggering.
 
A three-level trigger~\cite{PERF-2011-02} was used to select  events for offline analysis.
The first-level trigger is implemented in hardware and used a subset of the detector information to reduce the accepted rate to at most \SI{75}{\kilo\hertz}.
This was followed by two software-based trigger levels that
together reduced the average accepted event rate to \SI{400}{\hertz}.

 
\section{Data sample\label{sec:data}}
The data used for this analysis are from proton--proton collisions at $\sqrt{s}=8$~\TeV{} that were  collected by the ATLAS detector in 2012 during stable beam conditions.
Events recorded when any of the ATLAS subsystems were defective or non-operational are excluded.
Data were selected with a dielectron trigger, which required two reconstructed electron candidates with transverse momenta greater than 12~\GeV.
Only events with electron energy leakage of less than 1~\GeV{} into the hadronic calorimeter were accepted.
The trigger required that reconstructed electron candidates were identified using the \looseone{} criteria~\cite{triggermenu2012}.

The integrated luminosity of the analysis data sample after the trigger selection is \LUMI\ measured with an uncertainty of $\pm1.9$\%~\cite{DAPR-2013-01}.
The average number of simultaneous proton--proton interactions per bunch crossing is $20.7$.
 
In addition, a special data sample was selected for a data-driven study of multijet and \Wjets{} backgrounds.
For this purpose, the analysis data sample was enlarged by including auxiliary events selected by a logical OR of two single-electron triggers.
 
The first single-electron trigger required events with at least one reconstructed electron candidate with a transverse momentum greater than 24~\GeV{}
and hadronic energy leakage less than 1~\GeV.
The electron candidate satisfied the \mediumone{} identification criteria~\cite{triggermenu2012},
a tightened subset of \looseone{}.
The reconstructed electron track was required to be isolated from other tracks in the event.
The isolation requirement rejected an event if the scalar sum of reconstructed track transverse momenta in a cone of size $\Delta R=0.2$
around the electron track exceeded 10\%  of the electron track's transverse momentum.
 
The second single-electron trigger accepted events with at least one electron candidate with a transverse
momentum greater than 60~\GeV{} and identified as \mediumone{}.
This trigger reduced inefficiencies in events with high-\pT\ electrons  that resulted from the isolation requirement
used in the first trigger.
 
Events selected by single-electron triggers include a large number of background events that are normally rejected by the \Zjets{} selection requirements,
but these events are used in the data-driven background studies.
\FloatBarrier
 
\section{Monte Carlo simulations\label{sec:MC}}
Simulated \Zjets{} signal events were generated using the \sherpav{}~\cite{sherpa} multi-leg matrix element   MC generator.
The MEs were calculated at NLO accuracy for the inclusive \Zboson{} production process, and additionally with LO accuracy for up to five partons in the final state, using \amegic{}~\cite{amegic}.
\sherpa{} MEs were convolved with the CT10~\cite{ct10} PDFs.
\sherpa{} parton showers were matched to MEs following the CKKW scheme~\cite{ckkw}.
The \menlops{}~\cite{menlops} prescription was used to combine different parton multiplicities from matrix elements and parton showers.
\sherpa{} predictions were normalised to the inclusive \Zboson{} boson production \xs{} calculated at NNLO~\cite{fewzold,fewz,fewztwo}
and are used for the unfolding to particle level and for the evaluation of systematic uncertainties.

An additional \Zjets{} signal sample with up to five partons in the final state at LO was generated using \alpgenv{}~\cite{alpgen}.
The parton showers were generated using \pythiav{}~\cite{pythia6} with the Perugia 2011C~\cite{perugia2011c} set of tuned parameters to model the underlying event's contribution.
The \alpgen{} MEs were matched to the parton showers following the MLM prescription~\cite{alpgen}.
The proton structure was described by the CTEQ6L1~\cite{cteq} PDF.
Referred to as the  \alpgenpythia{} sample, these predictions were normalised to the NNLO \xs{}.
This sample is used in the analysis for the unfolding uncertainty evaluation and for comparisons with the measurement.

The five-flavour scheme with massless quarks was used in both the  \sherpa{} and  \alpgenpythia{} predictions.
 
Backgrounds from the \Ztt, diboson ($WW$, $WZ$ and $ZZ$), \ttbar{} and single-top-quark events are estimated using MC simulations.
The \Ztt{} events were generated using \powhegboxv{}~\cite{powheg,powhegz} interfaced to \pythiaev{}~\cite{pythia8} for parton showering
using the CT10 PDFs and the AU2~\cite{au2} set  of tuned parameters.
The \Ztt prediction was normalised to the NNLO \xs{}~\cite{fewzold,fewz,fewztwo}.
The $WW$, $WZ$ and $ZZ$ events were generated using \herwigv~\cite{herwig} with
the CTEQ6L1 PDFs and the AUET2~\cite{auet2} set of tuned parameters.
The diboson predictions were normalised to the NLO \xss{}~\cite{MCFM,dibosonnlo}.
Samples of single-top-quark events, produced via the $s$-, $t$- and $\Wboson t$-channels, and \ttbar{} events were generated with \powhegboxv{} interfaced to \pythiav{}, which used the CTEQ6L1 PDFs and the Perugia2011C set of tuned parameters.
The prediction for single-top-quark production in $s$-channel were normalised to the NNLO calculations
matched to the next-to-next-to-leading-logarithm (NNLL) calculations (NNLO+NNLL)~\cite{singletchs},
while predictions in $t$- and $\Wboson t$-channel are normalised to the NLO+NNLL calculations~\cite{singletcht,singletchWt}.
The \ttbar{} samples were  normalised to the NNLO+NNLL calculations~\cite{topplusplus}.
 
The Photos~\cite{photos} and Tauola~\cite{tauola} programs were interfaced to the MC generators, excluding Sherpa,
to model electromagnetic final-state radiation and $\tau$-lepton decays, respectively.
 
Additional proton--proton interactions, generally called pile-up,
were simulated using the \pythiaev{} generator with the MSTW2008~\cite{MSTW} PDFs and the A2~\cite{au2} set of tuned parameters.
The pile-up events were overlaid onto the events from the hard-scattering physics processes.
MC simulated events were reweighted to match the distribution of the average number of interactions per bunch crossing in data.
 
All MC predictions were obtained from events processed with the ATLAS detector simulation~\cite{SOFT-2010-01} that is based on \GEANT4~\cite{geant}.
 
\FloatBarrier
 
\section{Object definitions and event selection\label{sec:objects}}
The measured objects are the electrons and jets reconstructed in ATLAS.
The methods used to reconstruct, identify and calibrate electrons are presented in~\Cref{subsec:electronreco}.
The reconstruction of jets, their calibration, and background suppression methods are discussed in~\Cref{subsec:jetreco}.
Finally, all selection requirements are summarised in~\Cref{subsec:selection}.
\subsection{Electron reconstruction and identification\label{subsec:electronreco}}
Electron reconstruction in the central region, $|\eta|<2.5$, starts from energy deposits in calorimeter cells.
A sliding-window algorithm scans the central volume of the electromagnetic calorimeter in order to seed three-dimensional clusters.
The window has a  size of $3\times5$ in units of $0.025\times0.025$ in $\eta$--$\phi$ space. Seeded cells
have an energy sum of the constituent calorimeter cells greater than $2.5\GeV$.
An electron candidate  is reconstructed if the cluster is matched to at least one track assigned to the primary vertex, as measured in the inner detector.
The energy of a reconstructed electron candidate is given by the energy of a cluster that is
enlarged to a size of $3\times7$ ($5\times5$)  in $\eta$--$\phi$ space in the central (endcap) electromagnetic calorimeter in order
to take into account the shape of electromagnetic shower energy deposits in different calorimeter regions.
The $\eta$ and $\phi$ coordinates of a reconstructed electron candidate are taken from the matched track.
The details of the electron reconstruction are given in Ref.~\cite{PERF-2016-01}.

A multistep calibration is used to correct the electron energy scale to that of simulated electrons~\cite{PERF-2010-04}.
Cluster energies in data and in MC simulation are corrected for energy loss in the material upstream of the electromagnetic calorimeter,
energy lost outside of the cluster volume and energy leakage beyond the electromagnetic calorimeter.
The reconstructed electron energy in data is corrected as a function of electron pseudorapidity
using a multiplicative scale factor obtained from a comparison of \Zee{} mass distributions between data and simulation.
In addition, the electron energy in the MC simulation is scaled by a random number taken from a
Gaussian distribution with a mean value of one and an $\eta$-dependent width,
equal to the difference between the electron energy resolution in data and MC simulation, determined in situ using \Zee{} events.

A set of cut-based electron identification criteria, which use
cluster shape and track properties, is applied to reconstructed electrons to
suppress the residual backgrounds from photon conversions, jets misidentified as electrons and semileptonic heavy-hadron decays.
There are three types of identification criteria, listed in the order of increasing background-rejection strength but diminishing electron selection efficiency:
\loose{}, \medium{} and \tight{}~\cite{PERF-2016-01}.
The \loose{} criteria identify electrons using a set of thresholds applied to cluster shape properties measured in the first and second LAr calorimeter layers,
energy leakage into the  hadronic calorimeter, the number of hits in the pixel and SCT detectors, and the
angular distance between the cluster position in the first LAr calorimeter layer and the extrapolated track.
The \medium{} selection tightens \loose{} requirements on shower shape variables. In addition, the \medium{} selection sets conditions on
the energy deposited in the third calorimeter layer, track properties in the TRT detector and the vertex position.
The \tight{} selection tightens the \medium{} identification criteria thresholds, sets conditions on the measured ratio of cluster energy to track momentum
and rejects reconstructed electron candidates matched to photon conversions.
 
Each MC simulated event is reweighted by  scale factors that make the  trigger, reconstruction and  identification efficiencies
the same in data and MC simulation.
The scale factors are generally close to one and are calculated in bins of electron transverse momenta and pseudorapidity~\cite{triggermenu2012,PERF-2016-01}.
\subsection{Jet reconstruction, pile-up suppression and quality criteria\label{subsec:jetreco}}
Jets are reconstructed using the \antikt{} algorithm~\cite{antikt} with a radius parameter $R=0.4$, as implemented in the FastJet software package~\cite{FastJet}.
Jet reconstruction uses topologically clustered cells from both the electromagnetic and hadronic calorimeters~\cite{topoclusters}.
The topological clustering algorithm groups cells with statistically significant energy deposits as a method to suppress noise.
The energy scale of calorimeter cells is initially established for electromagnetic particles.
The local cell weighting (LCW)~\cite{lcw} calibration is applied to topological clusters
to correct for the difference between the detector responses to electromagnetic and hadronic particles,
energy losses in inactive material and out-of-cluster energy deposits.
The LCW corrections are derived using the MC simulation of the detector response to single pions.
 
The jet energy scale (JES) calibration~\cite{PERF-2012-01} corrects the energy scale of reconstructed jets to that of simulated particle-level jets.
The JES calibration includes origin correction, pile-up correction, MC-based correction of the jet energy and pseudorapidity (MCJES),
global sequential calibration (GSC) and residual in situ calibration.
 
The origin correction forces the four-momentum of the jet to point to the hard-scatter primary vertex rather than to the centre of the detector, while keeping the jet energy constant.
 
Pile-up contributions to the measured jet energies are accounted for by using a two-step procedure.
First, the reconstructed jet energy is corrected for the effect of pile-up by using the average energy density in the event and the area of the jet~\cite{PERF-2014-03}.
Second, a residual correction is applied to remove the remaining dependence of the jet energy on the number of reconstructed primary vertices, $N_\textrm{PV}$,
and the expected average number of interactions per bunch crossing, $\langle\mu\rangle$.
 
The MCJES corrects the reconstructed jet energy to the particle-level jet energy using MC simulation.
In addition, a correction is applied to the reconstructed jet pseudorapidity
to account for the biases caused by the transition between different calorimeter regions and the differences in calorimeter granularity.
 
Next, the GSC corrects the jet four-momenta to reduce the response's dependence on the flavour of the parton that initiates the jet. 
The GSC is determined using
the number of tracks assigned to a jet, the \pt{}-weighted transverse distance in the $\eta$--$\phi$ space
between the jet axis and all tracks assigned to the jet (track width), and the number of muon track segments assigned to the jet.
 
Finally, the residual in situ correction makes the jet response the same in data and MC simulation
as a function of detector pseudorapidity by using dijet events ($\eta$-intercalibration),
and as a function of jet transverse momentum by using well-calibrated reference objects in $\Zboson/\gamma$ and multijet events.
 
Jets originating from pile-up interactions are suppressed using the jet vertex fraction (JVF)~\cite{PERF-2014-03}.
The JVF is calculated for each jet and each primary vertex in the event as a ratio of the scalar sum of \pt{} of tracks,
matched to a jet and assigned to a given vertex, to the scalar sum of \pt{} of all tracks matched to a jet.
 
Applying  jet quality criteria suppresses jets from non-collision backgrounds that arise from proton interactions with the residual gas in the beam pipe,
beam interactions with the collimator upstream of the ATLAS detector,
cosmic rays overlapping in time with the proton--proton collision and noise in the calorimeter.
Jet quality criteria are used to distinguish jets by using the information about the quality of the energy reconstruction in calorimeter cells,
the direction of the shower development and the properties of tracks matched to jets.
There are four sets of selection criteria that establish jet quality: \looser{}, \loose{}, \medium{} and \tight{}~\cite{PERF-2012-01}.
They are listed in the order of increasing suppression of non-collision jet background  but decreasing jet selection efficiency.
 
\subsection{Event selection\label{subsec:selection}}
 
Events are required to have a primary vertex with at least three assigned tracks that have a  transverse momentum greater than $400\MeV$.
When several reconstructed primary vertices satisfy this requirement,
the hard-scatter vertex is taken to be the one with the highest sum of the squares of the transverse momenta of its assigned tracks.
 
Each event is required to have exactly two reconstructed electrons, each with transverse momentum greater than $20\GeV$
and an absolute pseudorapidity less than $2.47$, excluding the detector transition region, $1.37 < |\eta_{e}|< 1.52$,
between barrel and endcap electromagnetic calorimeters.
The electrons  are required to have opposite charges,  be identified using the \medium{}~\cite{PERF-2016-01} criteria and
be matched to electron candidates that were selected by the trigger.
The \medium{} identification ensures the electrons originate from the hard-scatter vertex.
The electron-pair invariant mass, \mee, is required to be in the $66\GeV < \mee < 116\GeV$ range.
 
Jets are required to have a  transverse momentum greater than $25\GeV$ and an absolute jet rapidity less than $3.4$.
Jets with $\ptj<50\GeV$,  $|\eta_\text{det}|<2.4$, where $\eta_\text{det}$ is reconstructed
relative to the detector centre, and  $|\text{JVF}|<0.25$
are considered to be from pile-up. Jets originating from pile-up are removed from the measurement.
MC simulations poorly describe the effects of high pile-up in the $\ptj < 50\GeV$ and $\absyj > 2.4$ region, so this region is not included in the measurement.
Jets reconstructed within $\Delta R = 0.4$  of selected electrons are rejected in order to avoid overlap.
Jets are required to satisfy the \medium{}~\cite{PERF-2012-01} quality criteria.
In addition, jets in regions of the detector that are poorly modelled are rejected in data and MC simulations in order to avoid biasing the measured jet energy~\cite{PERF-2012-01}.
Each jet that meets the selection requirements is used in the measurement.
 
As a result, 1\,486\,415 events with two electrons and at least one jet were selected for the analysis.
\FloatBarrier

\section{Backgrounds\label{sec:backgrounds}}
The majority of irreducible backgrounds in this measurement are studied using MC samples that simulate $\Zboson{} \rightarrow \tau \tau$, diboson, \ttbar{} and single top-quark production.
The $\Zboson{} \rightarrow \tau \tau$ process is a background if both $\tau$-leptons decay into an electron and neutrino.
Diboson production constitutes a background to the \Zjets{} signal if the \Wboson and/or \Zboson boson decays into  electrons.
Since the top-quark decays predominantly via $t \rightarrow Wb$, the \ttbar{} and single top-quark constitute a background to the \Zjets{} signal
when \Wboson{} bosons decay into an electron or jets are misidentified as electrons.

Multijet production constitutes a background to the \Zjets{} signal when two jets are misidentified as electrons.
The \Wjets{} background is due to an electron from \WBoson{} decay and a jet misidentified as electron.
A combined background from multijet and \Wjets{} events is studied using a data-driven technique,
thus providing a model-independent background estimate.
 
A background-enriched data sample is used for the combined multijet and \Wjets{} background control region.
Its selection requires two reconstructed electrons with at least one electron that satisfies the \medium{} identification criteria, but not the \tight{} ones.
This allows selection of events with at least one jet misidentified as an electron. No identification criteria are applied to the second reconstructed electron,
in order to allow for the possibility of \Wjets{} events with a genuine electron from  \WBoson{} decay,
and  multijet events with   a jet misidentified as another electron.
Both selected electrons are required to have the same charge to suppress the \Zjets{} signal events.
The combined multijet and \Wjets{} background template is constructed by subtracting the MC simulated \Zjets{} signal events
and $\Zboson{} \rightarrow \tau \tau$, diboson, \ttbar{} and single-top-quark background events in the control region from data.

The purity of the template is calculated as the fraction of multijet and \Wjets{} events in the data control region.
The purity is about 98\% in the tails of the \mee{} distribution and is about 80\% near the \mee{} peak at $91\GeV$.
The template purity is above $90\%$ in all \absyj{} and \ptj{} bins.

The combined multijet and \Wjets{} background template is normalised to data using the invariant mass distribution of reconstructed electron pairs.
A maximum-likelihood fit is used to adjust the normalisation of the combined multijet and \Wjets{} background template
relative to the measured \Zjets{} distribution. The normalisations of  MC simulated samples are
fixed in the fit:
the $\Zboson{} \rightarrow \tau \tau$, diboson, \ttbar{} and single-top-quark distributions are
normalised by their  fixed-order \xss{}, whereas
the normalisation of the MC simulated \Zjets{} signal events is scaled to data to give
the same total number of events near the peak of the \Zboson{} mass spectrum in  the $90\GeV < \mee < 92\GeV$ range.
The combined multijet and  \Wjets{} background is fit to data in an extended \mee{} region, $60\GeV < \mee < 140\GeV$, excluding
the bins under the \Zboson{} peak within the $80\GeV < \mee < 100\GeV$ region.
The extended \mee{} region is used for the normalisation extraction only, as it allows more background events
in the tails of the \Zboson{} mass spectrum. The normalisation of the multijet and \Wjets{}
background template, calculated in the fit, is used to adjust the templates, obtained in the \absyj{} and \ptj{}   bins, to the \Zjets{} signal region.

\begin{figure}[h]
\centering
\subfloat[$25\GeV < \ptj < 50\GeV$]       {\includegraphics[width=0.49\linewidth]{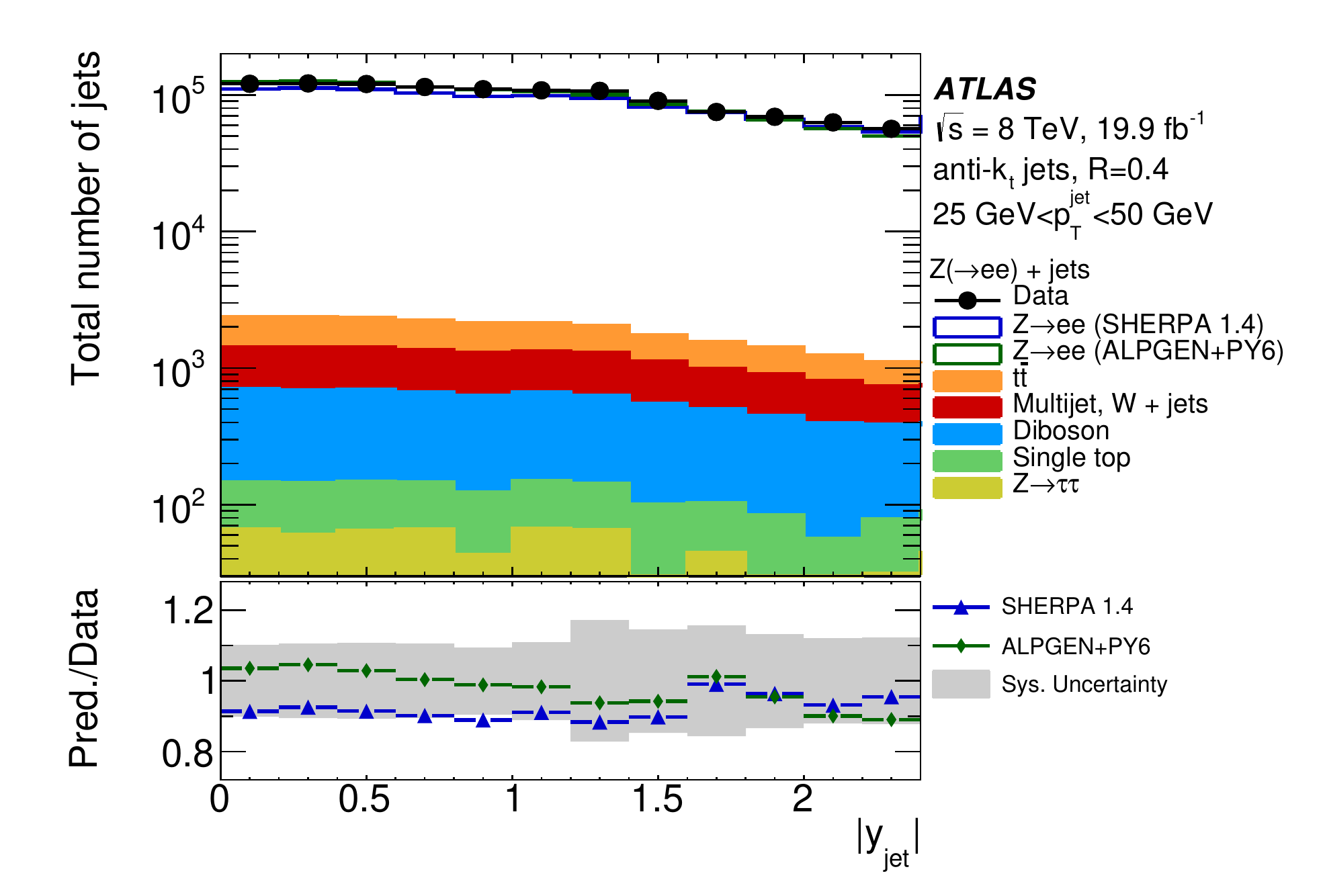}}
\subfloat[$50\GeV < \ptj < 100\GeV$]     {\includegraphics[width=0.49\linewidth]{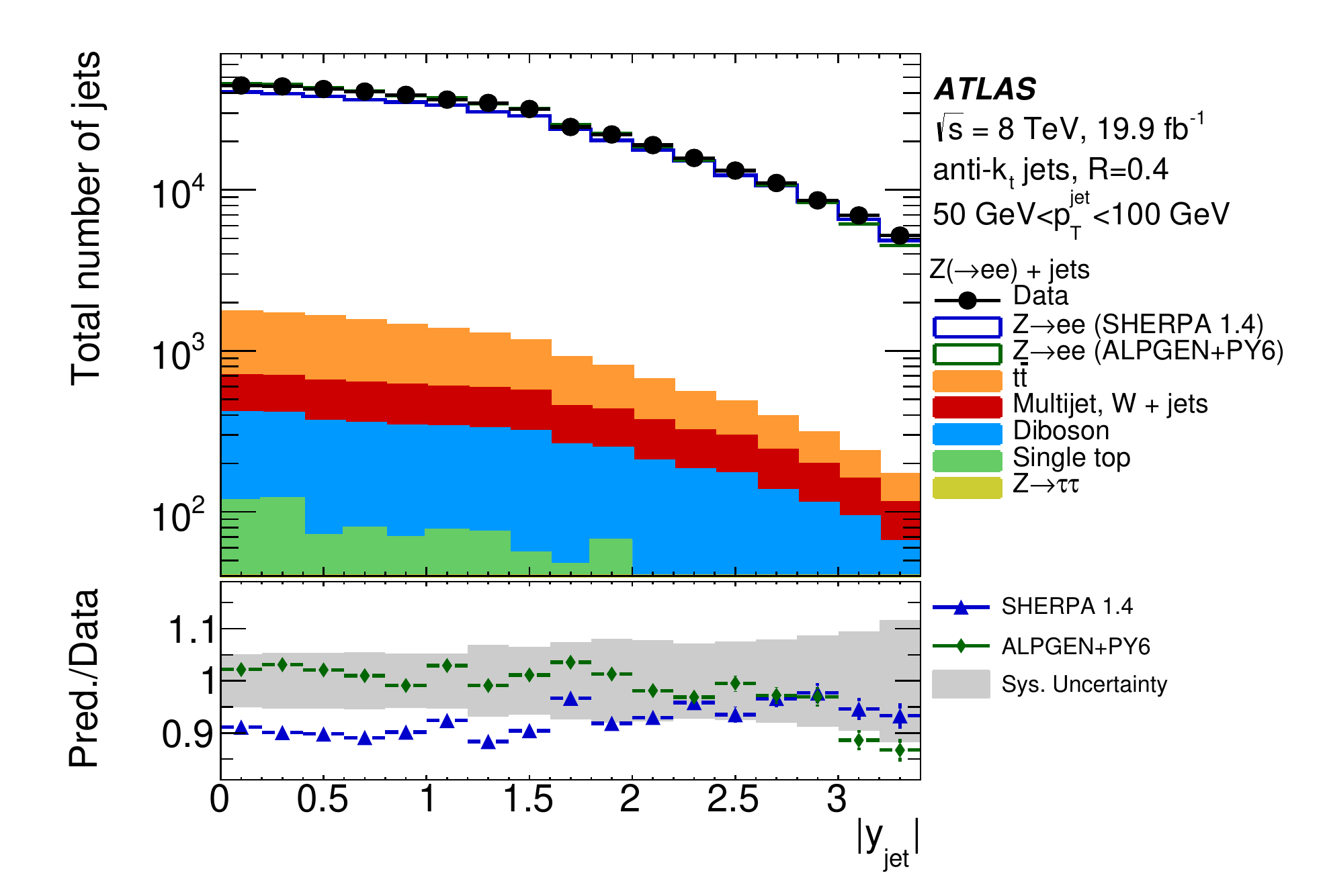}}\\
 
\subfloat[$100\GeV < \ptj < 200\GeV$]   {\includegraphics[width=0.49\linewidth]{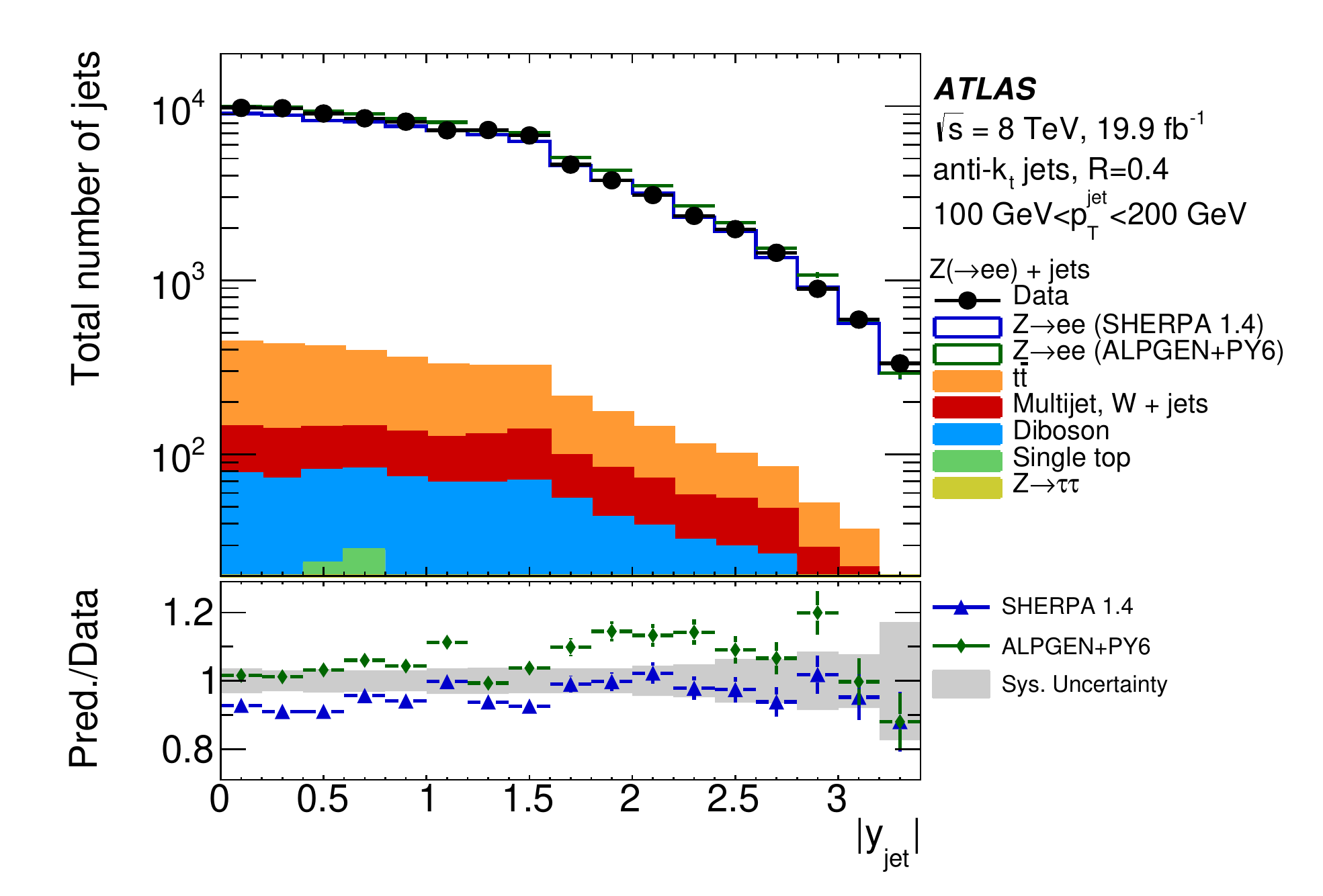}}
\subfloat[$200\GeV < \ptj < 300\GeV$]   {\includegraphics[width=0.49\linewidth]{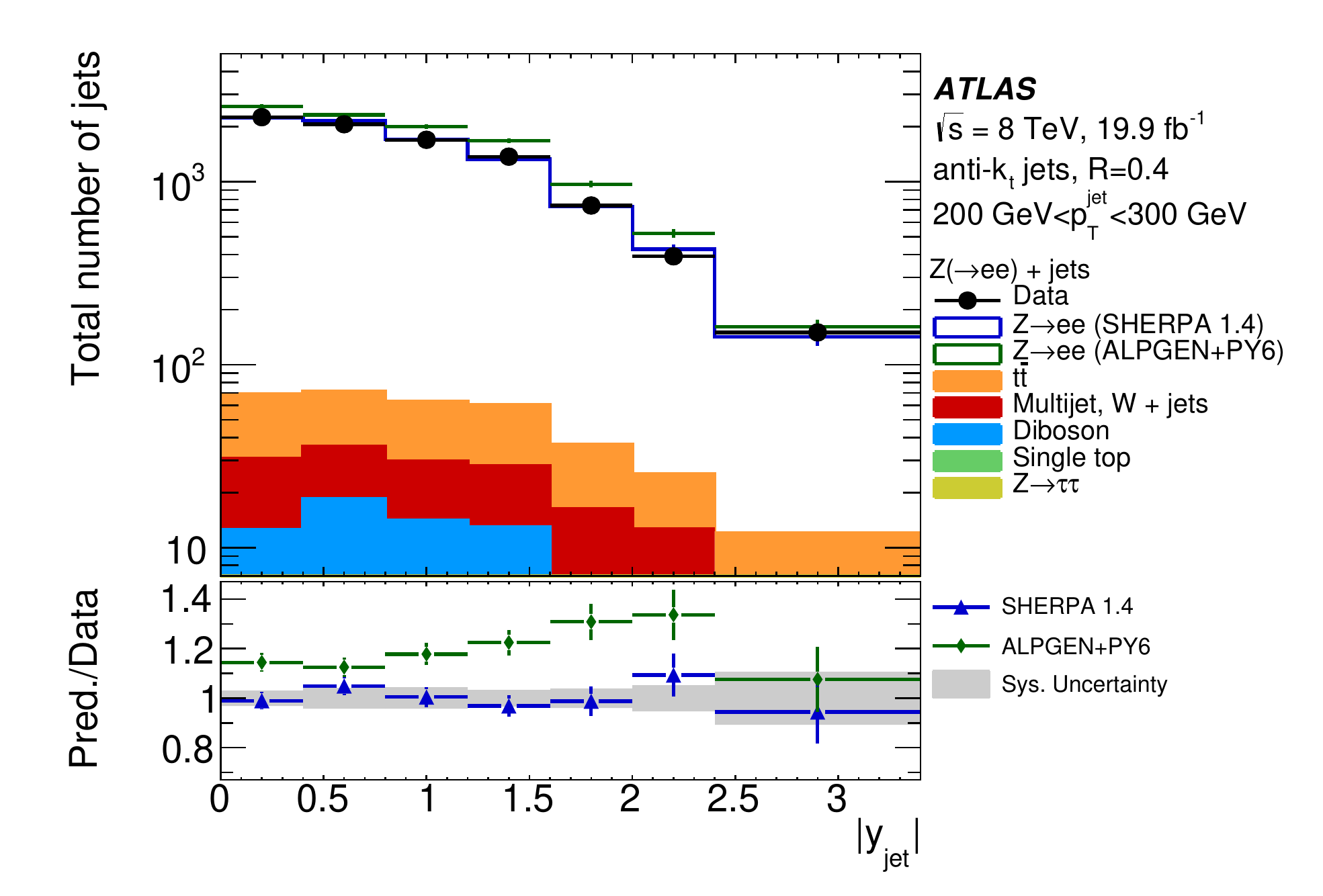}}\\
 
\subfloat[$300\GeV < \ptj < 400\GeV$]   {\includegraphics[width=0.49\linewidth]{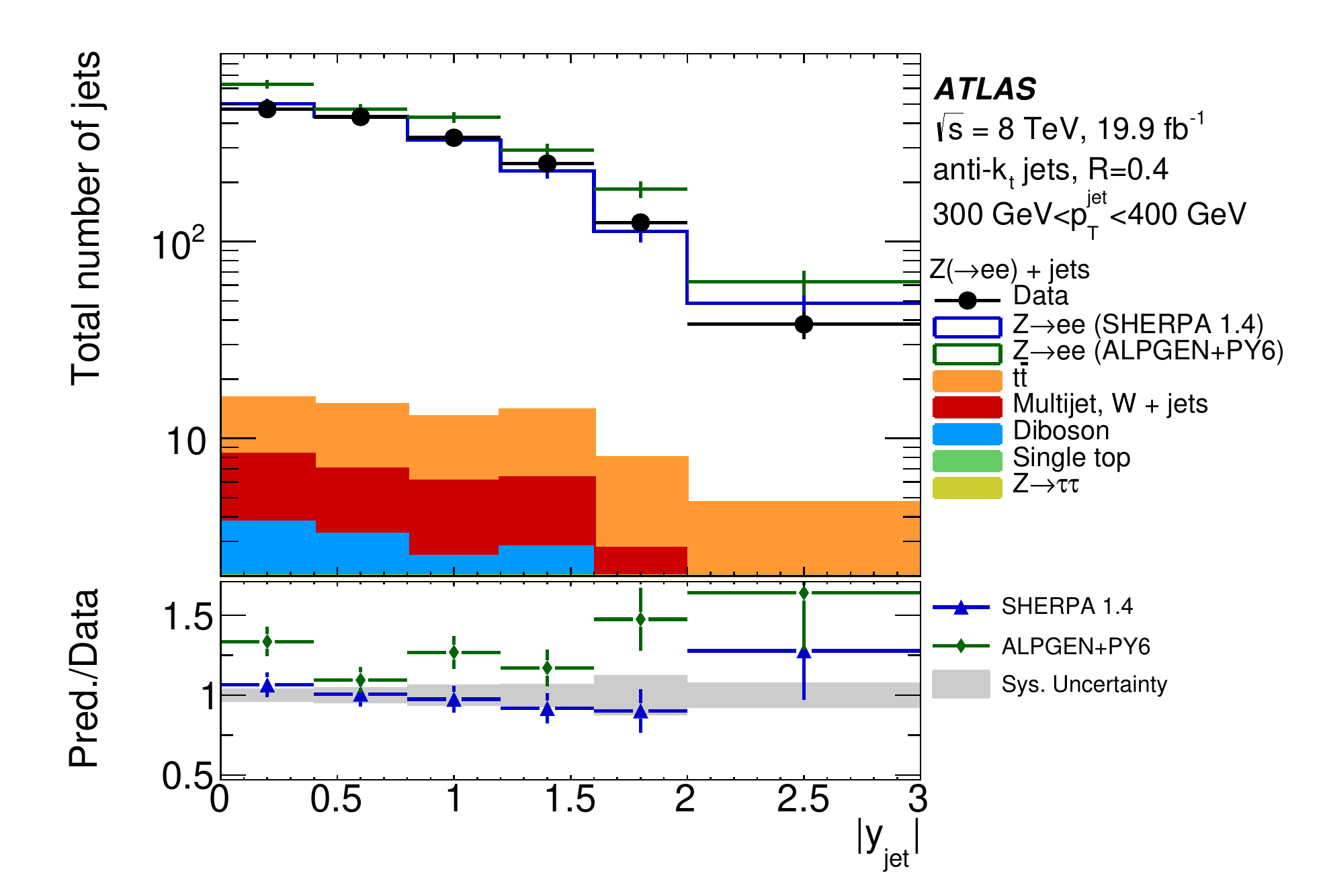}}
\subfloat[$400\GeV < \ptj < 1050\GeV$] {\includegraphics[width=0.49\linewidth]{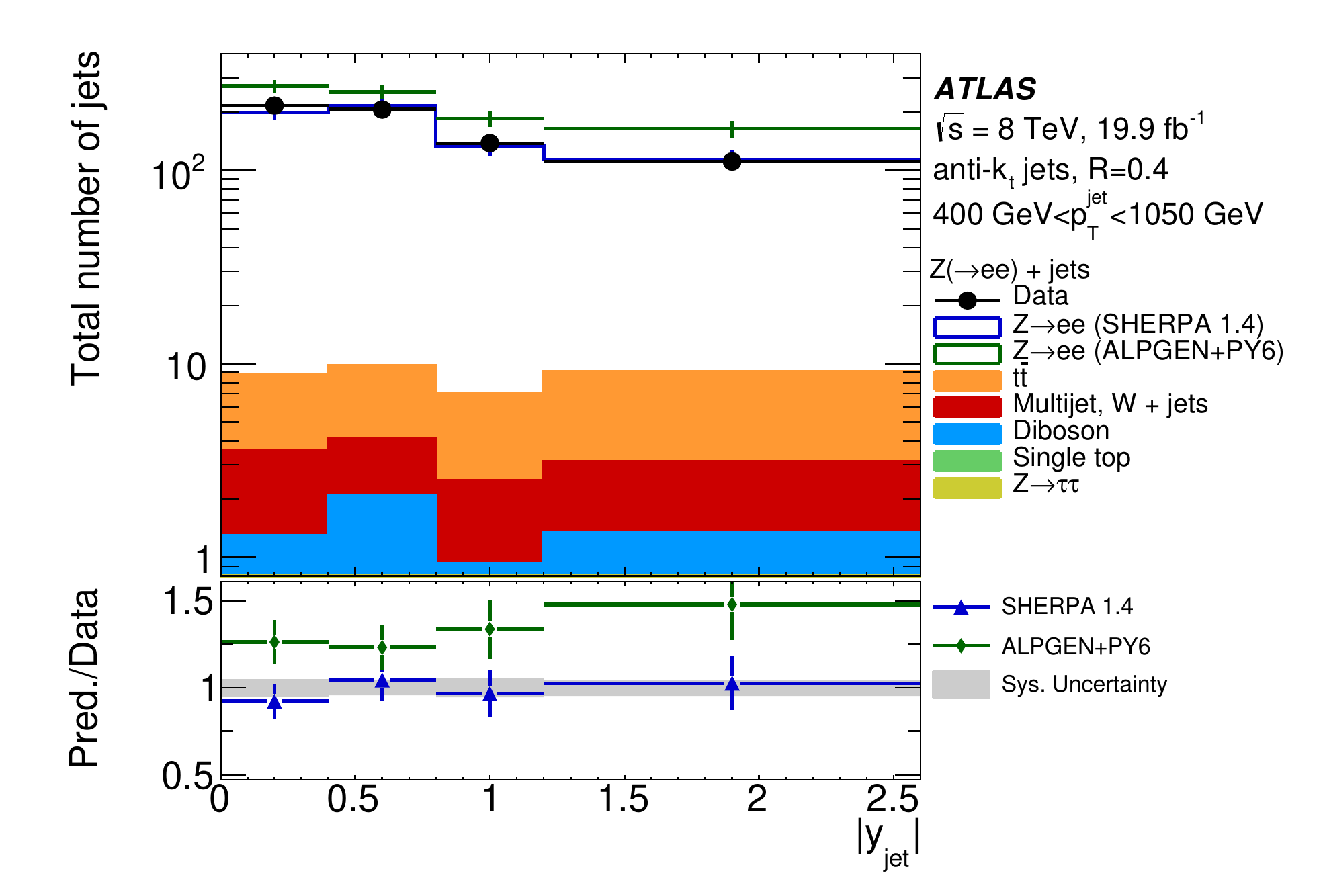}}\\
\caption{The total number of jets in \Zjets{} events as a function of \absyj{} in \ptj{} bins for the integrated luminosity of \LUMI.
Data are presented with markers. The filled areas correspond to the backgrounds stacked.
All backgrounds are added to the \Zjets{} \sherpa{} and \alpgenpythia{} predictions.
Lower panels show ratios of MC predictions to data.
The grey band shows the sum in quadrature of the electron and jet uncertainties.
The statistical uncertainties are shown with vertical error bars.
In the lower panels the total data + MC statistical uncertainty is shown.
\label{fig:fit_zmass_eta_stacked} }
\end{figure}

The total number of jets in \Zjets{} events are shown as a function of \absyj{} and \ptj{} bins in~\Cref{fig:fit_zmass_eta_stacked}.
Data are compared with the sum of signal MC events and all backgrounds.
 
The \sherpa{} \Zjets{} simulation, normalised to the NNLO cross-section, is lower than data by about 10\% in the  $\ptj < 200\GeV$ region.
These differences are mostly covered by the variations within electron and jet uncertainties introduced in~\Cref{sec:uncert}.
In the $\ptj >200\GeV$ region, agreement with data is within the statistical uncertainties.

The \alpgenpythia{} predictions are in agreement with data within 10\% for jets with transverse momenta below $100\GeV$.
However, the level of disagreement increases as a function of the jet transverse momenta, reaching 30\% in the $400\GeV < \ptj{} < 1050\GeV$ region.
 
The  dominant background in the measurement is from \ttbar{} events.
It is 0.3\%--0.8\% in the $25\GeV < \ptj < 50\GeV$ region and 1\%--2.5\% in the $50\GeV < \ptj < 100\GeV$ region, with the largest contribution in the central rapidity region.
In the $100\GeV < \ptj < 200\GeV$ region, this background is approximately 3\%,
while in the $200\GeV < \ptj < 1050\GeV$ region it is 1.8\%--8\%, increasing for forward rapidity jets.
 
The combined multijet and \Wjets{} background and the diboson background are similar in size.
The contributions of these backgrounds are 0.5\%--1\%.
 
The $\Zboson \rightarrow \tau \tau$ and single-top-quark backgrounds are below 0.1\%.

\FloatBarrier
 
\section{Unfolding of detector effects\label{sec:unfolding}}
The experimental measurements are affected by the detector resolution and reconstruction inefficiencies.
In order to compare the measured \xss{} with the theoretical \Zjets{} predictions at the particle level,
the reconstructed spectrum 
is corrected for detector effects using the iterative Bayesian unfolding method~\cite{UnfoldingBayes}.
The unfolding is performed using the  \sherpa{} \Zjets{} simulation.

The particle-level phase space in the MC simulation is defined using two dressed electrons and at least one jet.
For the dressed electron,  the  four-momenta of  any photons within a cone of $\Delta R=0.1$ around its axis are added   to the four-momentum of the electron.
Electrons are required to have $|\eta|<2.47$ and $\pt>20\GeV$. The electron pair's invariant mass is required to be within the range $66\GeV<\mee<116\GeV$.
 
Jets at the particle level are built by using the \antikt{} jet algorithm with a radius parameter $R=0.4$
to cluster stable final-state particles with a decay length of $c\tau>10$~mm,
excluding muons and neutrinos.
Jets are selected in the $\absyj<3.4$ and $\ptj>25\GeV$ region.
Jets within $\Delta R = 0.4$ of electrons are rejected.
 
The closest reconstructed and particle-level jets are considered matched if $\Delta R$ between their axes satisfies $\Delta R < 0.4$.
 
The input for the unfolding is the transfer matrix, which maps reconstructed jets to the particle-level jets in the \absyj{}--\ptj{} plane,
taking into account the bin-to-bin migrations that arise from limited detector resolution.
An additional \ptj{} bin, $17\GeV<\ptj<25\GeV$, is included in the reconstructed and particle-level jet spectra to account for the migrations from the low \ptj{} range.
This bin is not reported in the measurement.
 
Given the significant amount of migration between jet transverse momentum bins,
the unfolding is performed in all \absyj{} and \ptj{} bins simultaneously.
The migration between adjacent \absyj{} bins is found to be small.

The transfer matrix is defined for matched jets. Therefore, the reconstructed jet spectrum must be corrected to account for matching efficiencies
prior to unfolding.
The reconstruction-level matching efficiency is calculated as the fraction of reconstructed jets matching particle-level jets.
This efficiency is 80\%--90\% in the  $25\GeV<\ptj<100\GeV$ region and is above 99\% in the $\ptj>100\GeV$ region.
The particle-level jet matching efficiency is calculated as the fraction of particle-level jets matching reconstructed jets.
This efficiency is 45\%--55\% in all bins of the measurement due to the inefficiency of the \ZBoson{} reconstruction.

The backgrounds are subtracted from data prior to unfolding.
The unfolded number of jets in data, $N^{\mathcal{P}}_{i}$, in each bin $i$ of the measurement is obtained as
\begin{linenomath*}
\begin{equation}
N^{\mathcal{P}}_{i} = \frac{1}{\mathcal{E}^{\mathcal{P}}_{i}} \sum_{j} U_{ij} \mathcal{E}^{\mathcal{R}}_{j} N^{\mathcal{R}}_{j},
\label{eq:NJetsPerBin}
\end{equation}
\end{linenomath*}
where $N^{\mathcal{R}}_{j}$ is the number of jets reconstructed in bin $j$ after the background subtraction, $U_{ij}$ is an element of the unfolding matrix, and
$\mathcal{E}^{\mathcal{R}}_{j}$ and $\mathcal{E}^{\mathcal{P}}_{i}$ are the reconstruction-level and particle-level jet matching efficiencies, respectively.

The transfer matrix and the matching efficiencies are improved using three unfolding iterations to reduce the impact
of the particle-level jet spectra mis-modelling on the unfolded data.
 
\FloatBarrier
 
\section{Experimental uncertainties\label{sec:uncert}}
\subsection{Electron uncertainties}
 
The electron energy scale has associated statistical uncertainties and systematic uncertainties arising from a possible bias in the calibration method, the choice of generator,
the presampler energy scale, and imperfect knowledge of the material in front of the EM calorimeter.
The total energy-scale uncertainty is calculated as the sum in quadrature of these components.
It is varied by $\pm1\sigma$ in order to propagate the electron energy scale uncertainty into the measured \Zjets{} \xss{}.

The electron energy resolution has uncertainties associated to the extraction of the resolution difference between data and simulation using \Zee{} events,
to the knowledge of the sampling term of the calorimeter energy resolution and to the pile-up noise modelling.
These uncertainties are evaluated in situ using the \Zee{} events,
and the total uncertainty is calculated as the sum in quadrature of the different uncertainties.
The scale factor for  electron energy resolution in MC simulation  is varied by $\pm1\sigma$  in the total uncertainty
in order to propagate this uncertainty into the \Zjets{} \xs{} measurements.

The uncertainties in calculations of the electron trigger, reconstruction and identification efficiencies are propagated into the measurements
by $\pm1\sigma$ variations of the scale factors, used to reweight the MC simulated events, within the total uncertainty of each efficiency~\cite{triggermenu2012,PERF-2016-01}.

For each systematic variation a new transfer matrix and new matching efficiencies are calculated, and data unfolding is performed.
The deviation from the nominal unfolded result is assigned as the systematic uncertainty in the measurements.
\subsection{Jet uncertainties\label{subsec:jetunc}}
The uncertainty in the jet energy measurement is described by 65 uncertainty components~\cite{PERF-2012-01}.
Of these, 56 JES uncertainty components are related to detector description, physics modelling and sample sizes
of the $\Zboson/\gamma$ and multijet MC samples used for JES in situ measurements.
The single-hadron response studies are used to describe the JES uncertainty in high-\pt{} jet regions, where the in situ studies have few events.
The MC non-closure uncertainty takes into account the differences in the jet response due to different shower models used in MC generators.
Four uncertainty components are due to the pile-up corrections of the jet kinematics,
and take into account mis-modelling of $N_\textrm{PV}$ and $\langle\mu\rangle$ distributions, the average energy density
and the residual $\pt$ dependence of the $N_\textrm{PV}$ and $\langle\mu\rangle$ terms.
Two flavour-based uncertainties take into account the difference between the calorimeter responses to the quark- and gluon-initiated jets.
One uncertainty component describes the correction for the energy leakage beyond the calorimeter (`punch-through' effect).
All JES uncertainties are treated as bin-to-bin correlated and independent of each other.
 
A reduced set of uncertainties, which combines the uncertainties of the in situ methods into six components with a small loss of correlation information, is used in this measurement.
The JES uncertainties are propagated into the measurements in the same way as done for electron uncertainties.

The uncertainty that accounts for the difference in JVF requirement efficiency between data and MC simulation
is evaluated by varying the nominal JVF requirement in MC simulation to represent a few percent change in efficiency~\cite{ATLAS-CONF-2013-083}.
The unfolding transfer matrix and the matching efficiencies are re-derived, and the results of varying the JVF requirement are propagated to the unfolded data.
The deviations from the nominal results are used as the systematic uncertainty.
 
Pile-up jets are effectively suppressed   by the selection requirements. The jet yields in events with low $\langle\mu\rangle$ and  high $\langle\mu\rangle$  are compared with the jet yields in events without any requirements on $\langle\mu\rangle$.
These jet yields agree with each other within the statistical uncertainties. The same result is achieved by comparing the jet yields in events that have low or high numbers of reconstructed primary vertices with the jet yields in events from the nominal selection.
Consequently, no additional pile-up uncertainty is introduced.

The jet energy resolution (JER) uncertainty accounts for the mis-modelling of the detector jet energy resolution by the MC simulation.
To evaluate the JER uncertainty in the measured \Zjets{} \xss{}, the energy of each jet in MC simulation is
smeared by a random number taken from a Gaussian distribution
with a mean value of one and a width equal to the quadratic difference between the varied resolution and the nominal resolution~\cite{PERF-2011-04}.
The smearing is repeated 100 times and then averaged. The transfer matrix determined from the averaged
smearing is used for unfolding.
The result is compared with the nominal measurement and the symmetrised difference is used as the JER uncertainty.

The uncertainty that accounts for the mis-modelling of the \medium{} jet quality criteria is evaluated using jets, selected with the \loose{} and \tight{} criteria.
The data-to-MC ratios of the reconstructed \Zjets{} distributions, obtained with different jet quality criteria, is compared with the nominal.
An uncertainty of 1\%, which takes the observed differences into account, is assigned to the measured \Zjets{} \xs{} in all bins of \absyj{} and \ptj{}.
\subsection{Background uncertainties}
The uncertainties in each background estimation
are propagated to the measured \Zjets{} \xss{}.
 
The data contamination by the $\Zboson{} \rightarrow \tau \tau$, diboson, \ttbar{} and single-top-quark backgrounds is estimated using  simulated spectra scaled to  the corresponding total \xss. Each of these background \xss{} has an  uncertainty.
The normalisation of each background is independently varied up and down  by its  uncertainty and  propagated to the final result.
The MC simulation of the dominant \ttbar{} background  describes  the shapes of the  jet \ptj{} and \yj{} distributions in data to within a few percent~\cite{TOPQ-2013-04}, such that  possible shape mis-modellings  of the jet kinematics in \ttbar{} events  are covered by the uncertainty in the total \ttbar{} \xs.
The  shape mis-modellings in other backgrounds have negligible effect on the final results.
Therefore, no dedicated uncertainties due to the background  shape mis-modelling  are assigned.

The uncertainties in the combined multijet and \Wjets{} background arise from assumptions about the template shape and normalisation.
The shape of the template depends on the control region selection and the control region contaminations by the other backgrounds.
The template normalisation depends on the \mee{} range, used to fit the template to the measured \Zjets{} events,
due to different amounts of background contamination in the tails of the \mee{} distribution.
 
To evaluate the template shape uncertainty due to the control region selection, a different set of electron identification criteria is used to derive a modified template.
The selection requires two reconstructed electrons with at least one electron that satisfies the \loose{} identification criteria, but not the \medium{} ones.
The difference between the nominal and modified templates is used to create a symmetric template to provide up and down variations of this systematic uncertainty.
 
To estimate the template shape uncertainty due to the control region contaminations by the other backgrounds,
the $\Zboson{} \rightarrow \tau \tau$, diboson, \ttbar{} and single-top-quark \xss{} are varied within their uncertainties.
The dominant change in the template shape is due to \ttbar{} \xs{} variation,
while the contributions from the variation of the other background \xss{}  are small.
The templates varied within the \ttbar{} \xs{} uncertainties are used to propagate this uncertainty into the measurement.

The uncertainty in the multijet and \Wjets{} background template normalisation to the measured \Zjets{} events is evaluated
by fitting the template in the $66\GeV<\mee<140\GeV$ and $60\GeV<\mee<116\GeV$ regions,
excluding the bins under the \ZBoson{} peak within the $80\GeV<\mee<100\GeV$ region, and
in the $60\GeV<\mee<140\GeV$ region, excluding the bins within the $70\GeV<\mee<110\GeV$.
As a result, the normalisation varies up and down depending on the number of background events in both tails of the \mee{} distribution.
The templates with the largest change in the normalisation are used to propagate this uncertainty into the measurement.
 
The data unfolding is repeated for each systematic variation of the backgrounds.
The differences relative to the nominal \Zjets{} \xs{} are used as the systematic uncertainties.
\subsection{Unfolding uncertainty\label{subsec:unfunc}}
The accuracy of the unfolding procedure depends on the quality of the description of the measured spectrum in the MC simulation used to build the unfolding matrix.
Two effects are considered in order to estimate the influence of MC modelling on the unfolding results:   the shape of the particle-level spectrum   and
the  parton shower description.
 
The impact of the particle-level shape mis-modelling on the unfolding is estimated using a data-driven closure test.
For this test, the particle-level (\absyj{},~\ptj{}) distribution in \sherpa{} is reweighted using the transfer matrix,
such that the shape of the matched reconstructed  (\absyj{},~\ptj{}) distribution agrees with the measured spectrum corrected for the matching efficiency.
The reweighted reconstructed (\absyj{},~\ptj{}) distributions are then unfolded using the nominal \sherpa{} transfer matrix.
The results are compared with the reweighted particle-level (\absyj{},~\ptj{}) spectrum and the relative differences  are assigned as the uncertainty.
 
The impact of the differences in the parton shower description between \sherpa{} and \alpgenpythia{} on the unfolding results is estimated using the following test.
The \alpgenpythia{} particle-level (\absyj{},~\ptj{}) spectrum  is reweighted using the \alpgenpythia{} transfer matrix, such that its reconstruction-level distribution agrees with the one  in \sherpa{}.
The original reconstructed (\absyj{},~\ptj{}) distribution in \sherpa{} is then unfolded using the reweighted \alpgenpythia{} transfer matrix.
The results are compared with the original particle-level (\absyj{},~\ptj{}) spectrum in \sherpa{} and the differences are assigned as the uncertainty.
 
Both unfolding uncertainties are symmetrised at the \xs{} level.
 
\subsection{Reduction of statistical fluctuations in systematic uncertainties}
 
The systematic uncertainties suffer from fluctuations of a statistical nature.
 
The statistical components in the electron and jet uncertainties are estimated using toy MC simulations with 100 pseudo-experiments.
Each \Zjets{} event in the systematically varied configurations is reweighted by a random number taken from a Poisson distribution with a mean value of one.
As a result, 100 replicas of transfer matrix and matching efficiencies are created for a given systematic uncertainty variation, and are used to unfold the data.
The replicas of unfolded spectra are then divided by the nominal \Zjets{} distributions to create an ensemble of systematic uncertainty spectra.
The statistical component in the systematic uncertainties is calculated as the RMS across all replicas in an ensemble.
 
The pseudo-experiments are not performed for the JER systematic uncertainty.
The statistical errors in the JER systematic uncertainty are calculated,
considering the unfolded data in the nominal and JER varied configurations to be independent of each other.

Each component of the unfolding uncertainty is derived using 100 pseudo-experiments to calculate the statistical error.
 
To reduce the statistical fluctuations, the bins are combined iteratively starting from both the right and
left sides of each systematic uncertainty spectrum until their significance satisfies $\sigma>1.5$. The result with the most bins remaining is used as the systematic uncertainty.
A Gaussian kernel is then applied to regain the fine binning and smooth out any additional statistical fluctuations.
 
If up and down systematic variations within a bin result in  uncertainties with the same sign, then the smaller uncertainty is set to zero.

\subsection{Statistical uncertainties}
Statistical uncertainties are derived using toy MC simulations with 100 pseudo-experiments performed in both data and MC simulation.
The data portion of the statistical uncertainty is evaluated by unfolding the replicas of the data using the nominal transfer matrix and matching efficiencies.
The MC portion is calculated using the replicas of the transfer matrix and matching efficiencies to unfold the nominal data.
To calculate the total statistical uncertainty in the measurement, the \Zjets{} distributions, obtained from pseudo-experiments drawn from the data yields,
are unfolded using the transfer matrices and efficiency corrections, calculated using pseudo-experiments in the MC simulation.
The covariance matrices between bins of the measurement are computed using the unfolded results.
The total statistical uncertainties are calculated using the diagonal elements of the covariance matrices.

\subsection{Summary of experimental uncertainties}

The \Zjets{} \xs{} measurement has 39 systematic uncertainty components.
All systematic uncertainties are treated as being uncorrelated with each other and fully correlated among \absyj{} and \ptj{} bins.
 
The systematic uncertainties in the electron energy scale, electron energy resolution, and electron trigger, reconstruction and identification efficiencies are found to be below 1\%.
 
The JES in situ methods uncertainty is 2\%--5\% in most bins of the measurement.
The $\eta$-intercalibration uncertainty is below 1\% in the $\absyj<1.0$ and $\ptj<200\GeV$ regions,
but it increases with \absyj{}, reaching 6\%--14\% in the most forward rapidity bins.
The $\eta$-intercalibration uncertainty is below 1.5\% for jets with $\ptj>200\GeV$.
The flavour-based JES uncertainties are below 3\%.
The pile-up components of the JES uncertainty are 0.5\%--1.5\%.
Other components of the JES uncertainty are below 0.2\%.
 
The JVF uncertainty is below 1\%.  
 
The JER is the dominant source of uncertainty in the \Zjets{} \xs{} in the $25\GeV <  \ptj < 50\GeV$ region with a 3\%--10\% contribution.
In the $50\GeV<\ptj<100\GeV$ region the JER uncertainty is 1\%--3\%, and below 1\% for jets with higher transverse momenta.
 
The jet quality uncertainty is set constant at 1\%.

The unfolding uncertainty due to the shape of the particle-level spectrum
is 2\%--5\% in the first \ptj{} bin, $25\GeV <  \ptj < 50\GeV$.
In the $50\GeV <  \ptj < 200\GeV$ region, this uncertainty is about 1.5\% for central jets below $\absyj=2$, while for forward jets this uncertainty increases to 5\%.
In the $\ptj > 200\GeV$ region, this uncertainty is below 1.5\%.
The unfolding uncertainty due to the  parton shower description is 0.7\% in the $400\GeV<\ptj<1050\GeV$ region,
while for jets with smaller transverse momenta this uncertainty is negligible.

The \ttbar{} background uncertainty is 0.02\%--0.6\% in all bins of the measurement.
The $\Zboson{} \rightarrow \tau \tau$, diboson and single-top-quark background uncertainties are below 0.05\%.
 
The multijet and \Wjets{} background uncertainty is 0.1\%--1.2\% depending on \absyj{} and \ptj{}.
The uncertainty in the background template normalisation is asymmetric due to different background contributions in the tails of the \mee{} distribution
in the background normalisation evaluation procedure.
This uncertainty is $_{-0.4}^{+0.1}$\% in the low \ptj{} bins, increasing to $_{-1.2}^{+0.4}$\% in the high \ptj{} bins.
The uncertainty in the multijet and \Wjets{} background control region selection increases from 0.03\% to 0.6\% as a function of \ptj{}.
The contribution of the \ttbar{} \xs{} variation to the multijet and \Wjets{} background uncertainty is below 0.1\%.

The statistical uncertainties are 0.5\%--4\% in the $\ptj<100\GeV$ region, 2\%--14\% in the $100\GeV<\ptj<300\GeV$ region,
8\%--39\% in the $300\GeV<\ptj<400\GeV$ region and 11\%--18\% in the last \ptj{} bin, $400\GeV<\ptj<1050\GeV$.
The smallest statistical uncertainty corresponds to central rapidity regions, while the largest uncertainty corresponds to forward rapidity regions.
 
\begin{figure}[h]
\centering
\subfloat[$25\GeV < \ptj < 50\GeV$]      {\includegraphics[width=0.49\linewidth]{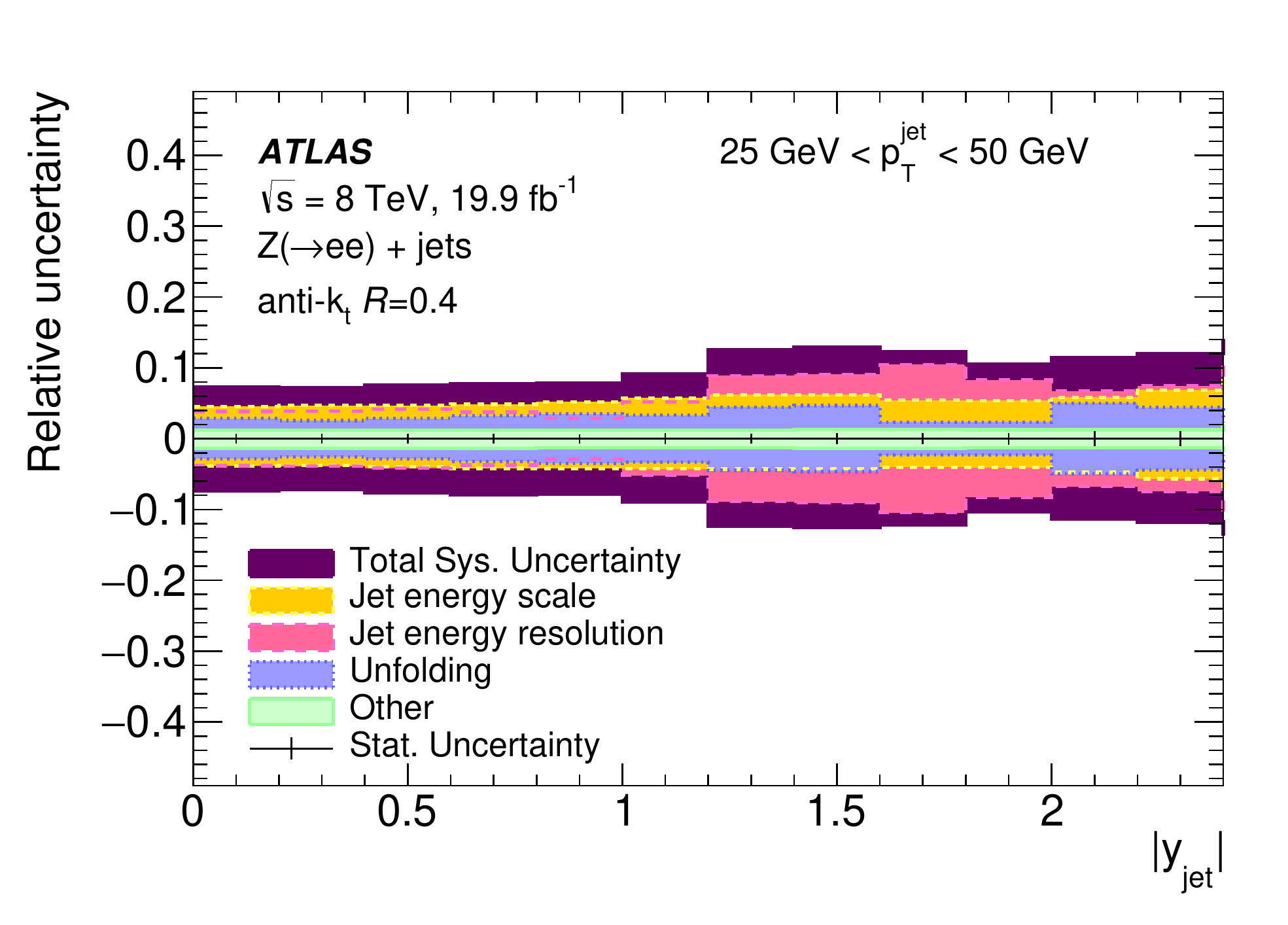}}
\subfloat[$50\GeV < \ptj < 100\GeV$]    {\includegraphics[width=0.49\linewidth]{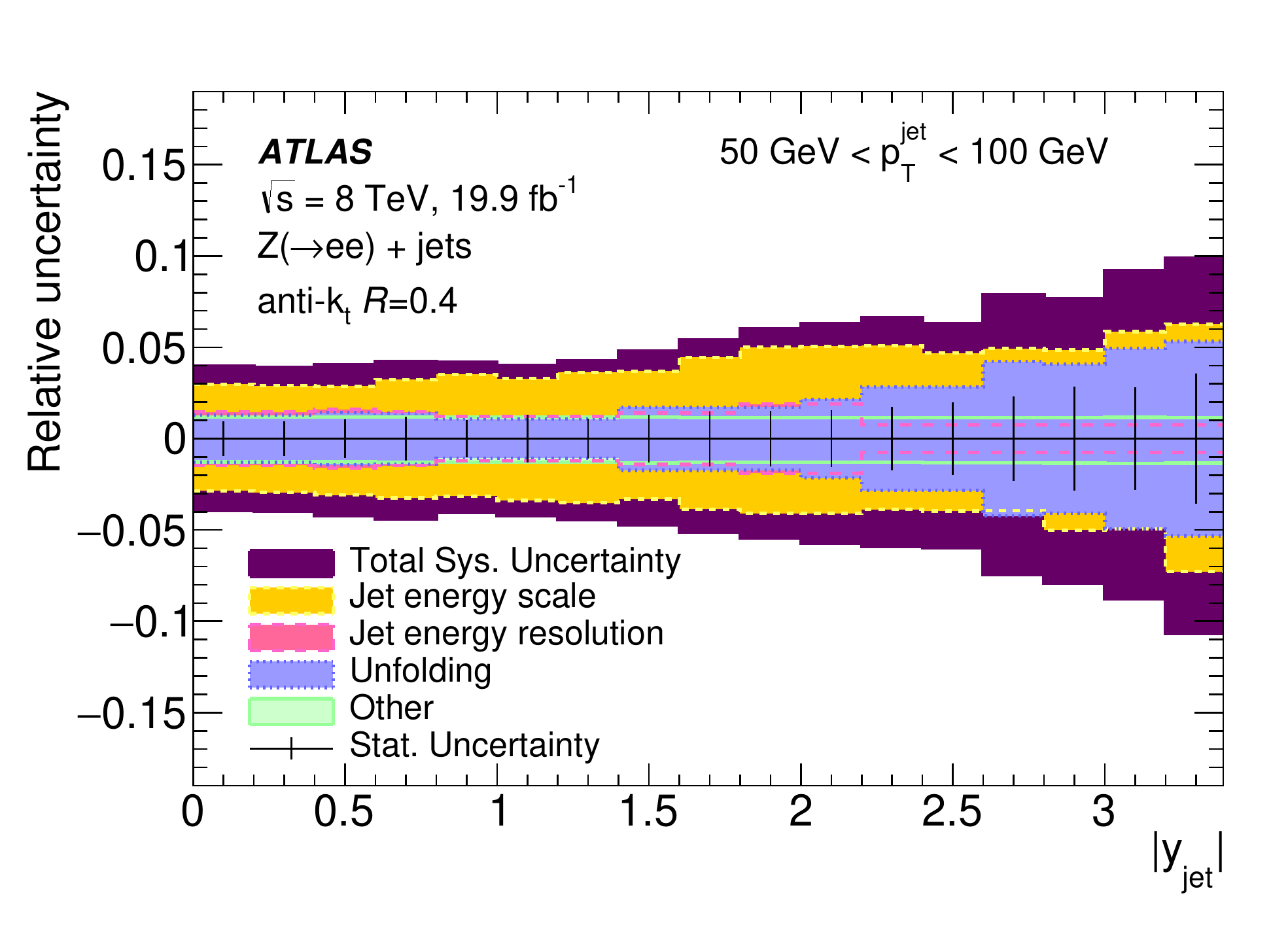}} \\
\subfloat[$100\GeV < \ptj < 200\GeV$]  {\includegraphics[width=0.49\linewidth]{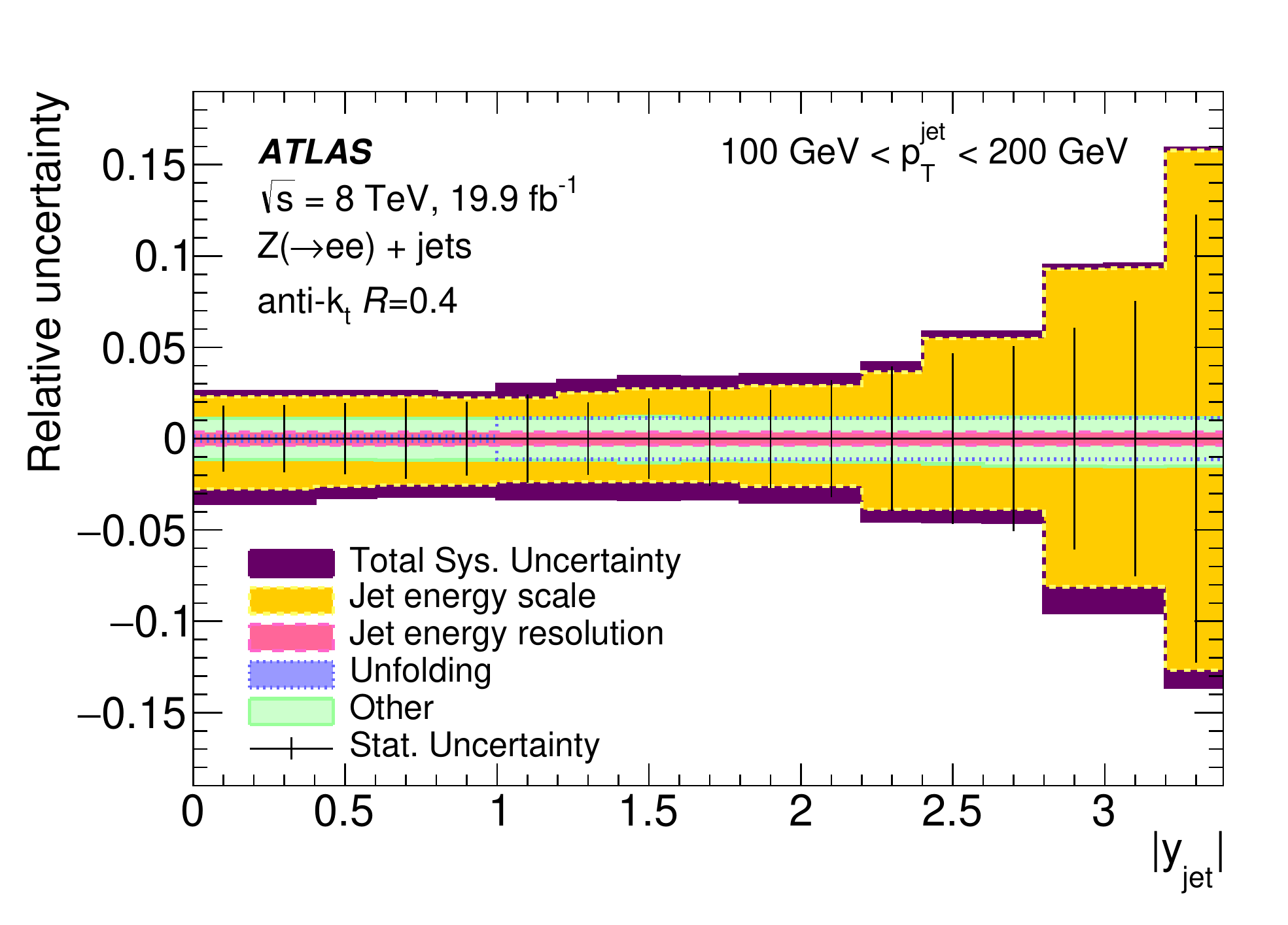}}
\subfloat[$200\GeV < \ptj < 300\GeV$]  {\includegraphics[width=0.49\linewidth]{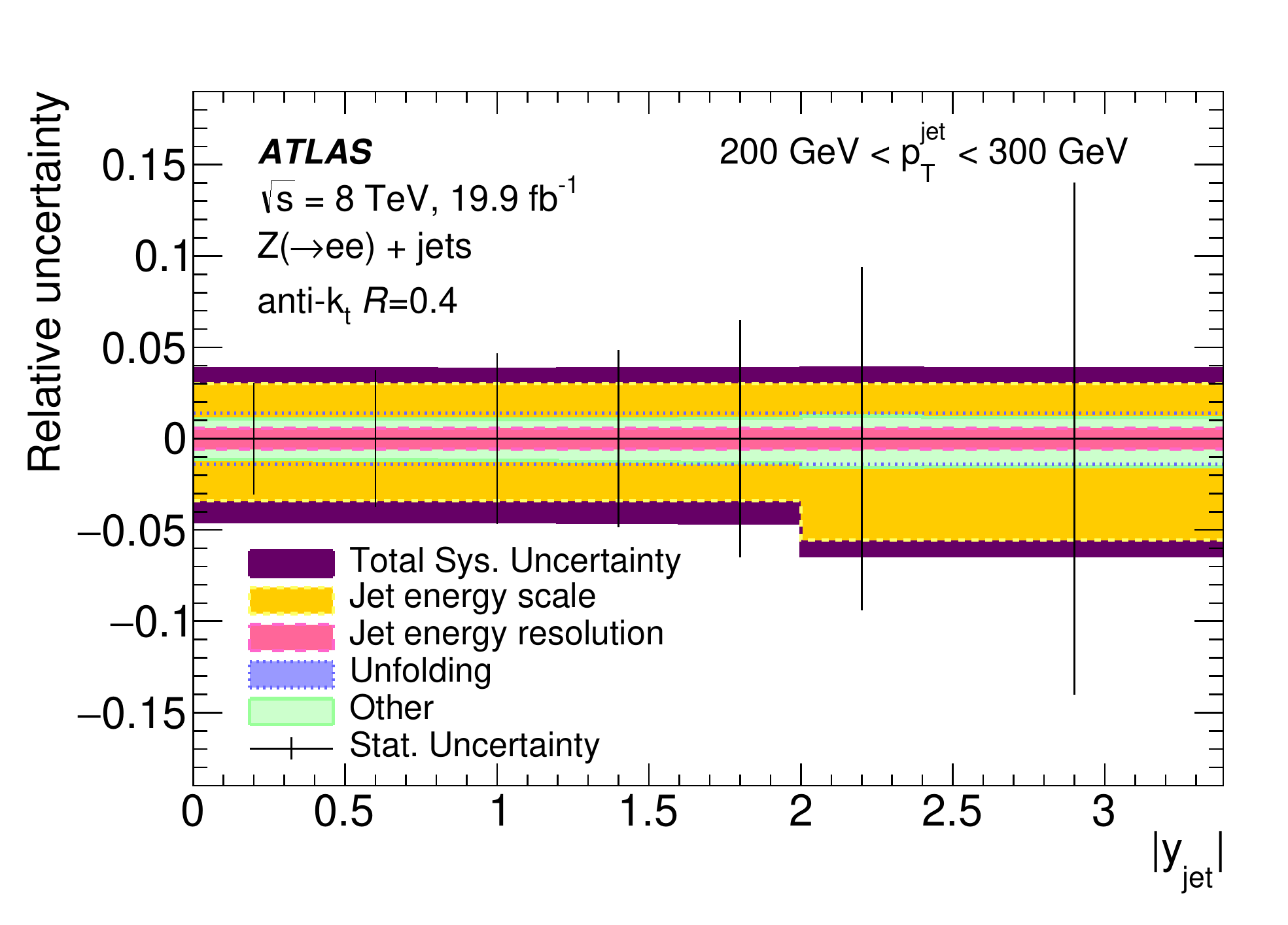}} \\
\subfloat[$300\GeV < \ptj < 400\GeV$]  {\includegraphics[width=0.49\linewidth]{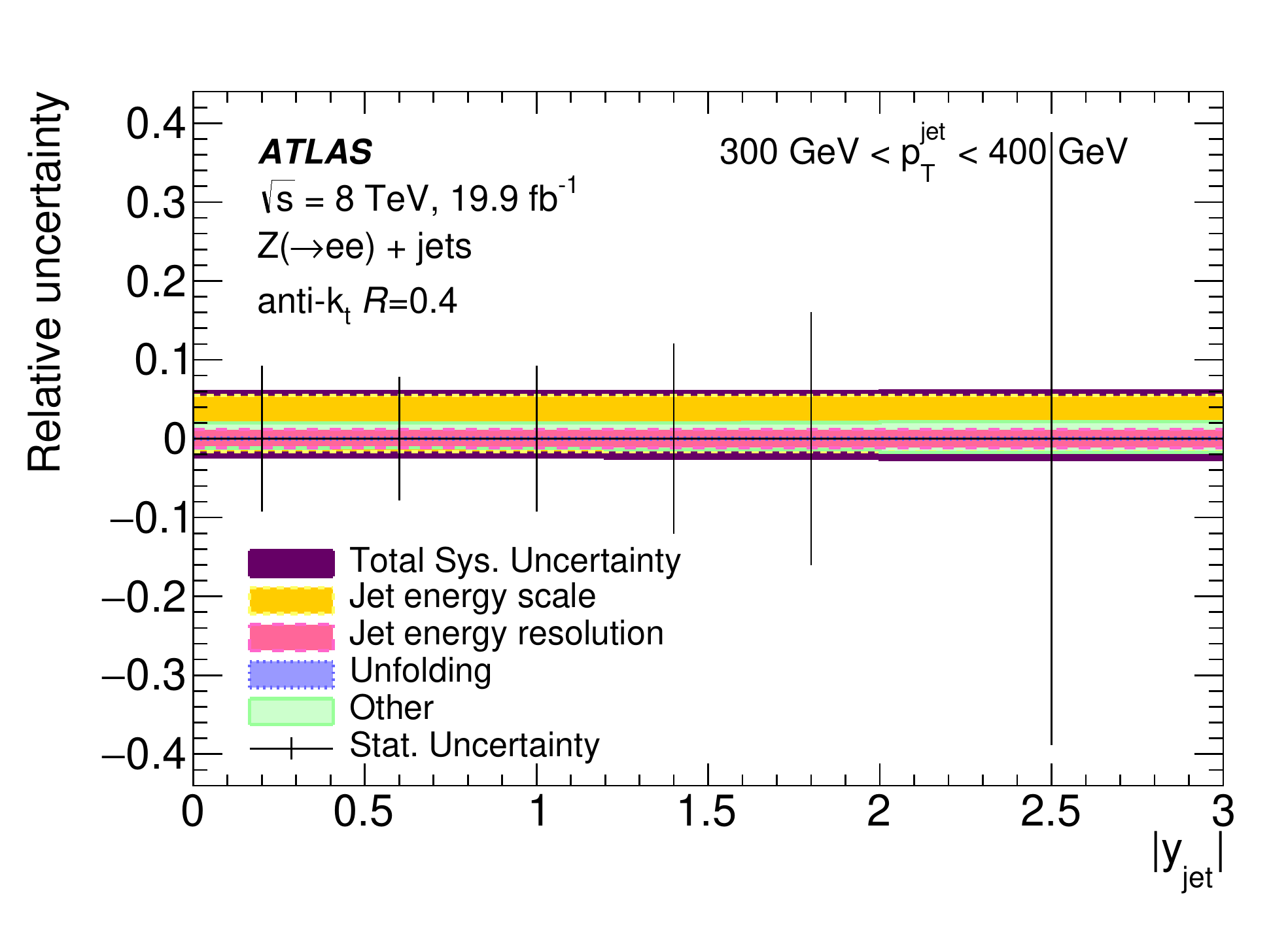}}
\subfloat[$400\GeV < \ptj < 1050\GeV$]{\includegraphics[width=0.49\linewidth]{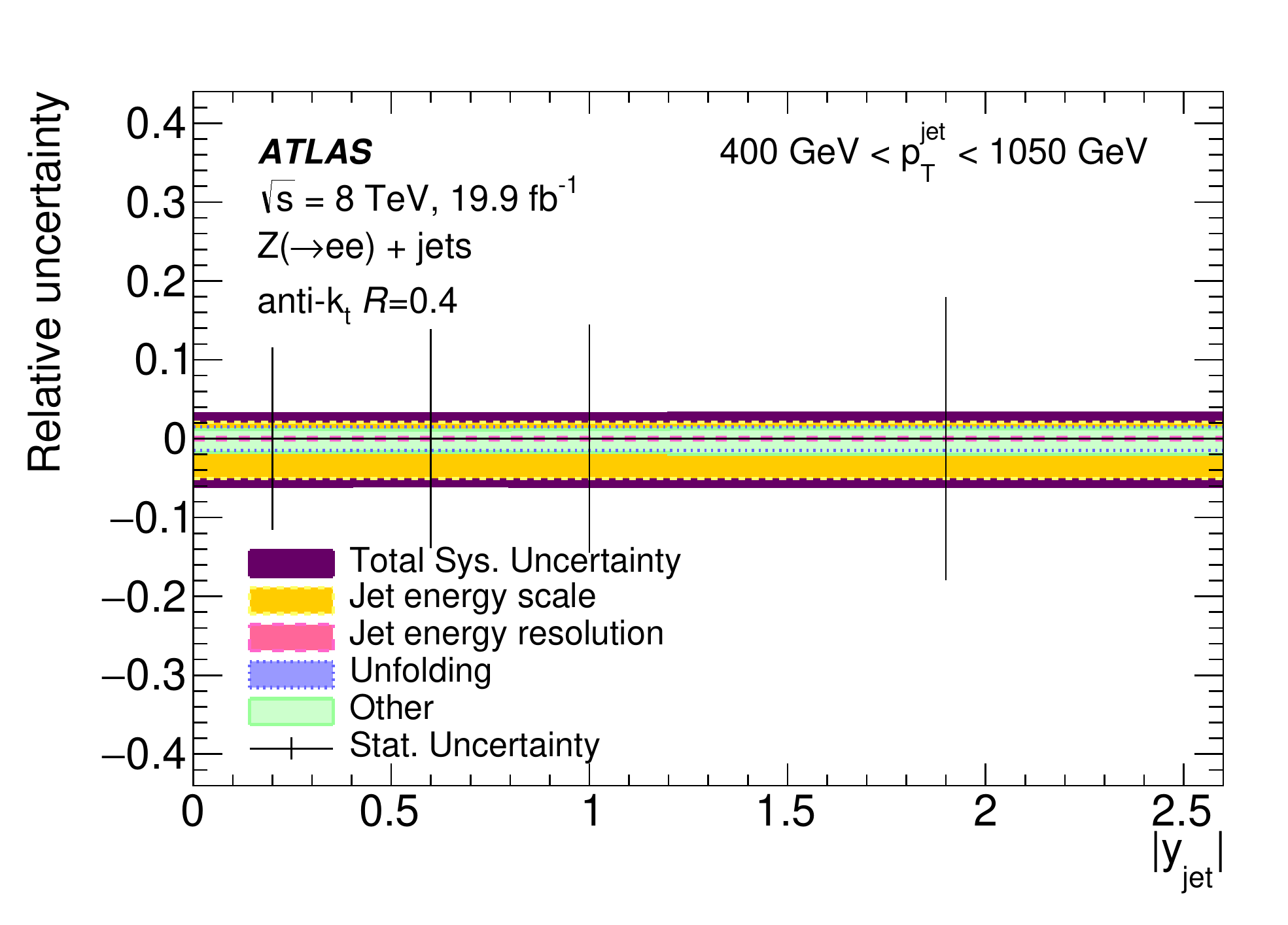}} \\
 
\caption[Uncertainty]{
Experimental uncertainties in the measured double-differential \Zjets{} production \xs{} as a function of \absyj{} in \ptj{} bins.
The jet energy scale, jet energy resolution, unfolding, `other' and total systematic uncertainties are shown with different colours overlaid.
The jet energy scale uncertainty is the sum in quadrature of the jet energy scale uncertainty components.
The unfolding uncertainty is the  sum in quadrature of two unfolding uncertainties.
The `other' systematic uncertainty is the sum in quadrature of the electron uncertainties, background uncertainties, JVF and jet quality uncertainties.
The total systematic uncertainty is the sum in quadrature of all systematic uncertainty components except for the luminosity uncertainty of 1.9\%.
The total statistical uncertainties are shown with vertical error bars.
}
\label{fig:uncall_summary}
\end{figure}

The experimental uncertainties are shown in~\Cref{fig:uncall_summary}.
The largest total systematic uncertainty of 7\%--12\% is in the $25\GeV<\ptj<50\GeV$ region, where the uncertainty increases from central rapidity jets to the forward rapidity jets,
and up to 15\% for the forward rapidity jets in the $100\GeV<\ptj<200\GeV$ region.
The total systematic uncertainty decreases with increasing \ptj{}. In the $400\GeV<\ptj<1050\GeV$ region the total systematic uncertainty is 2\%--5\%.
The luminosity uncertainty of 1.9\% is not shown and not included in the total uncertainty and its components.

\FloatBarrier
 
\section{Fixed-order predictions and theoretical uncertainties\label{sec:theory}}
 
\subsection{Fixed-order calculations}
 
Theoretical \Zjets{} predictions at NLO are calculated using MCFM~\cite{MCFM} interfaced to APPLgrid~\cite{Applgrid} for fast convolution with PDFs.
The renormalisation and factorisation scales, \mur{} and \muf{}, are set to
\begin{linenomath}
\begin{equation*}
\mur=\muf=\frac{\sqrt{m_{ee}^2 + p_{\textrm{T}, Z}^2}+\sum p_\textrm{T, partons}}{2},
\end{equation*}
\end{linenomath}
where \mee{} is the electron pair's invariant mass, $p_{\textrm{T}, Z}$ is the transverse momentum
of the \ZBoson{} and $\sum p_\textrm{T, partons}$ is the sum of the transverse momenta of the partons.
 
The NLO \Zjets{} predictions are obtained using the CT14 NLO~\cite{ct14}, NNPDF3.1~\cite{nnpf30},
JR14 NLO~\cite{jr14}, HERAPDF2.0~\cite{herapdf20}, MMHT2014~\cite{mmht2014}, 
ABMP16~\cite{abmp16} and ATLAS-epWZ16~\cite{atlasepwz16} PDF sets.
The PDFs are determined by various groups using the experimental data and are provided with the uncertainties.
The PDF uncertainties in the \Zjets{} \xss{} are calculated at the 68\% confidence level according to the prescription recommended by the PDF4LHC group~\cite{pdfforlhcrec}.

The following variations of factorisation and renormalisation scales are performed to assess the uncertainty due to missing higher-order terms:
$\{\mur/2, \muf\}$, $\{2\mur, \muf\}$, $\{\mur, \muf/2\}$, $\{\mur, 2\muf\}$, $\{\mur/2, \muf/2\}$, $\{2\mur, 2\muf\}$.
The envelope of the \xss{} calculated with different scales is used as the uncertainty.

The uncertainty due to the strong coupling is estimated using additional PDF sets, calculated with $\alphas(m_Z^2)=0.116$ and $\alphas(m_Z^2)=0.120$.
The resulting uncertainty is scaled to the uncertainty of the world average $\alphas(m_Z^2)= 0.118 \pm 0.0012$, as recommended by the PDF4LHC group~\cite{pdfforlhcrec}.

The state-of-the-art NNLO \Zjets{} \xs{} is calculated by the authors of Ref.~\cite{zjet_nnlo} using NNLOJET~\cite{nnlojet}.
The NNLO predictions are convolved with the CT14 PDF. The renormalisation and factorisation scales are set similarly to those in NLO calculations.
 
\subsection{Non-perturbative correction}
The fixed-order predictions are obtained at the parton level.
Bringing fixed-order predictions to the particle level for comparisons with the measured \Zjets{} \xss{} requires  a non-perturbative correction~(NPC) that accounts for both  the hadronisation and underlying-event effects.

The NPCs are studied using several MC generators to account for differences in the modelling of hadronisation and the underlying event.
The studies are done using the leading-logarithm parton shower MC generators \npcpythiaev{} with the A14~\cite{ATL-PHYS-PUB-2014-021} underlying-event tune
and \npcherwigppv{} with the UE-EE5 tune~\cite{ueee5}, and the multi-leg matrix element MC generators \npcsherpav{} with the CT10 PDF, \npcsherpatv{} with NNPDF~2.3~\cite{nnpf23}
and \npcmadgraphv{}~\cite{madgraph}, supplemented with parton showers from \npcpythiaev{} with the A14 tune.
 
\begin{figure}[htb]
\centering
\subfloat[$25\GeV < \ptj < 50\GeV$]       {\includegraphics[width=0.49\linewidth]{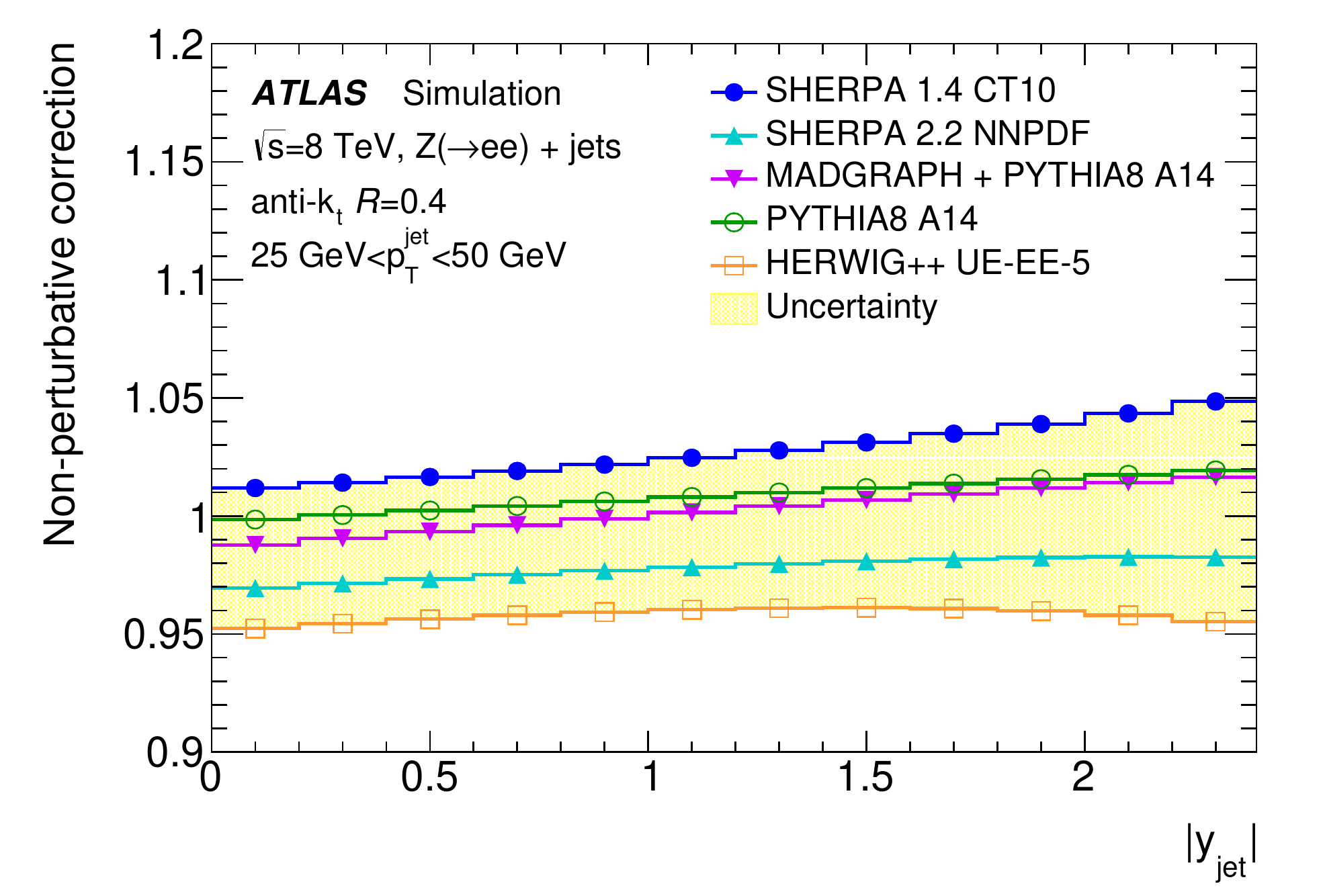}}
\subfloat[$50\GeV < \ptj < 100\GeV$]     {\includegraphics[width=0.49\linewidth]{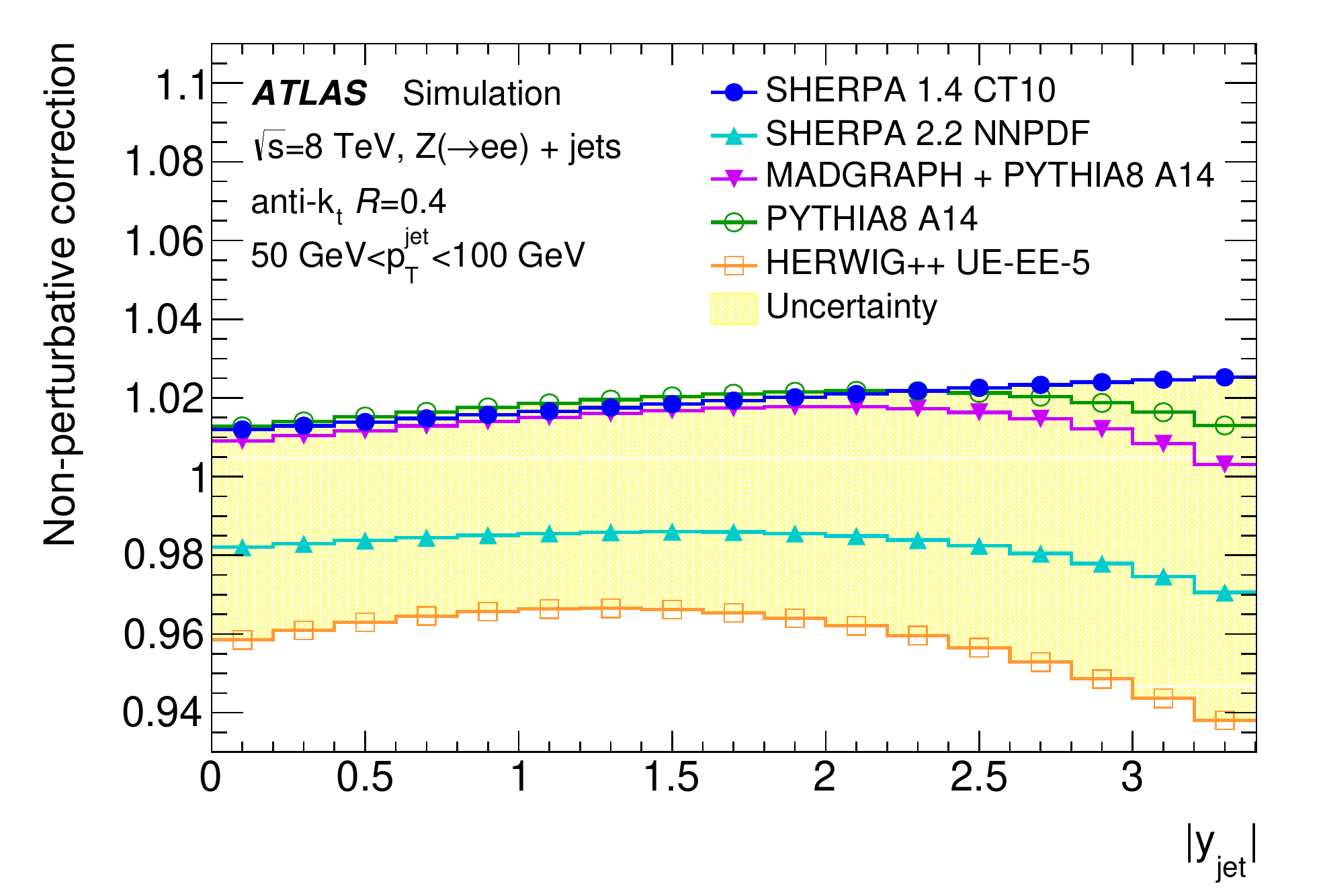}} \\
\subfloat[$100\GeV < \ptj < 200\GeV$]   {\includegraphics[width=0.49\linewidth]{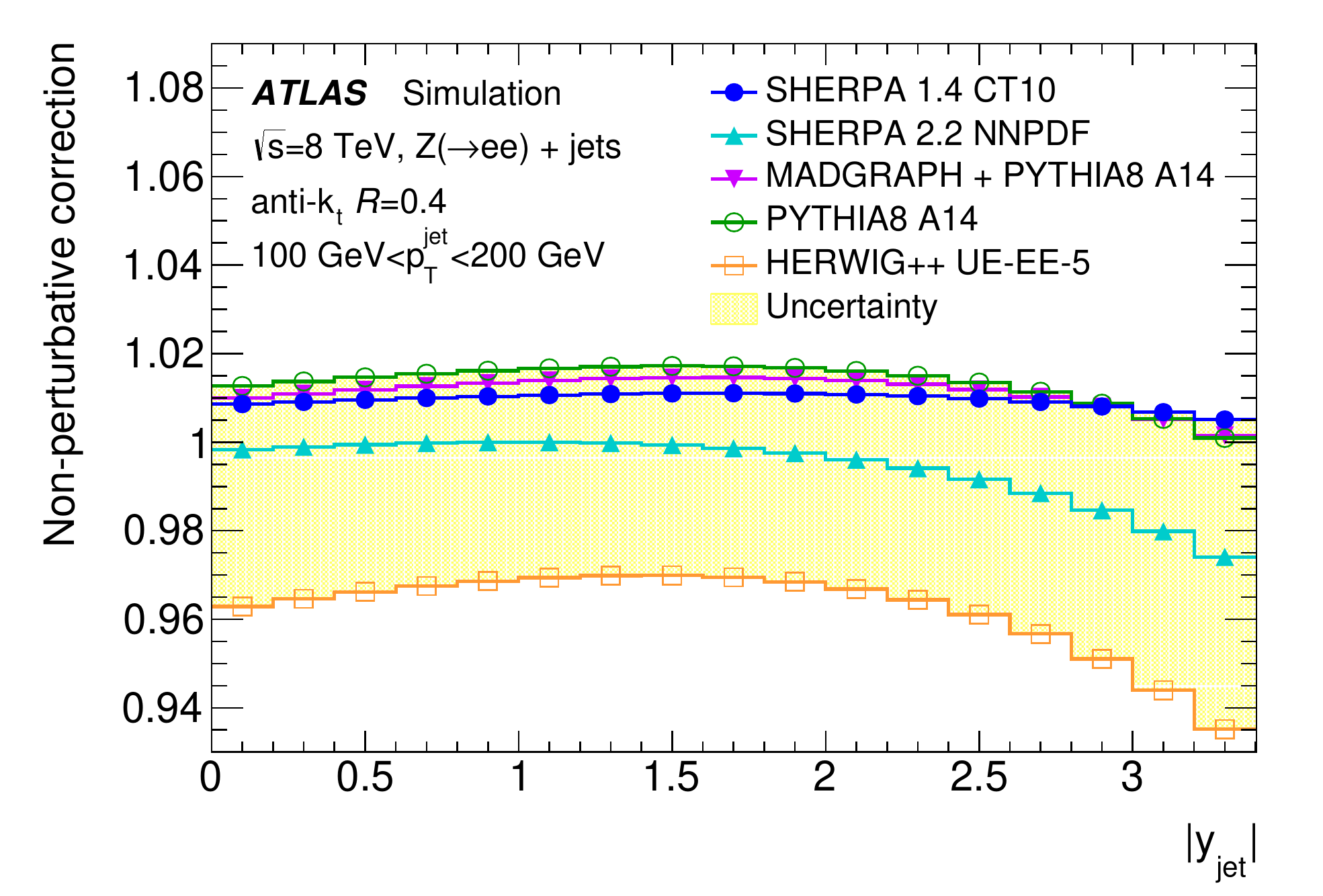}}
\subfloat[$200\GeV < \ptj < 300\GeV$]   {\includegraphics[width=0.49\linewidth]{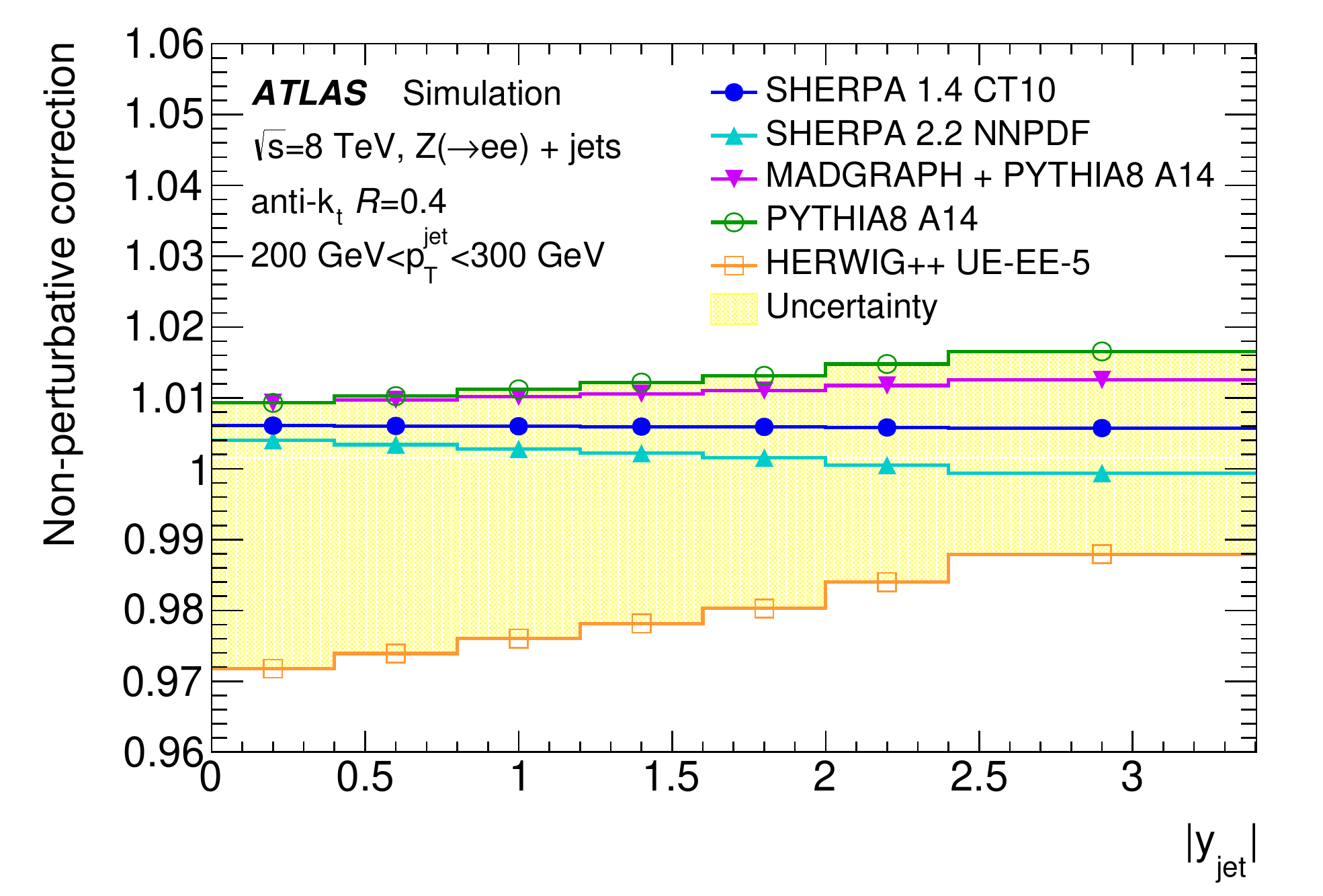}} \\
\subfloat[$300\GeV < \ptj < 400\GeV$]   {\includegraphics[width=0.49\linewidth]{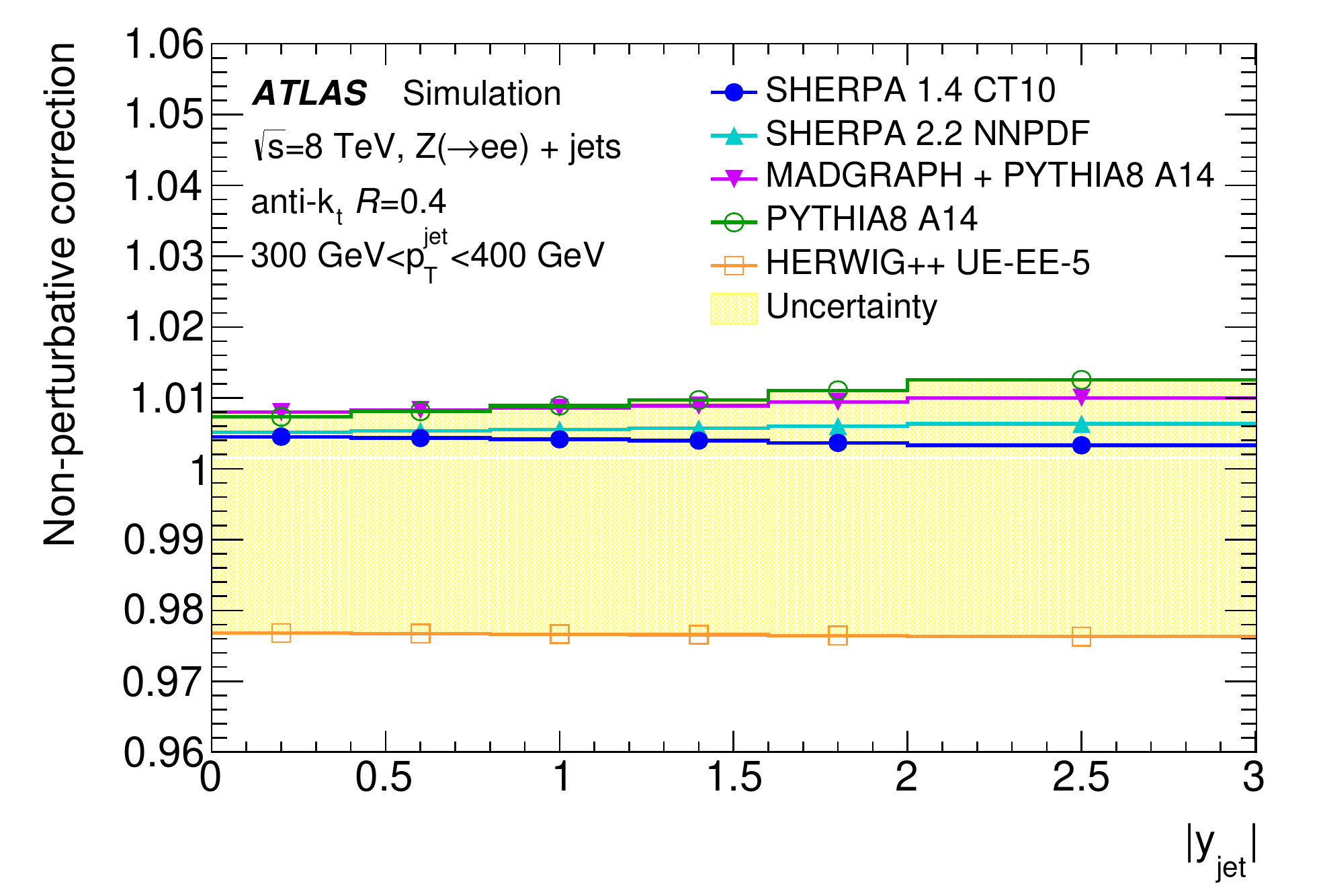}}
\subfloat[$400\GeV < \ptj < 1050\GeV$] {\includegraphics[width=0.49\linewidth]{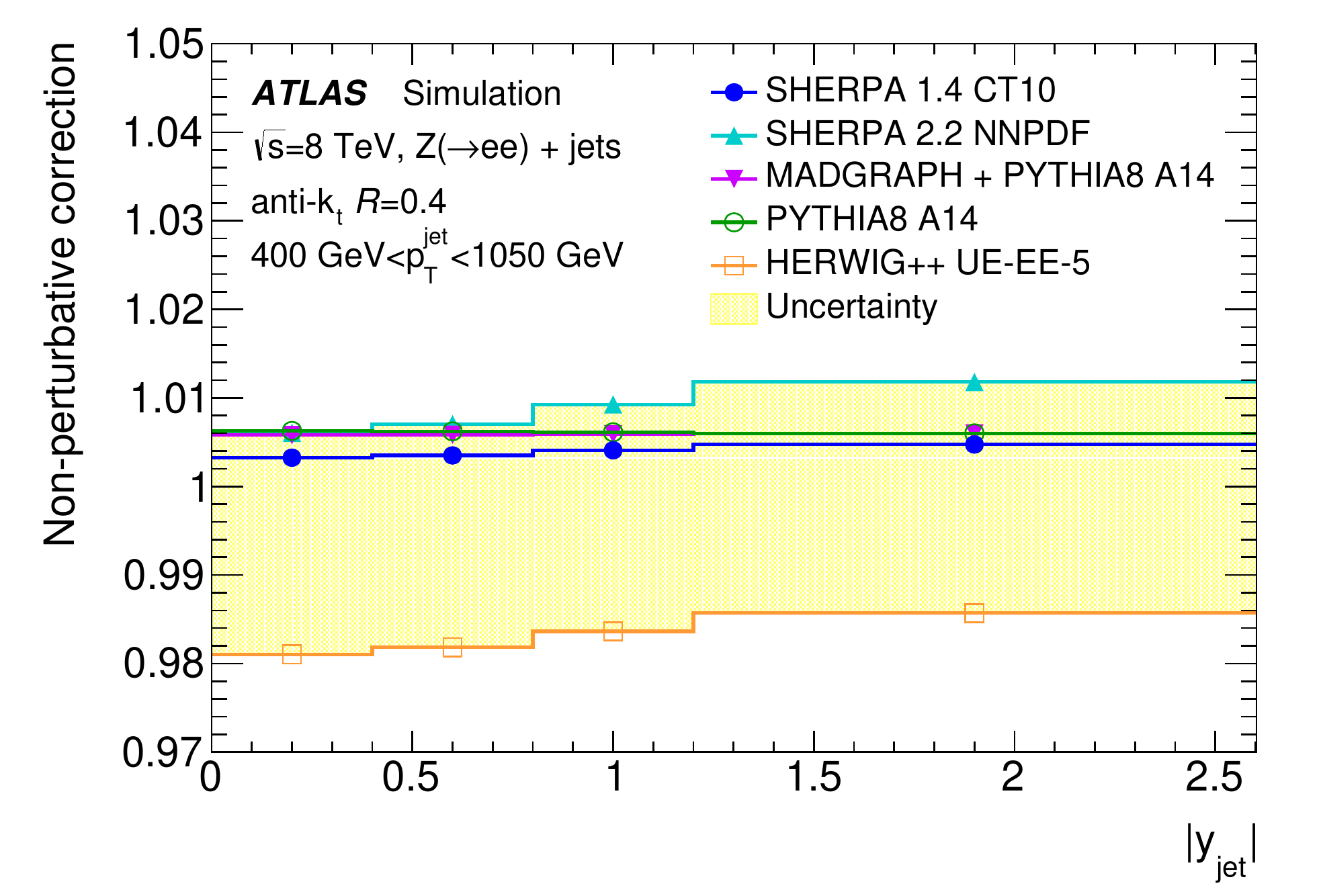}} \\
\caption[MCFM with NPC]{
The non-perturbative correction for the \Zjets{} production \xs{} as a function of \absyj{} in \ptj{} bins. The spread of predictions represents the uncertainty.}
\label{fig:npcfit_jeteta}
\end{figure}

The NPCs are calculated using the ratios of \Zjets{} \xss{} obtained at the particle level to those at the parton level.
The correction derived using \npcsherpav{} is the nominal one in this analysis.
The envelope of the non-perturbative corrections, calculated with other MC generators, is used as the systematic uncertainty.
The NPCs in different MC generators are shown in~\Cref{fig:npcfit_jeteta}.
The nominal correction for jets with low transverse momenta, $25\GeV<\ptj<50\GeV$, in the central rapidity regions, $\absyj<1.5$, is small,
but it increases to 5\% in the forward rapidity bins. The nominal correction for jets with higher transverse momenta is below 2\%.
These corrections together with uncertainties are provided in HEPData~\cite{hepdata}.

\subsection{QED radiation correction}
 
\begin{figure}[tb]
\centering
\subfloat[$25\GeV < \ptj < 50\GeV$]       {\includegraphics[width=0.49\linewidth]{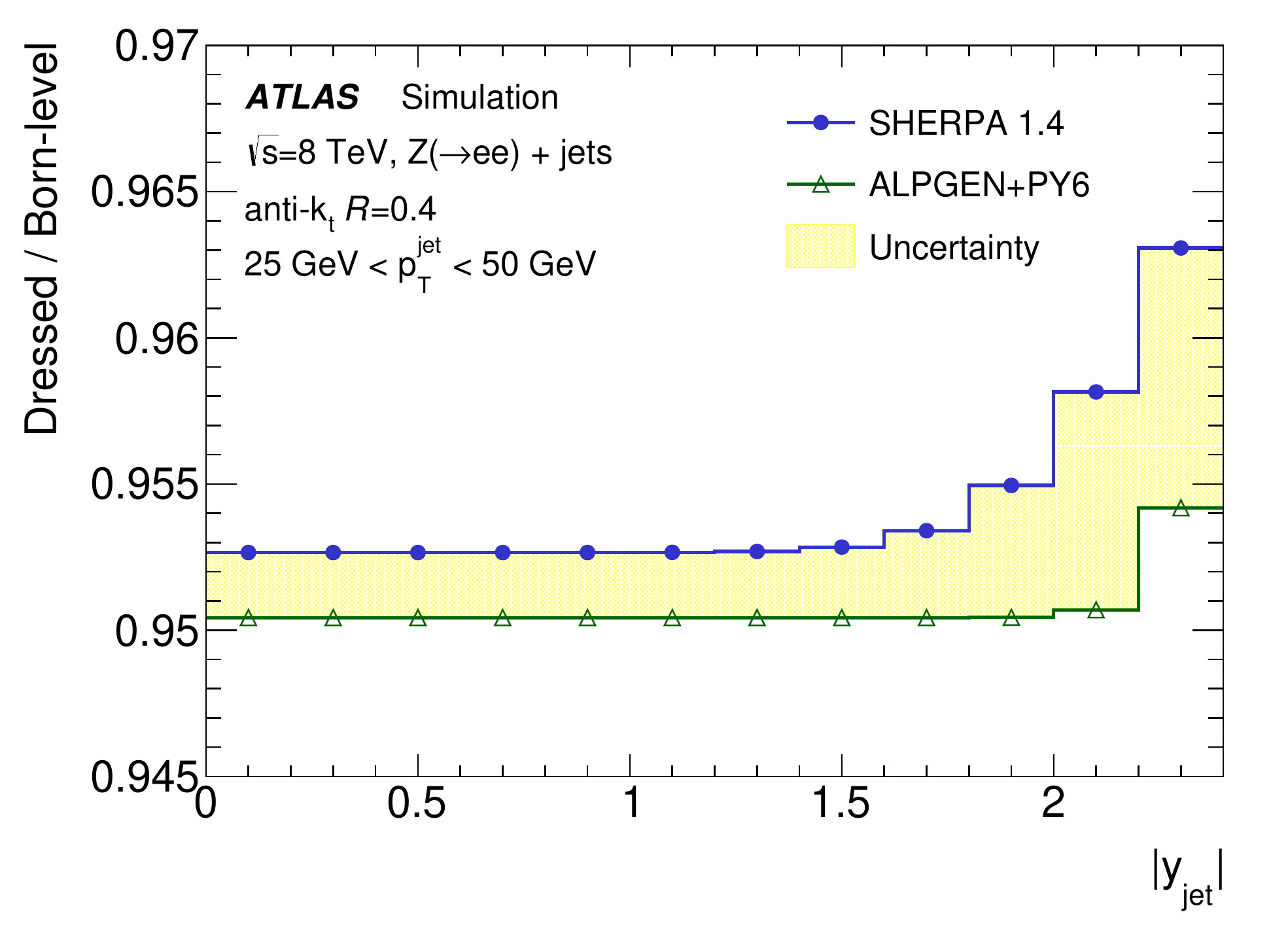} }
\subfloat[$50\GeV < \ptj < 100\GeV$]     {\includegraphics[width=0.49\linewidth]{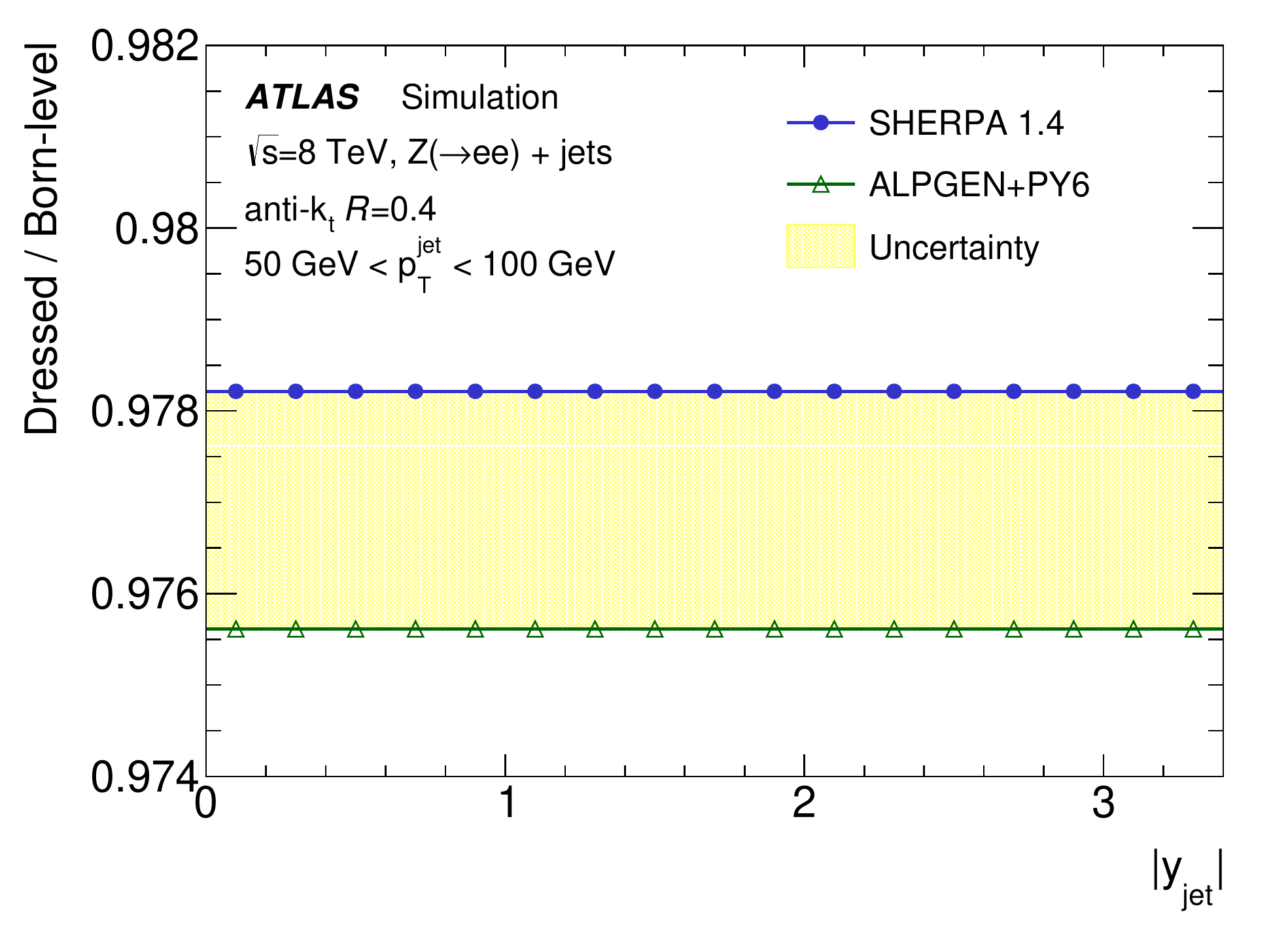} } \\
\subfloat[$100\GeV < \ptj < 200\GeV$]   {\includegraphics[width=0.49\linewidth]{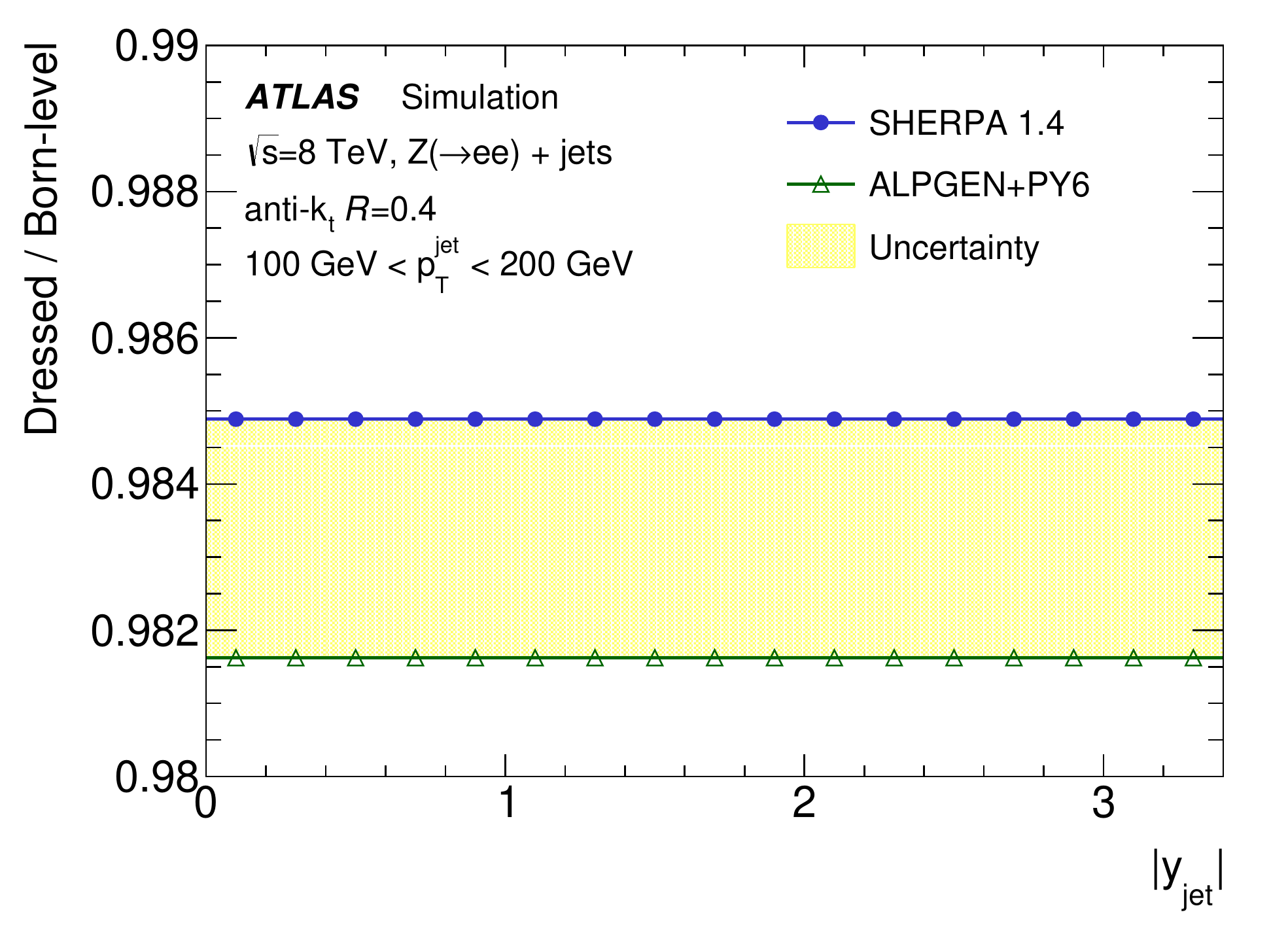} }
\subfloat[$200\GeV < \ptj < 300\GeV$]   {\includegraphics[width=0.49\linewidth]{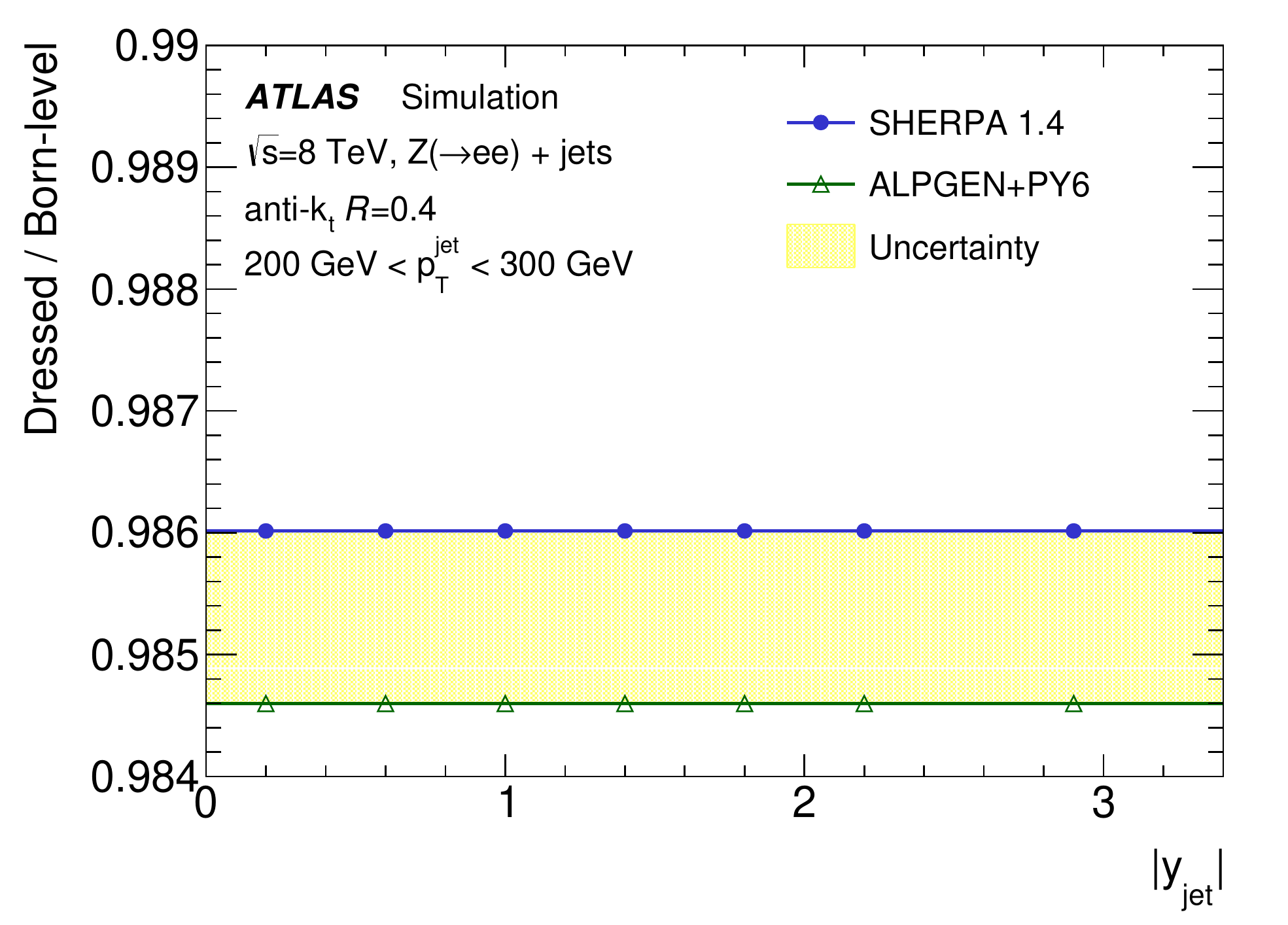} } \\
\subfloat[$300\GeV < \ptj < 400\GeV$]   {\includegraphics[width=0.49\linewidth]{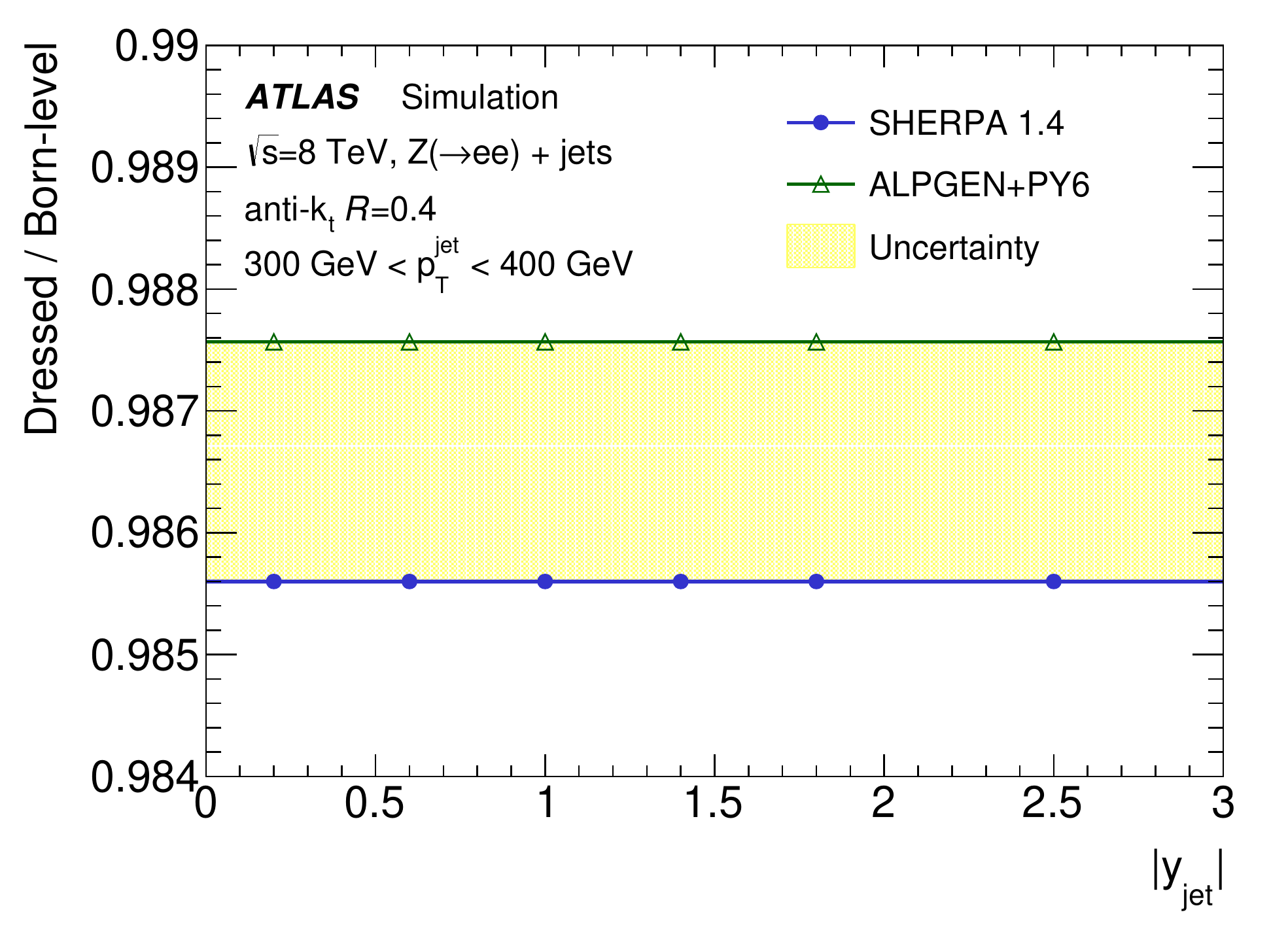} }
\subfloat[$400\GeV < \ptj < 1050\GeV$] {\includegraphics[width=0.49\linewidth]{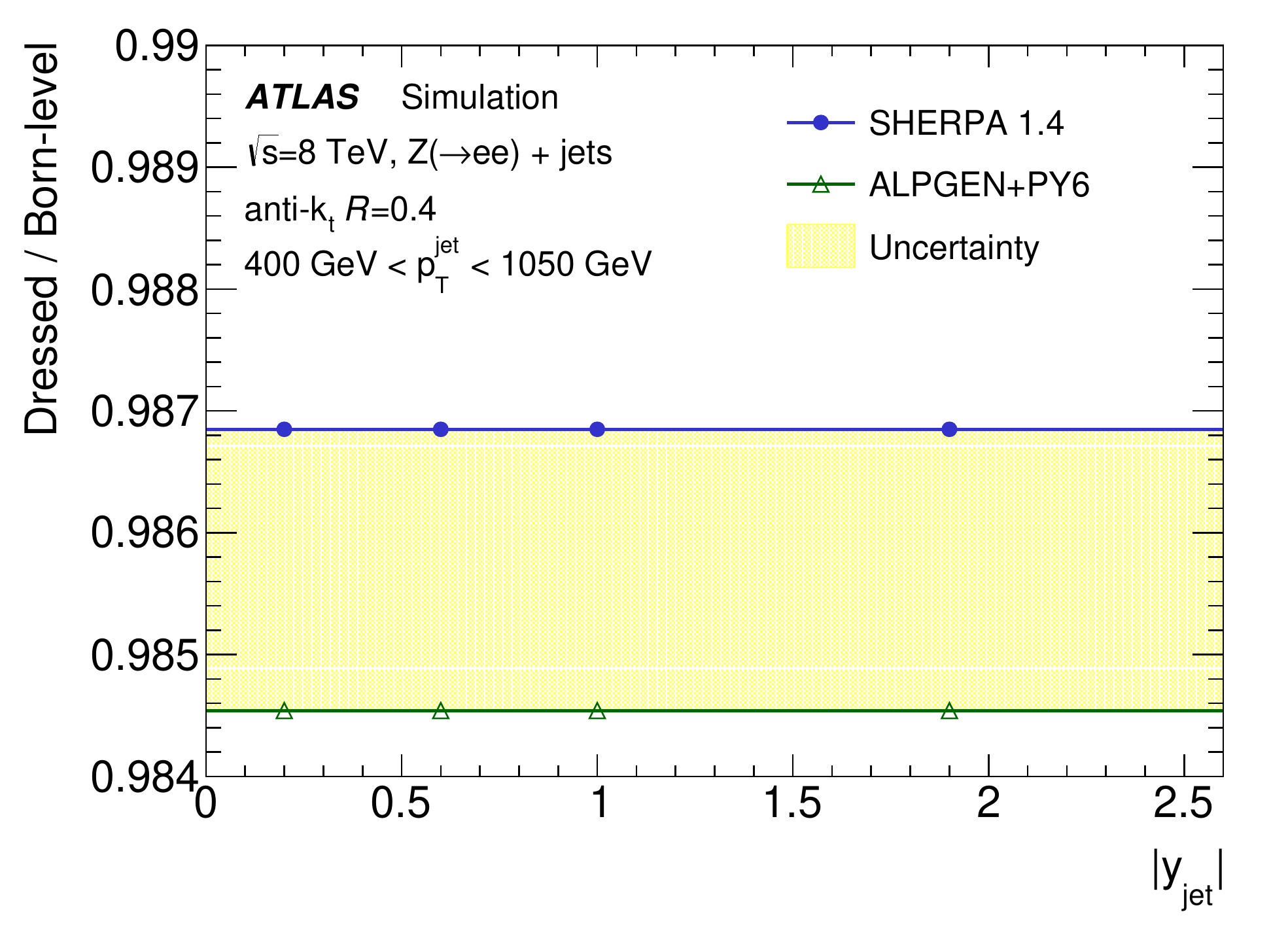} } \\
\caption[]{The correction for QED radiation effects for the \Zjets{} production \xs{}  as a function of \absyj{} in \ptj{} bins. The spread of predictions represents the uncertainty.}
\label{fig:qed_corr_fit}
\end{figure}

The fixed-order \Zjets{} \xs{} predictions must be corrected for the QED radiation in order to be compared with data.
The correction is determined as the ratio of two \Zjets{} \xss{}, one calculated using dressed electrons after QED final-state radiation (FSR),
with all photons clustered within a cone of $\Delta R = 0.1$ around the electron axis, and the other calculated using Born-level electrons at the lowest order in the electromagnetic coupling $\alpha_\text{QED}$ prior to QED FSR radiation.
The baseline correction is calculated using the \sherpa{} MC samples, while the correction calculated using  \alpgenpythia{} is used to estimate the uncertainty.
The uncertainty is calculated as the width of the envelope of corrections obtained with these two MC generators. The results are shown in~\Cref{fig:qed_corr_fit}.
The QED correction is largest in the $25\GeV<\ptj<50\GeV$ region. It is about 5\% for jets in the central absolute rapidity regions.
In the $\ptj>50\GeV$ regions the QED correction is 1.5\%--2.5\%, decreasing as a function of jet transverse momentum.
The QED corrections calculated using \alpgenpythia{} are in good agreement with those from \sherpa{}.
These corrections together with uncertainties are provided in HEPData.

\subsection{Summary of theoretical uncertainties}
 
The total theoretical uncertainties are calculated as the sum in quadrature of the effects of the PDF, scale, and \alphas{} uncertainties,
and the uncertainties due to non-perturbative and QED radiation effects.
 
\begin{figure}[h]
\centering
\subfloat[$25\GeV < \ptj < 50\GeV$]       {\includegraphics[width=0.49\linewidth]{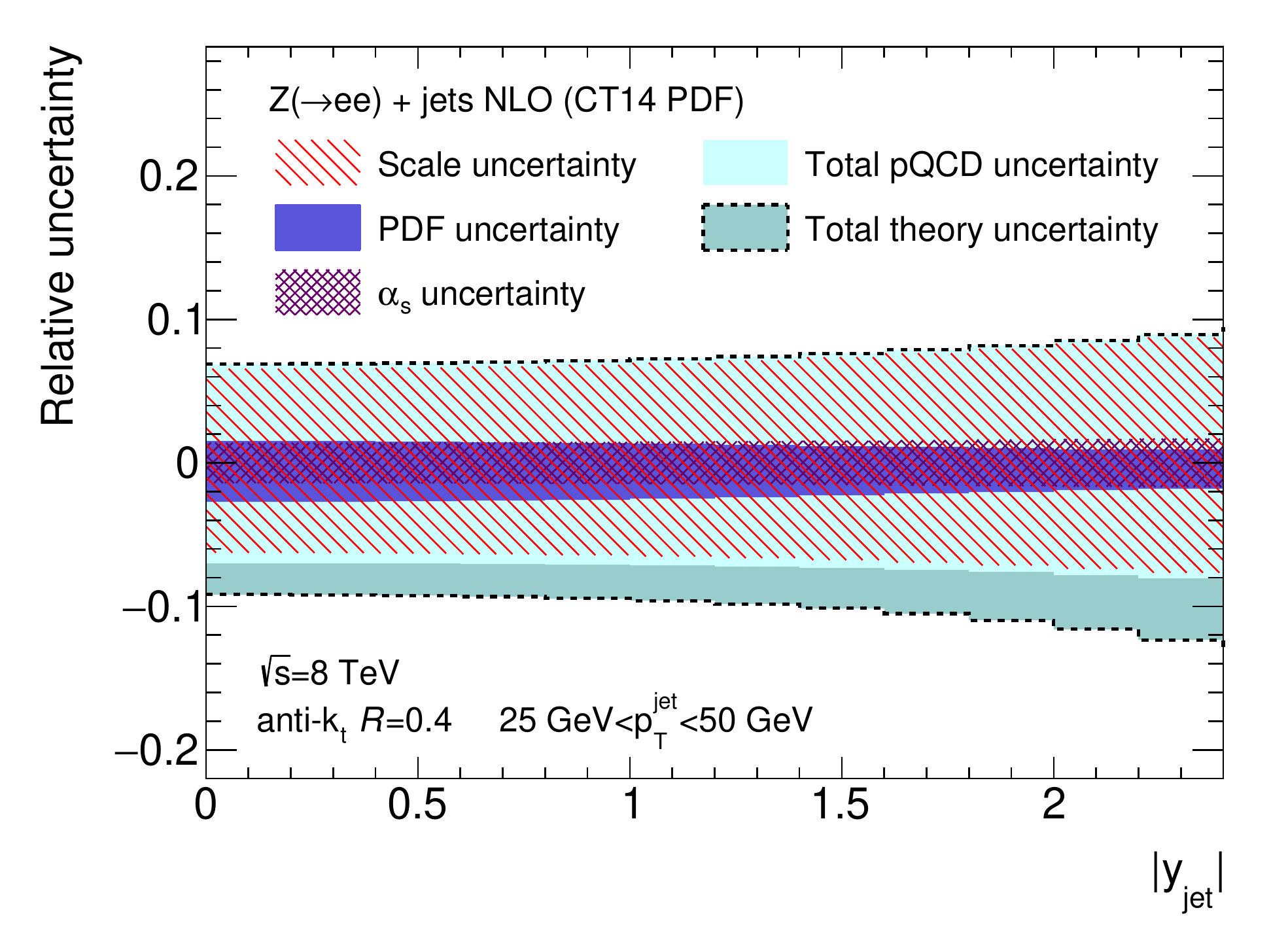}}
\subfloat[$50\GeV < \ptj < 100\GeV$]     {\includegraphics[width=0.49\linewidth]{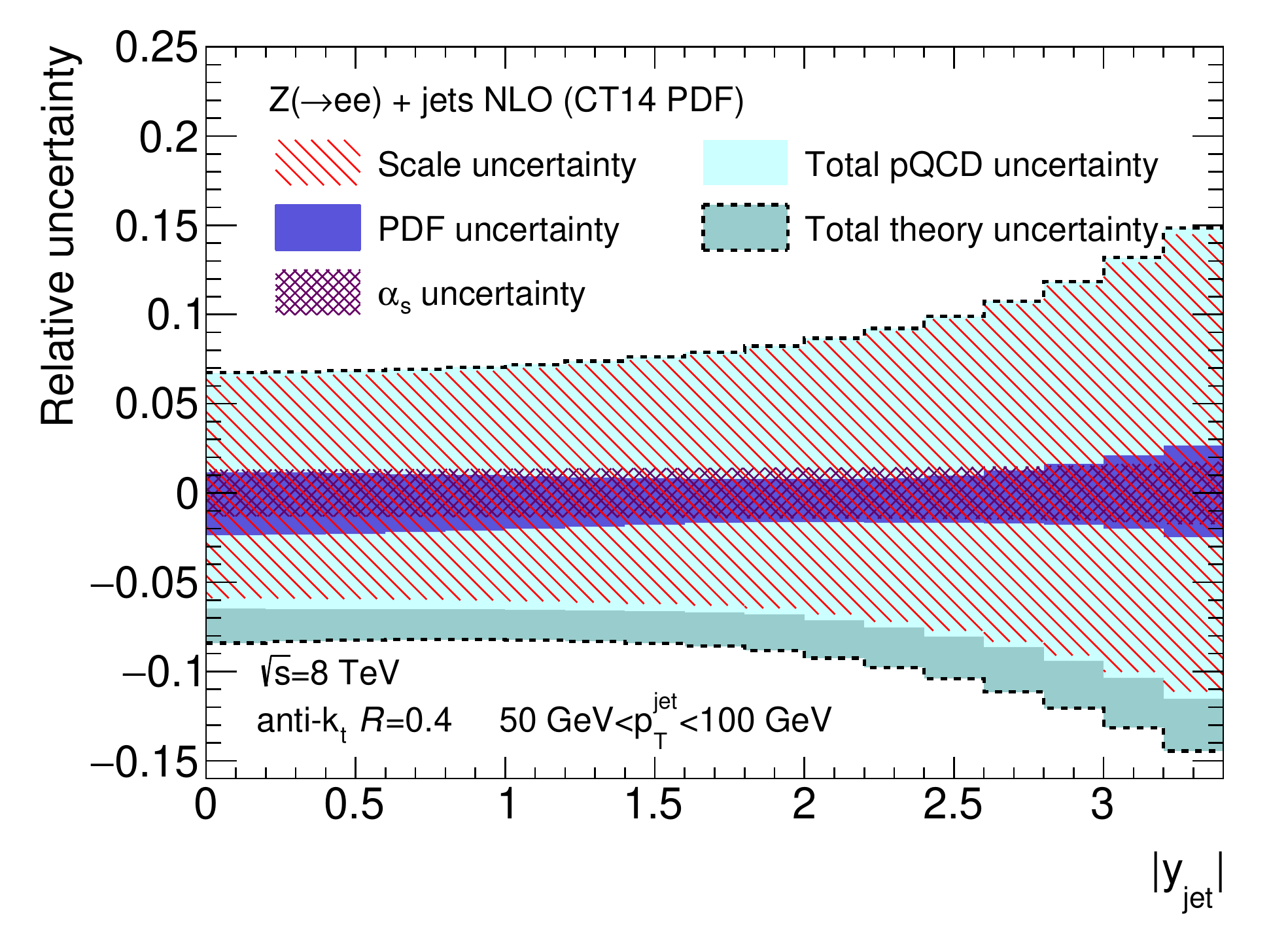}} \\
\subfloat[$100\GeV < \ptj < 200\GeV$]   {\includegraphics[width=0.49\linewidth]{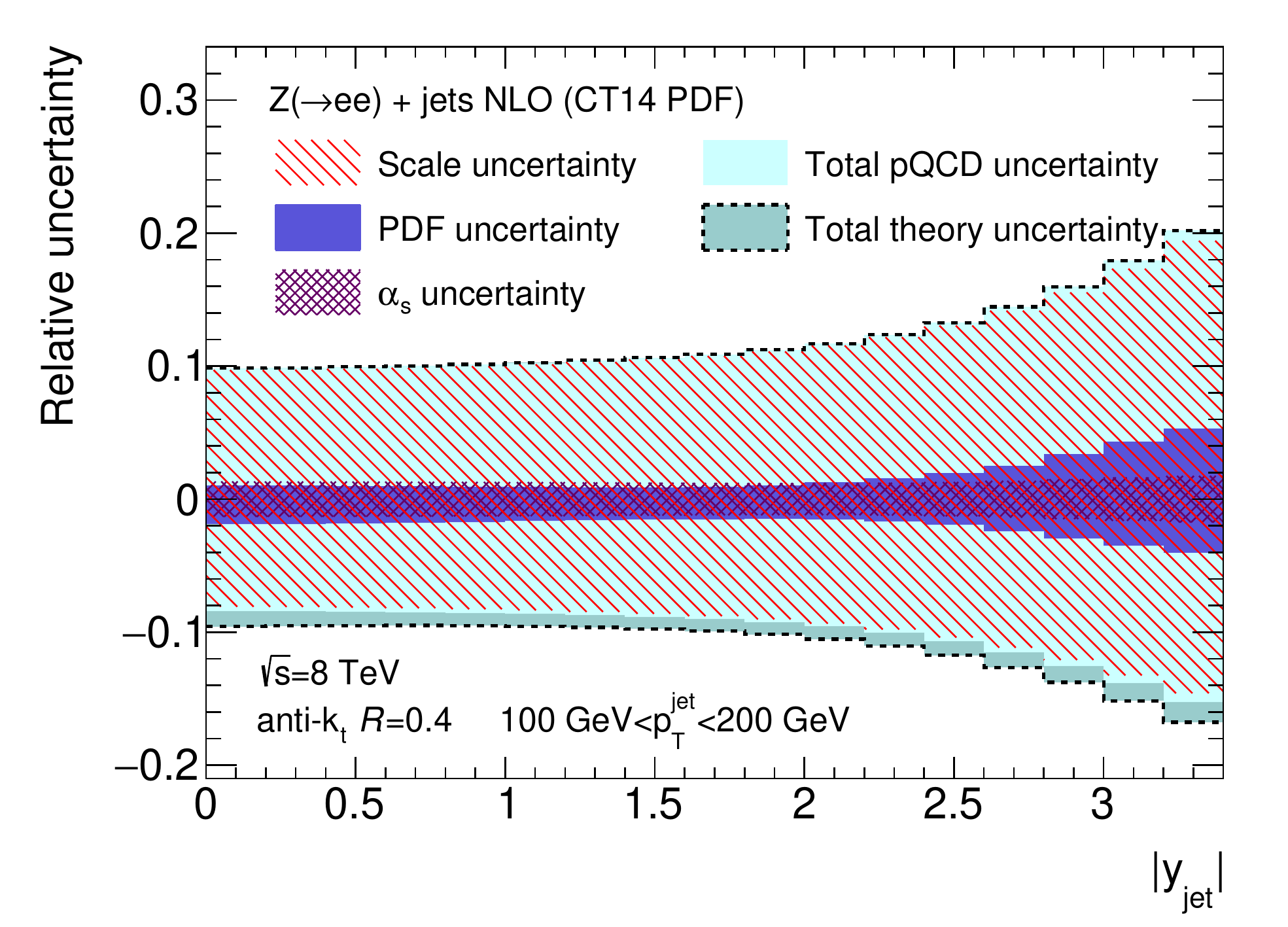}}
\subfloat[$200\GeV < \ptj < 300\GeV$]   {\includegraphics[width=0.49\linewidth]{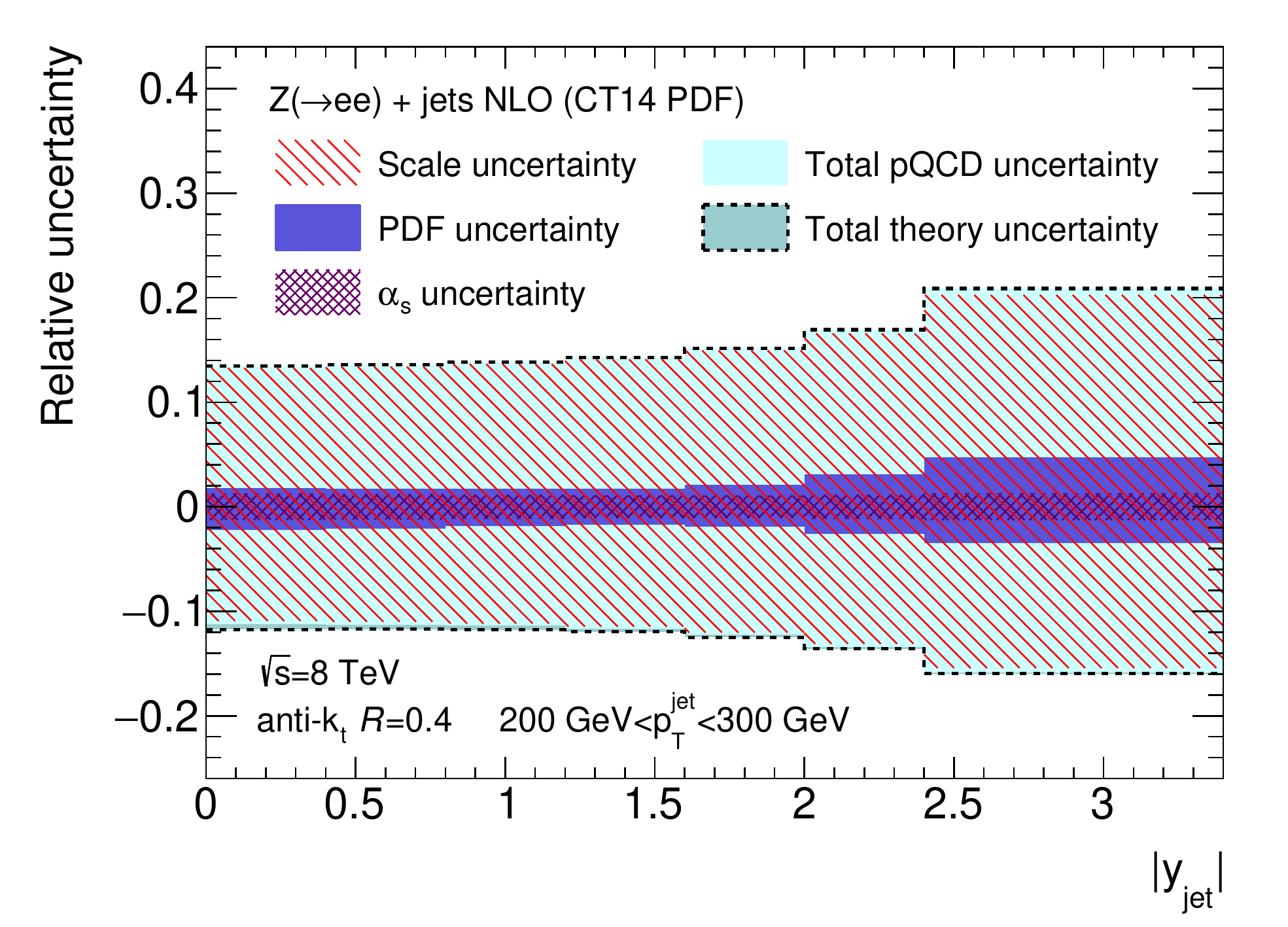}} \\
\subfloat[$300\GeV < \ptj < 400\GeV$]   {\includegraphics[width=0.49\linewidth]{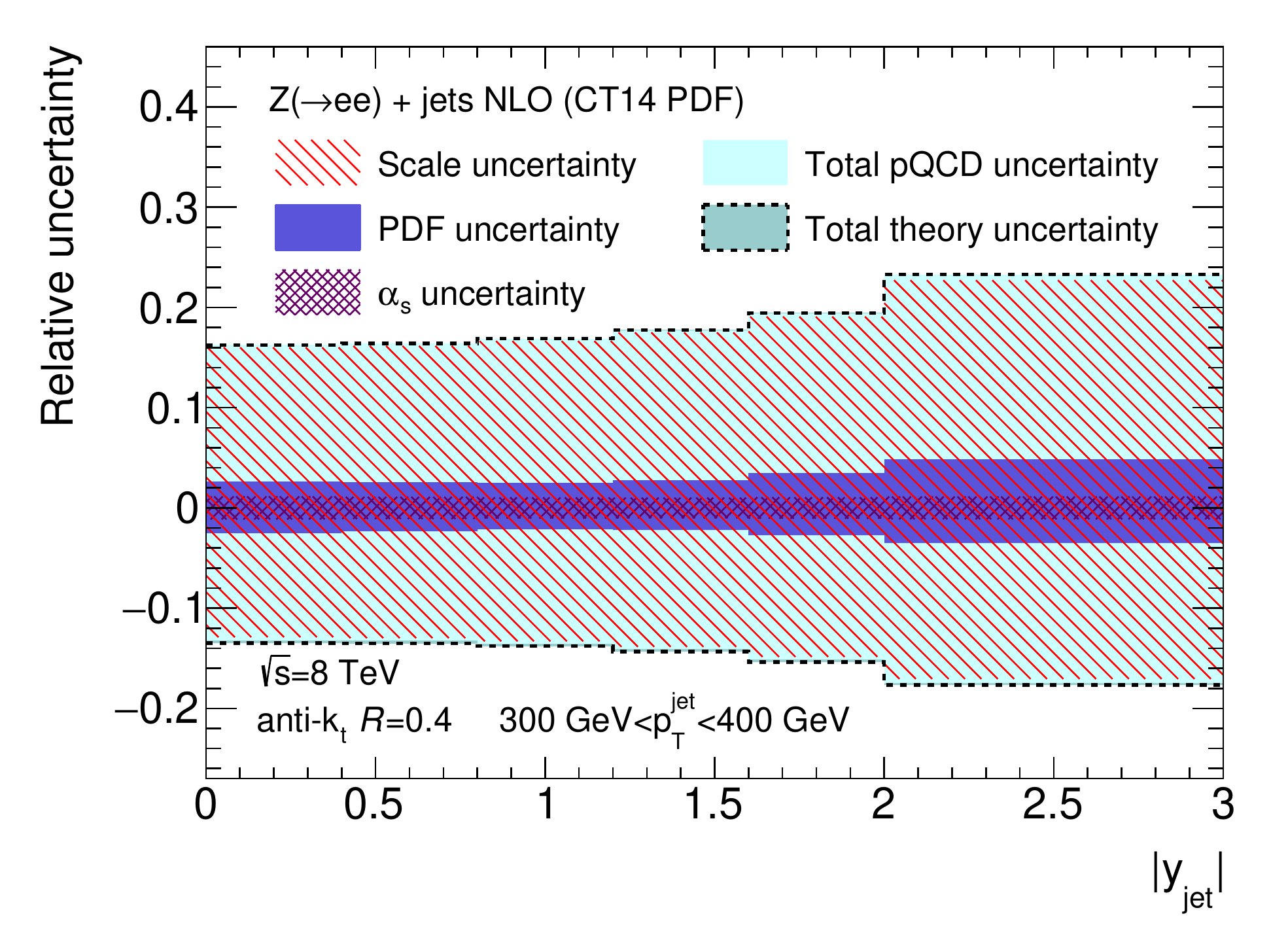}}
\subfloat[$400\GeV < \ptj < 1050\GeV$] {\includegraphics[width=0.49\linewidth]{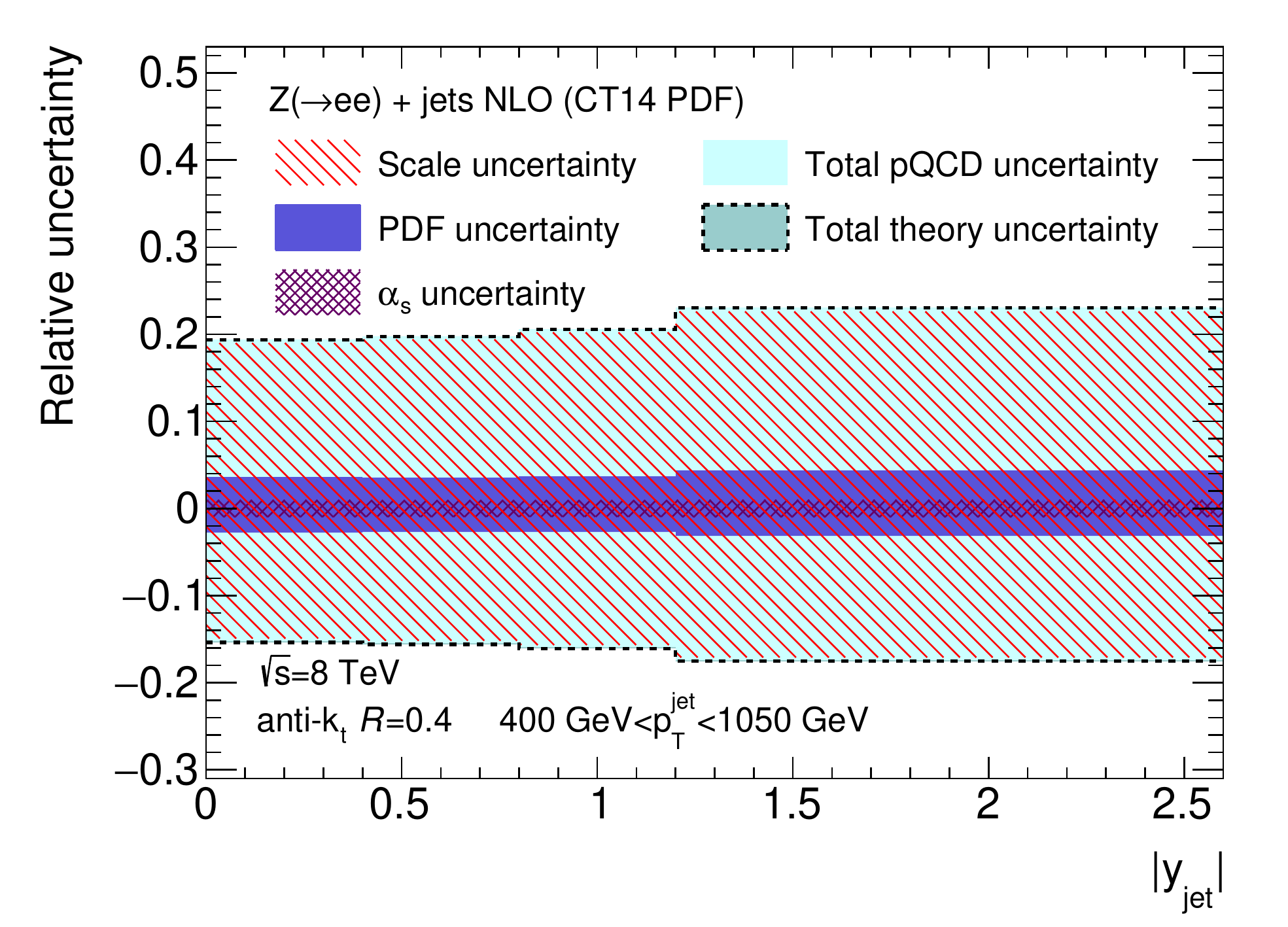}}  \\
\caption[Theory uncertainty]{The uncertainties in NLO pQCD predictions as a function of \absyj{} in \ptj{} bins.
Total pQCD uncertainty is the sum in quadrature of the PDF, scale and \alphas{} uncertainties.
Total theory uncertainty is the sum in quadrature of the total pQCD uncertainty and the uncertainties from the non-perturbative and QED radiation corrections.
The CT14 PDF is used in the calculations.
}
\label{fig:theory_uncertainty}
\end{figure}

The uncertainties for the \Zjets{} \xs{} calculated at NLO using the CT14 PDF as a function of \absyj{} in \ptj{} bins are shown in~\Cref{fig:theory_uncertainty}.
The total uncertainties are dominated by the scale and NPC uncertainties in the $\ptj<100\GeV$ region, where they reach $\pm15\%$ and $-10\%$, respectively.
In the $\ptj>100\GeV$ region, the scale uncertainty alone dominates, as the NPC uncertainty decreases for high jet transverse momenta.
The total uncertainty in this region is 10\%--20\%.  Other uncertainties are below 5\%.
 
\begin{figure}[ht]
\centering
\subfloat[$25\GeV < \ptj < 50\GeV$]       {\includegraphics[width=0.49\linewidth]{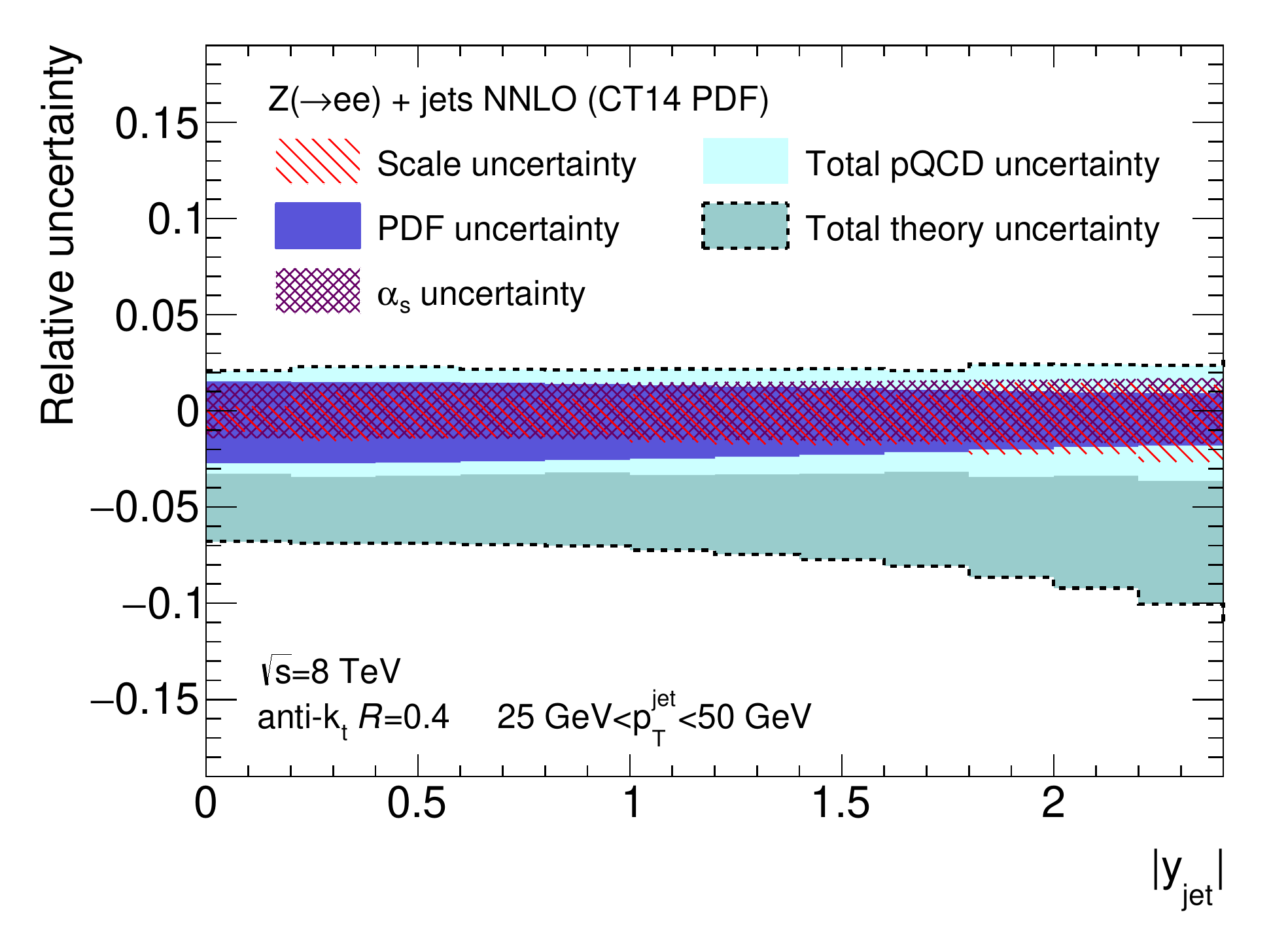}}
\subfloat[$50\GeV < \ptj < 100\GeV$]     {\includegraphics[width=0.49\linewidth]{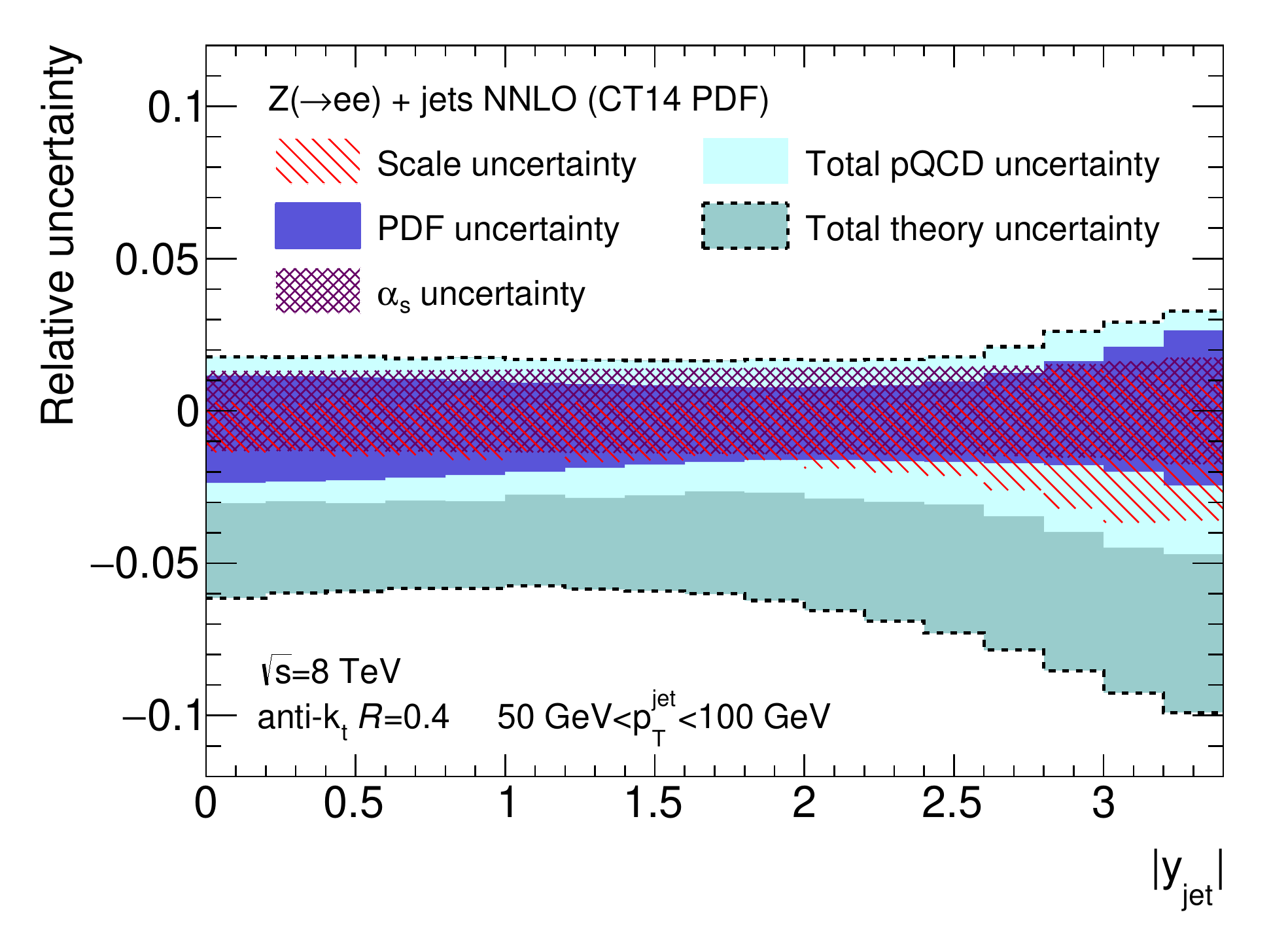}} \\
\subfloat[$100\GeV < \ptj < 200\GeV$]   {\includegraphics[width=0.49\linewidth]{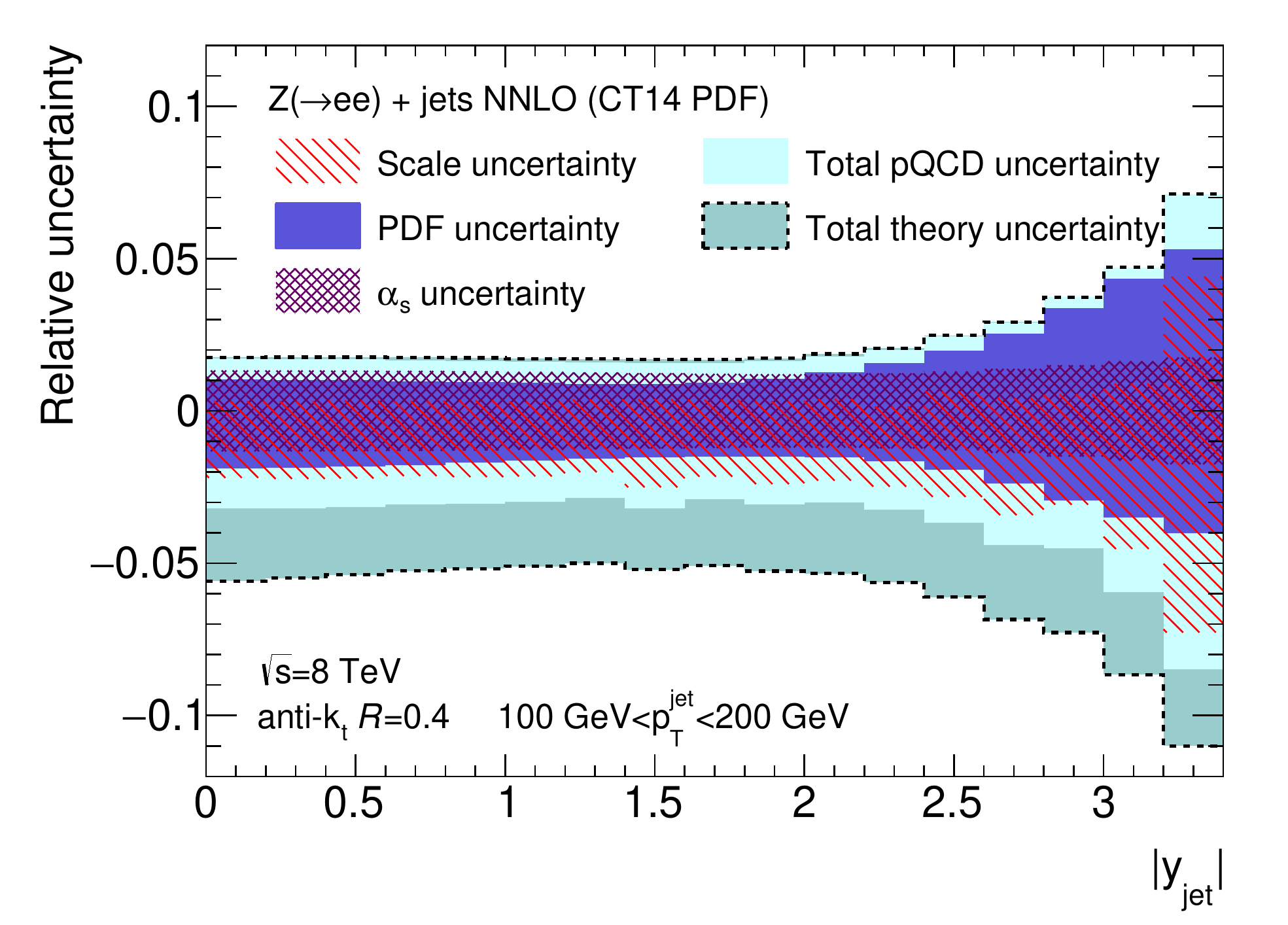}}
\subfloat[$200\GeV < \ptj < 300\GeV$]   {\includegraphics[width=0.49\linewidth]{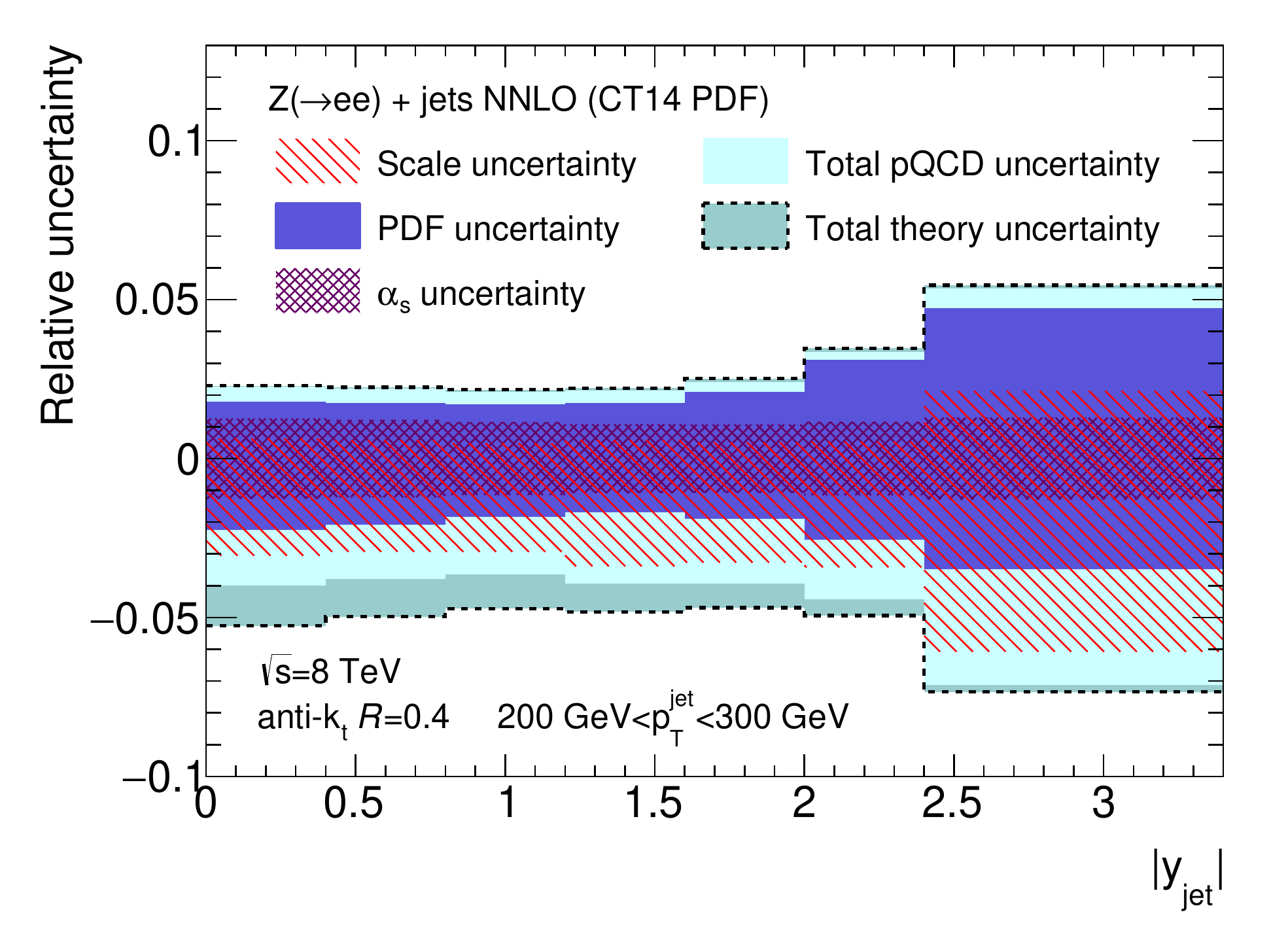}} \\
\subfloat[$300\GeV < \ptj < 400\GeV$]   {\includegraphics[width=0.49\linewidth]{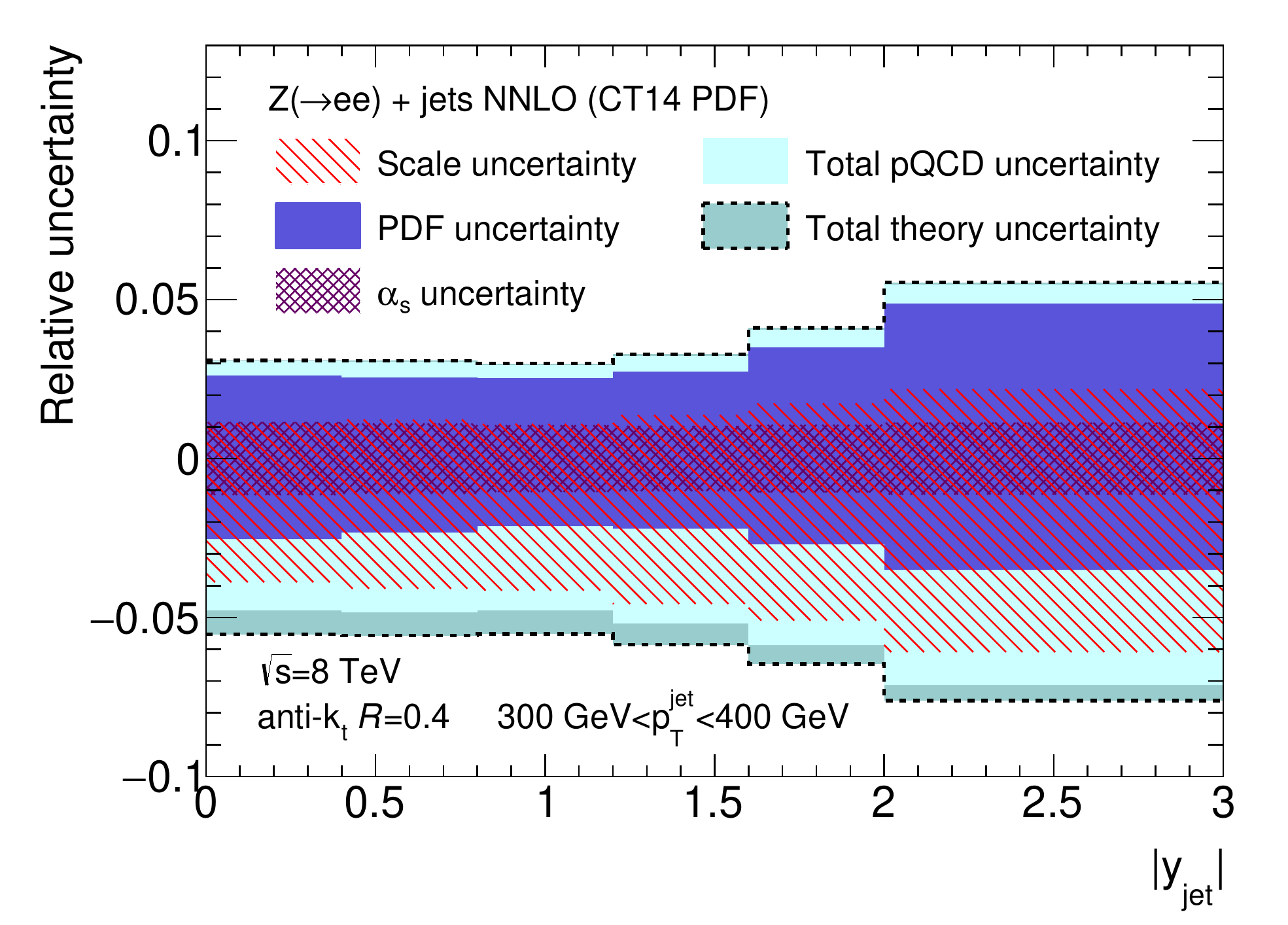}}
\subfloat[$400\GeV < \ptj < 1050\GeV$] {\includegraphics[width=0.49\linewidth]{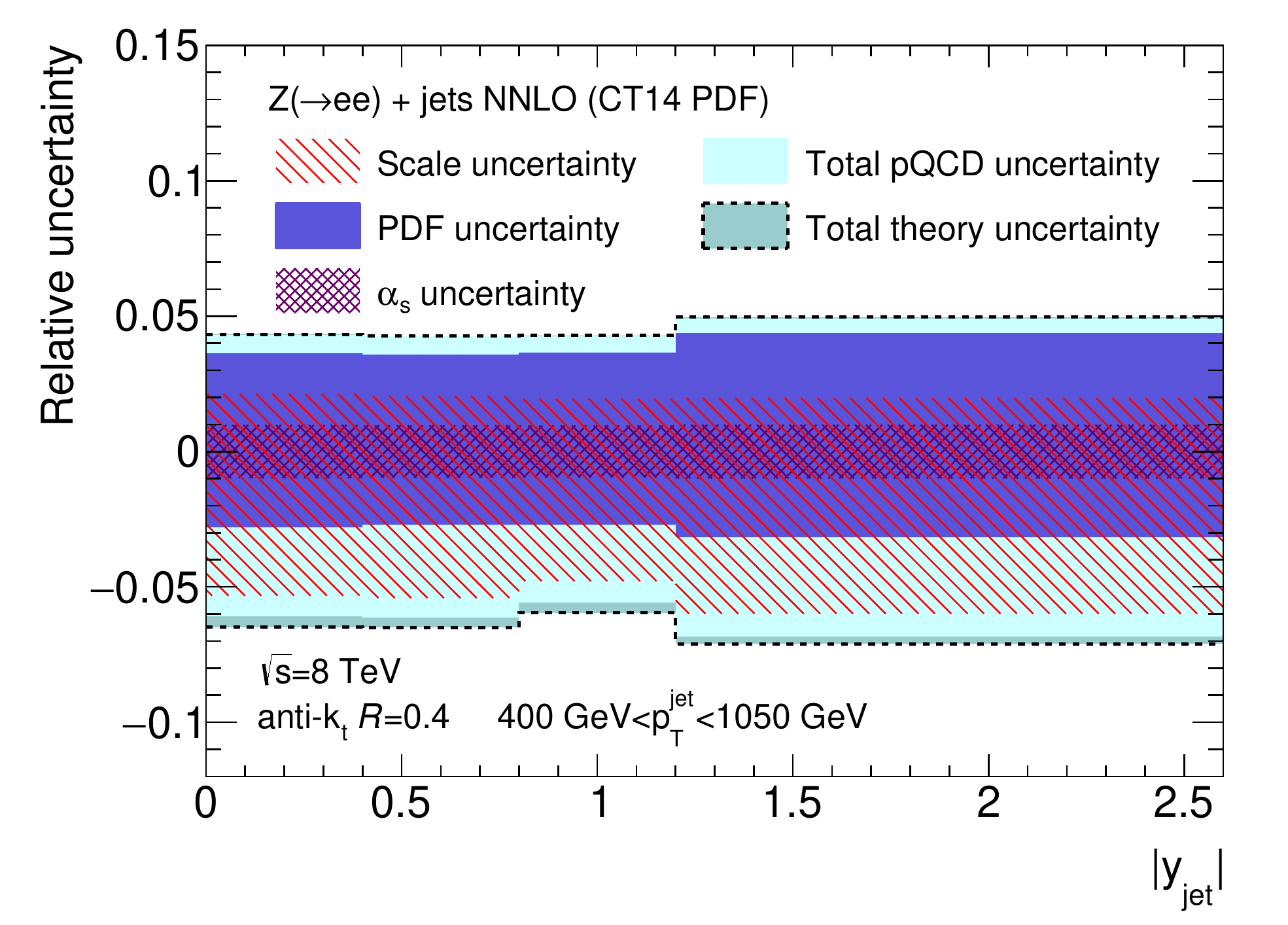}} \\
\caption[NNLOuncertainty]{The uncertainties in NNLO pQCD predictions as a function of \absyj{} in \ptj{} bins.
Total pQCD uncertainty is the sum in quadrature of the PDF, scale and \alphas{} uncertainties.
Total theory uncertainty is the sum in quadrature of the total pQCD uncertainty and the uncertainties from the non-perturbative and QED radiation corrections.
The CT14 PDF is used in the calculations.
}
\label{fig:nnlo_uncertainty}
\end{figure}

The NNLO uncertainties are shown in~\Cref{fig:nnlo_uncertainty}. The scale uncertainty at NNLO is significantly reduced.
This uncertainty is below 1\% in the $25\GeV<\ptj<50\GeV$ bin, increasing to 5\% in the $400\GeV<\ptj<1050\GeV$ bin.
In the $\ptj<200\GeV$ region, the negative part of the total uncertainty is dominated by the NPC uncertainty and its absolute value reaches 7\%--15\% depending on the jet rapidity.
The positive part of the total uncertainty is within 5\%, with about equal contributions from PDF, scale and \alphas{} uncertainties.
In the $\ptj>200\GeV$ region, both the negative and positive parts of the total uncertainty are within 6\% in most bins.
 
The uncertainty in the QED correction is below 0.5\% and is negligible in the fixed-order theory predictions.
 
\FloatBarrier
 
\section{Results\label{sec:results}}
The double-differential \Zjets{} \xs{} as a function of \absyj{} and \ptj{} is calculated as
 
\begin{equation*}
\frac{\textrm{d}^{2} \sigma}{\textrm{d}\ptj \textrm{d}\absyj } = \frac{1}{ \mathcal{L}} \frac{N^{\mathcal{P}}_{i}}{\Delta \ptj{} \Delta \absyj{}},
\end{equation*}
 
where $\mathcal{L}$ is the integrated luminosity, $N^{\mathcal{P}}_{i} $ is the number of jets in data at the particle level as given
in Eq.~\eqref{eq:NJetsPerBin}, and $\Delta\ptj$ and $\Delta\absyj$
are the widths of the jet transverse momentum and absolute jet rapidity ranges for bin $i$, respectively.
The backgrounds are subtracted before data unfolding is performed to obtain $N^{\mathcal{P}}_{i} $.
 
The measured \Zjets{} \xs{} covers five orders of magnitude and falls steeply as a function of \absyj{} and \ptj{}.
A summary of measured \xss{}, together with the systematic and statistical uncertainties, is provided in~\Cref{sec:tables}.
The measured \xss{} with the full breakdown of all uncertainties are provided in HEPData.
 
\begin{figure}
\centering
\includegraphics[width=0.85\textwidth]{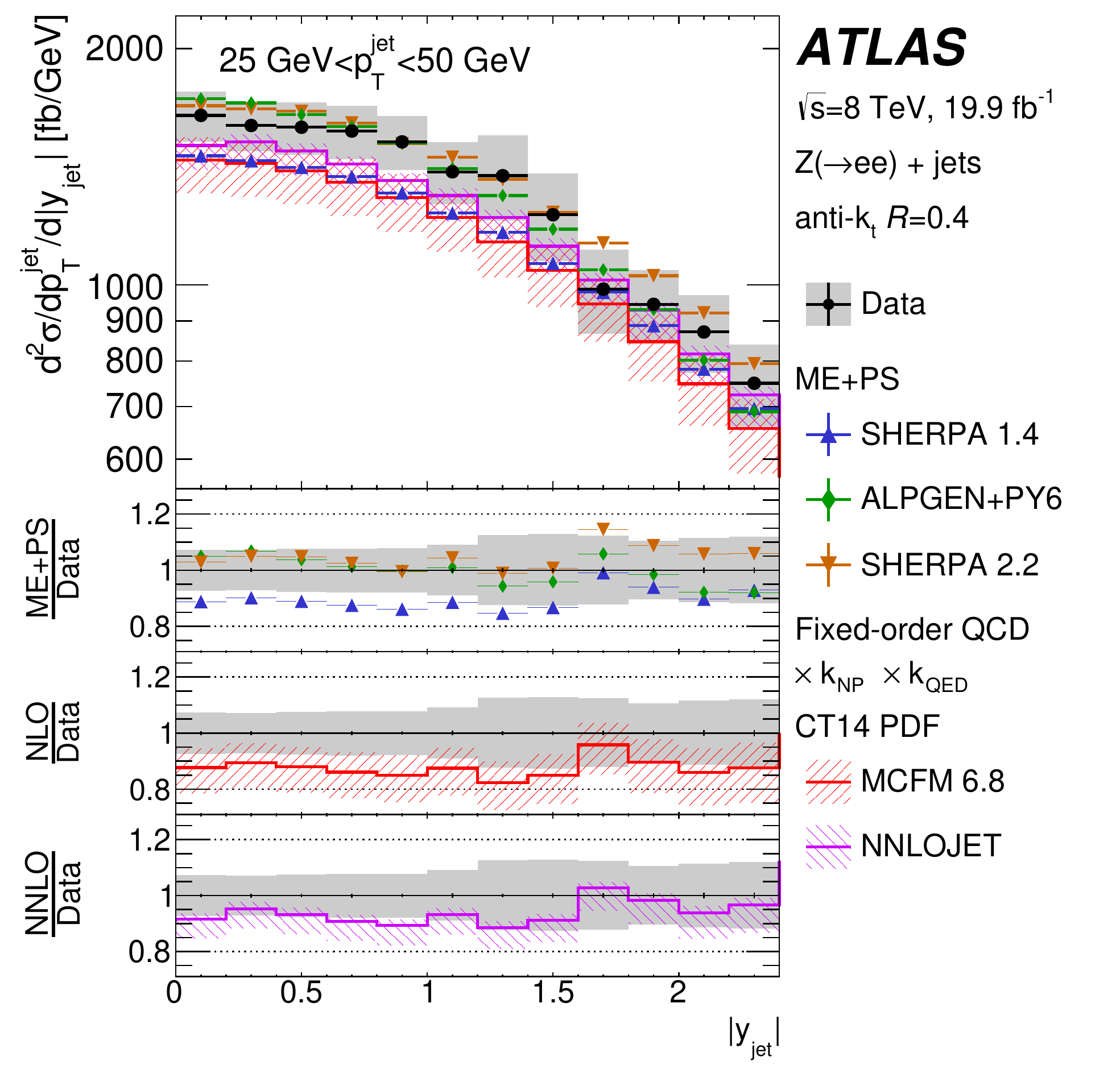}
\caption[CS]{
The double-differential \Zjets{} production \xs{} as a function of \absyj{} in the $25\GeV < \ptj < 50\GeV$ range.
The data are compared with the \sherpav{}, \sherpattv{} and \alpgenpythia{} parton shower MC generator predictions and with the fixed-order theory predictions.
The fixed-order theory predictions are corrected for the non-perturbative and QED radiation effects. The fixed-order calculations are performed using the CT14 PDF.
The total statistical uncertainties are shown with error bars.
The total uncertainties in the measurement and in the fixed-order theory predictions are represented with shaded bands.
The total uncertainty in the measurement is the sum in quadrature of the statistical and systematic uncertainties except for the luminosity uncertainty of 1.9\%.
The total uncertainty in the fixed-order theory predictions is the sum in quadrature of the effects of the PDF, scale, and
\alphas{} uncertainties, and the uncertainties from the non-perturbative and QED radiation corrections.
Lower panels show the ratios of predictions to data.
}
\label{fig:cs_bin1}
\end{figure}

\begin{figure}
\centering
\includegraphics[width=0.85\textwidth]{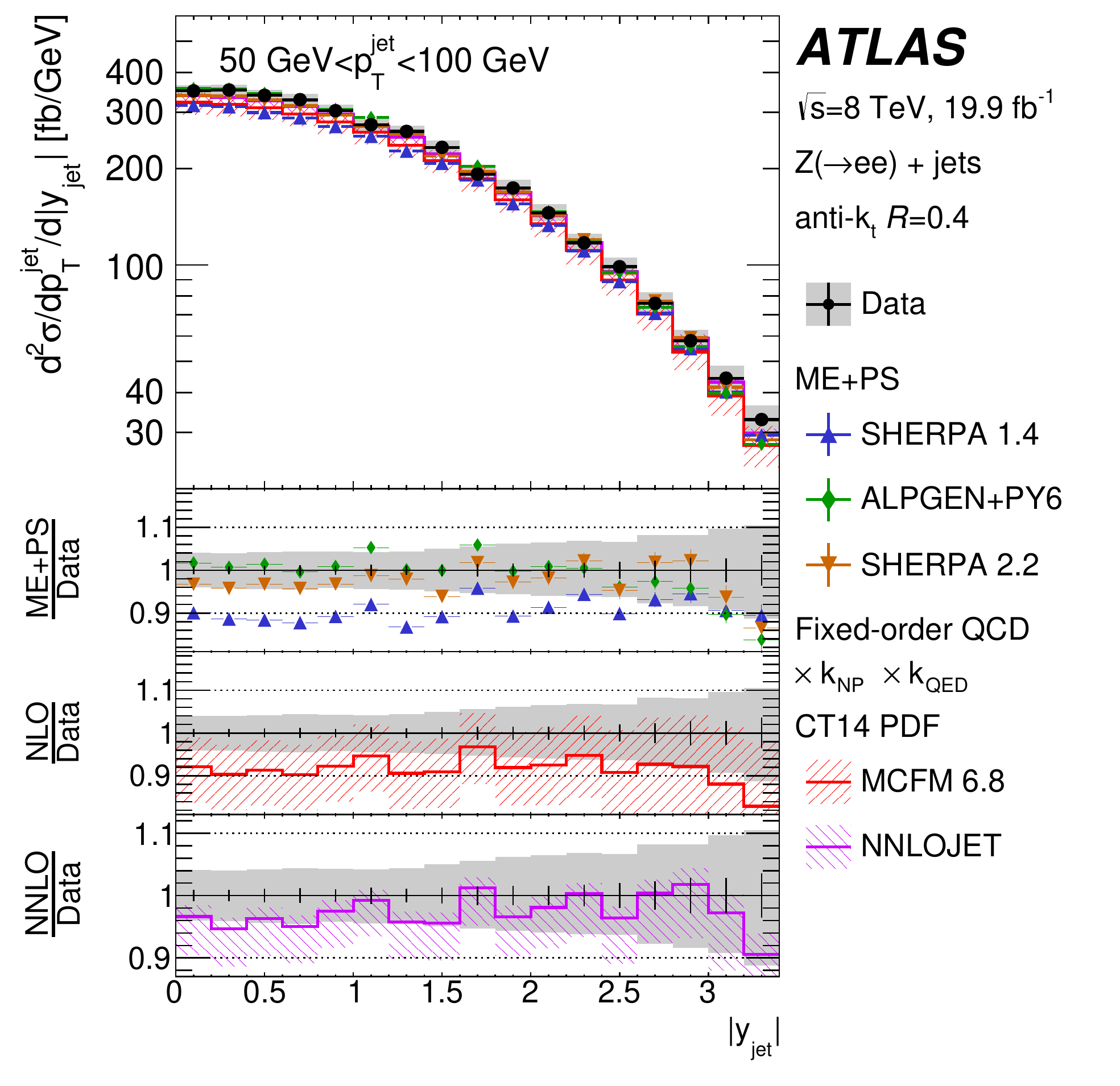}
\caption[CS]{
The double-differential \Zjets{} production \xs{} as a function of \absyj{} in the $50\GeV < \ptj < 100\GeV$ range.
The data are compared with the \sherpav{}, \sherpattv{} and \alpgenpythia{} parton shower MC generator predictions and with the fixed-order theory predictions.
The fixed-order theory predictions are corrected for the non-perturbative and QED radiation effects. The fixed-order calculations are performed using the CT14 PDF.
The total statistical uncertainties are shown with error bars.
The total uncertainties in the measurement and in the fixed-order theory predictions are represented with shaded bands.
The total uncertainty in the measurement is the sum in quadrature of the statistical and systematic uncertainties except for the luminosity uncertainty of 1.9\%.
The total uncertainty in the fixed-order theory predictions is the sum in quadrature of the effects of the PDF, scale, and \alphas{} uncertainties, and the uncertainties from the non-perturbative and QED radiation corrections.
Lower panels show the ratios of predictions to data.
}
\label{fig:cs_bin2}
\end{figure}

\begin{figure}
\centering
\includegraphics[width=0.85\textwidth]{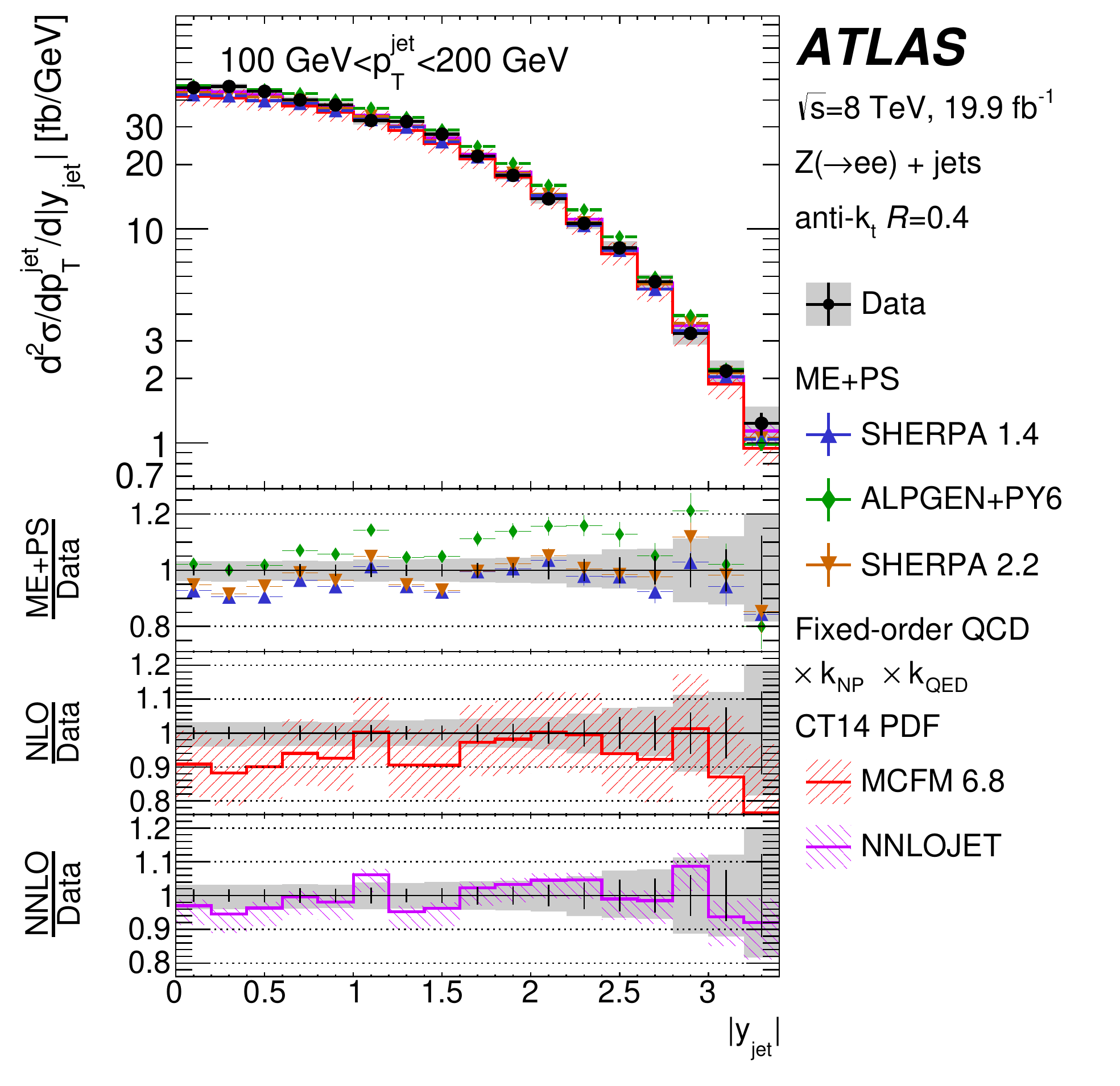}
\caption[CS]{
The double-differential \Zjets{} production \xs{} as a function of \absyj{} in the $100\GeV < \ptj < 200\GeV$ range.
The data are compared with the \sherpav{}, \sherpattv{} and \alpgenpythia{} parton shower MC generator predictions and with the fixed-order theory predictions.
The fixed-order theory predictions are corrected for the non-perturbative and QED radiation effects. The fixed-order calculations are performed using the CT14 PDF.
The total statistical uncertainties are shown with error bars.
The total uncertainties in the measurement and in the fixed-order theory predictions are represented with shaded bands.
The total uncertainty in the measurement is the sum in quadrature of the statistical and systematic uncertainties except for the luminosity uncertainty of 1.9\%.
The total uncertainty in the fixed-order theory predictions is the sum in quadrature of the effects of the PDF, scale, and \alphas{} uncertainties, and the uncertainties from the non-perturbative and QED radiation corrections.
Lower panels show the ratios of predictions to data.
}
\label{fig:cs_bin3}
\end{figure}

\begin{figure}
\centering
\includegraphics[width=0.85\textwidth]{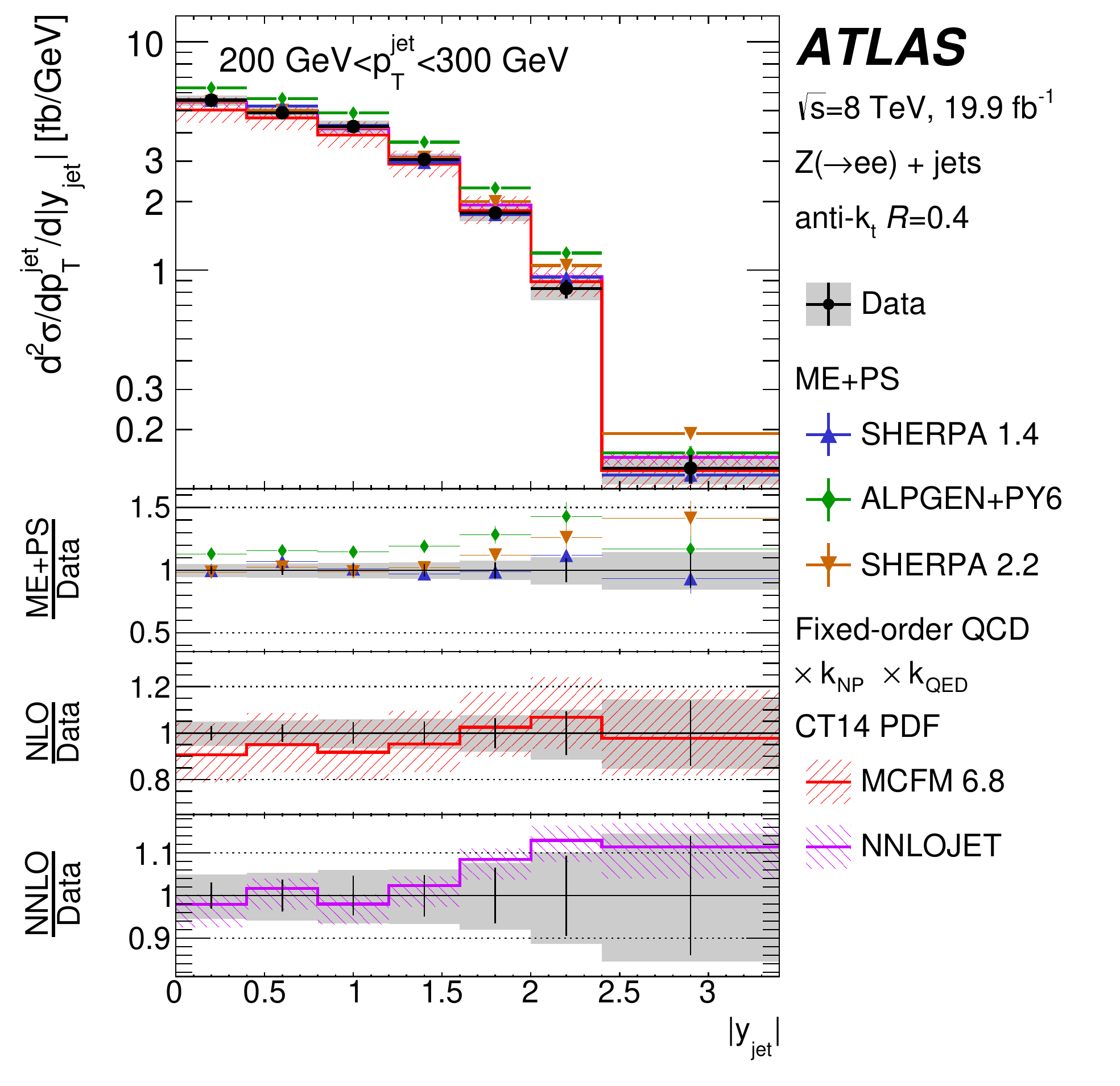}
\caption[CS]{
The double-differential \Zjets{} production \xs{} as a function of \absyj{} in the $200\GeV < \ptj < 300\GeV$ range.
The data are compared with the \sherpav{}, \sherpattv{} and \alpgenpythia{} parton shower MC generator predictions and with the fixed-order theory predictions.
The fixed-order theory predictions are corrected for the non-perturbative and QED radiation effects. The fixed-order calculations are performed using the CT14 PDF.
The total statistical uncertainties are shown with error bars.
The total uncertainties in the measurement and in the fixed-order theory predictions are represented with shaded bands.
The total uncertainty in the measurement is the sum in quadrature of the statistical and systematic uncertainties except for the luminosity uncertainty of 1.9\%.
The total uncertainty in the fixed-order theory predictions is the sum in quadrature of the effects of the PDF, scale, and \alphas{} uncertainties, and the uncertainties from the non-perturbative and QED radiation corrections.
Lower panels show the ratios of predictions to data.
}
\label{fig:cs_bin4}
\end{figure}

\begin{figure}
\centering
\includegraphics[width=0.85\textwidth]{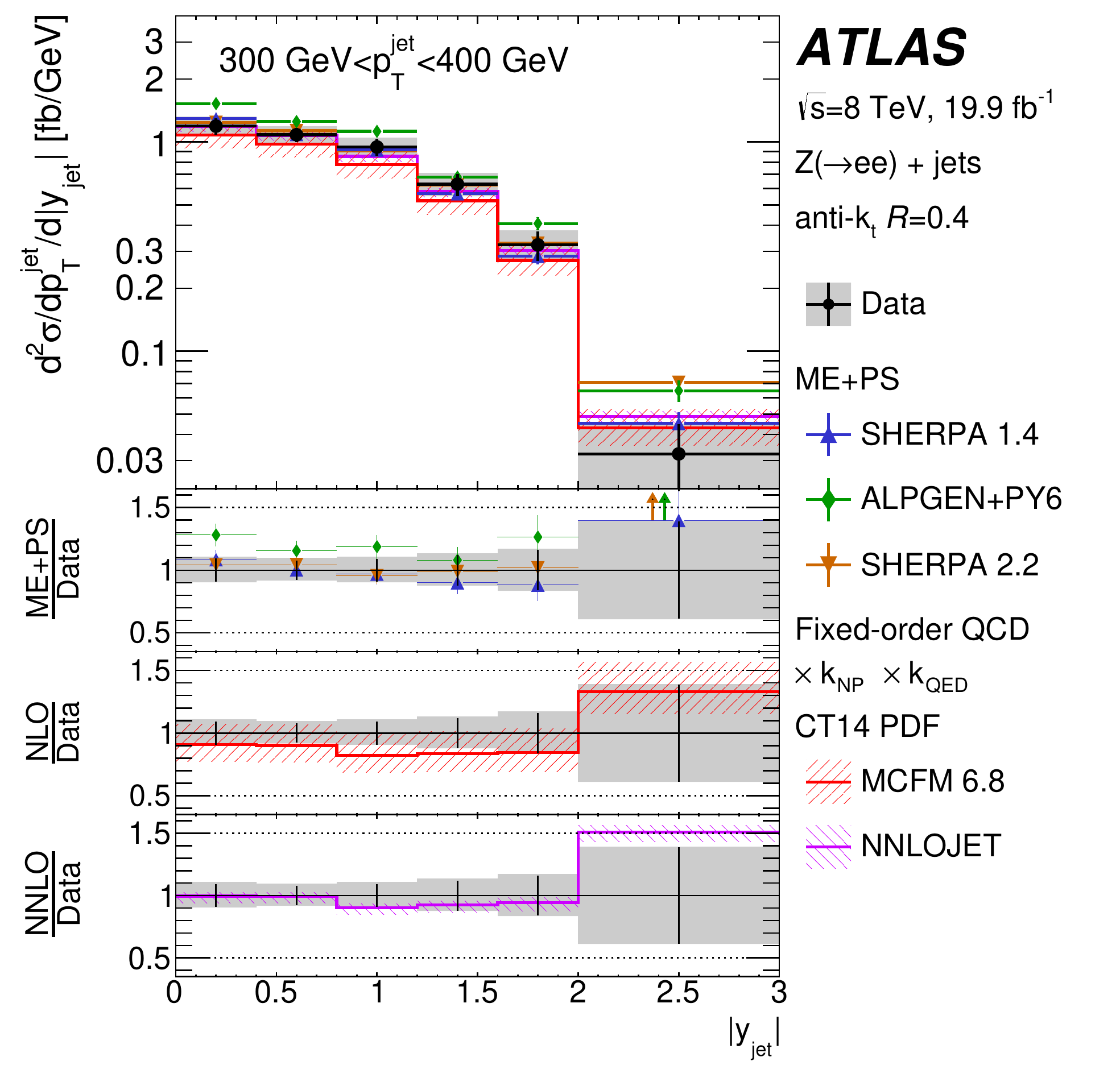}
\caption[CS]{
The double-differential \Zjets{} production \xs{} as a function of \absyj{} in the $300\GeV < \ptj < 400\GeV$ range.
The data are compared with the \sherpav{}, \sherpattv{} and \alpgenpythia{} parton shower MC generator predictions and with the fixed-order theory predictions.
The fixed-order theory predictions are corrected for the non-perturbative and QED radiation effects. The fixed-order calculations are performed using the CT14 PDF.
The total statistical uncertainties are shown with error bars.
The total uncertainties in the measurement and in the fixed-order theory predictions are represented with shaded bands.
The total uncertainty in the measurement is the sum in quadrature of the statistical and systematic uncertainties except for the luminosity uncertainty of 1.9\%.
The total uncertainty in the fixed-order theory predictions is the sum in quadrature of the effects of the PDF, scale, and \alphas{} uncertainties, and the uncertainties from the non-perturbative and QED radiation corrections.
Lower panels show the ratios of predictions to data.
}
\label{fig:cs_bin5}
\end{figure}

\begin{figure}
\centering
\includegraphics[width=0.85\textwidth]{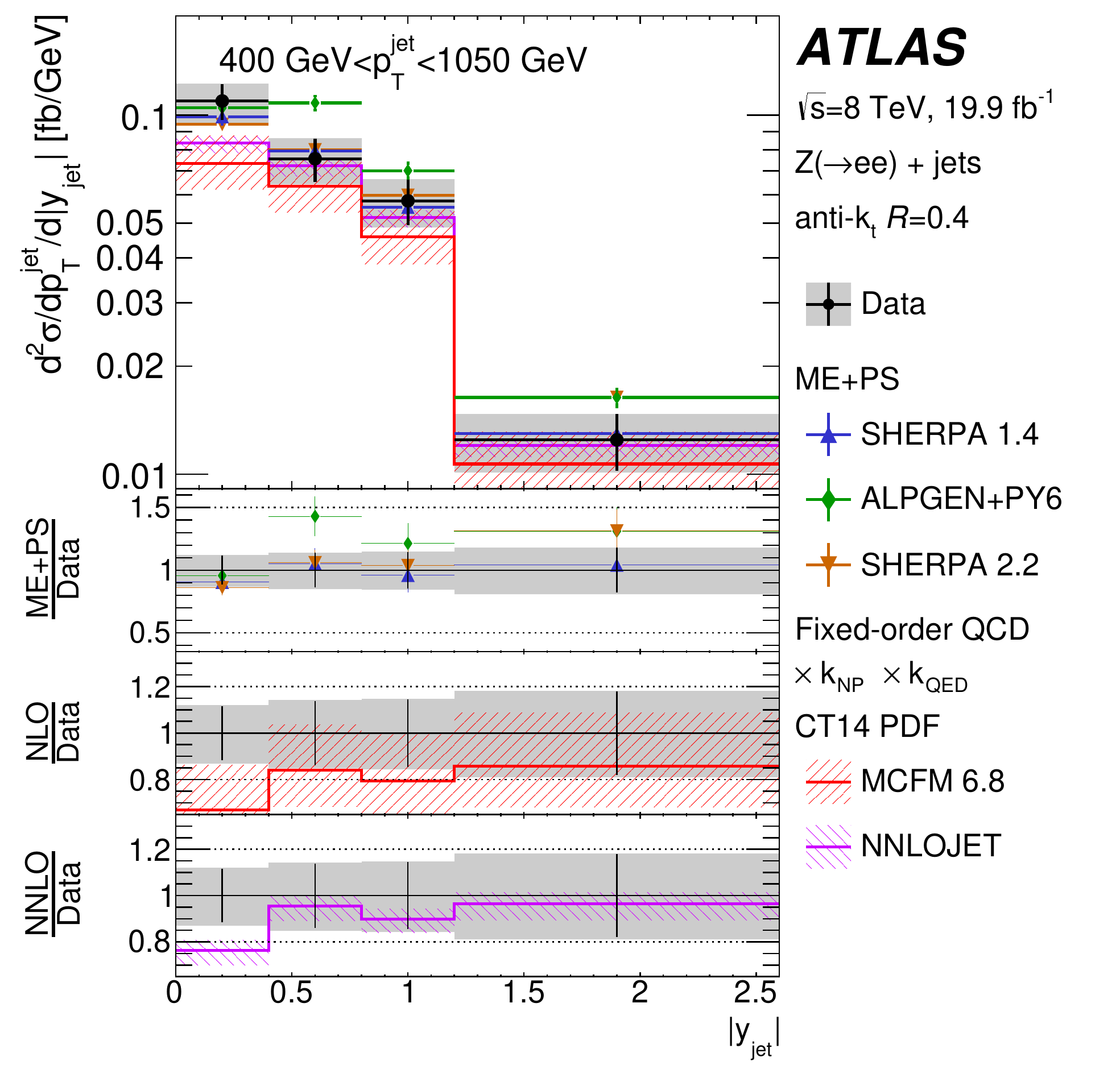}
\caption[CS]{
The double-differential \Zjets{} production \xs{} as a function of \absyj{} in the $400\GeV < \ptj < 1050\GeV$ range.
The data are compared with the \sherpav{}, \sherpattv{} and \alpgenpythia{} parton shower MC generator predictions and with the fixed-order theory predictions.
The fixed-order theory predictions are corrected for the non-perturbative and QED radiation effects. The fixed-order calculations are performed using the CT14 PDF.
The total statistical uncertainties are shown with error bars.
The total uncertainties in the measurement and in the fixed-order theory predictions are represented with shaded bands.
The total uncertainty in the measurement is the sum in quadrature of the statistical and systematic uncertainties except for the luminosity uncertainty of 1.9\%.
The total uncertainty in the fixed-order theory predictions is the sum in quadrature of the effects of the PDF, scale, and \alphas{} uncertainties, and the uncertainties from the non-perturbative and QED radiation corrections.
Lower panels show the ratios of predictions to data.
}
\label{fig:cs_bin6}
\end{figure}

The comparisons with the theoretical predictions are shown in~\Cref{fig:cs_bin1,fig:cs_bin2,fig:cs_bin3,fig:cs_bin4,fig:cs_bin5,fig:cs_bin6}.
The fixed-order theoretical predictions are corrected for the non-perturbative and QED radiation effects.
The NLO predictions are lower than the data by approximately 5\%--10\%. However, this difference is covered by the uncertainties.
The NNLO calculations compensate for the NLO-to-data differences in most bins of the measurement and show better agreement with the central values of the \xss{} in data.
The \sherpav{} and \alpgenpythia{} MC-to-data ratios are approximately constant across all \absyj{} bins, but a dependence on \ptj{} is observed.
The \sherpav{} predictions are lower than the data by about 10\% in the $25\GeV < \ptj < 200\GeV$ region,
but in the $\ptj > 200\GeV$ region they agree within a few percent.
The \alpgenpythia{} predictions agree with data in the $25\GeV < \ptj < 100\GeV$ region,
but exceed the data as a function of \ptj{}, the largest difference being about 20\% in the highest \ptj{} bin, $400\GeV < \ptj < 1050\GeV$.
 
Additionally, data is compared to the \sherpattv{} prediction.
In this prediction, the matrix elements are calculated with NLO accuracy for the inclusive Z production process up to two additional partons in the final state, and with LO accuracy in the final states with up to four partons.
\sherpattv{} MEs are convolved with the NNPDF~3.0~\cite{nnpf30} PDFs.
The MEs are merged with \sherpa{} parton shower using the ME+PS@NLO~\cite{mepsatnlo} prescription.
This prediction shows a good agreement with data in all bins of the measurement.
 
\begin{figure}
\centering
\includegraphics[width=1.0\textwidth]{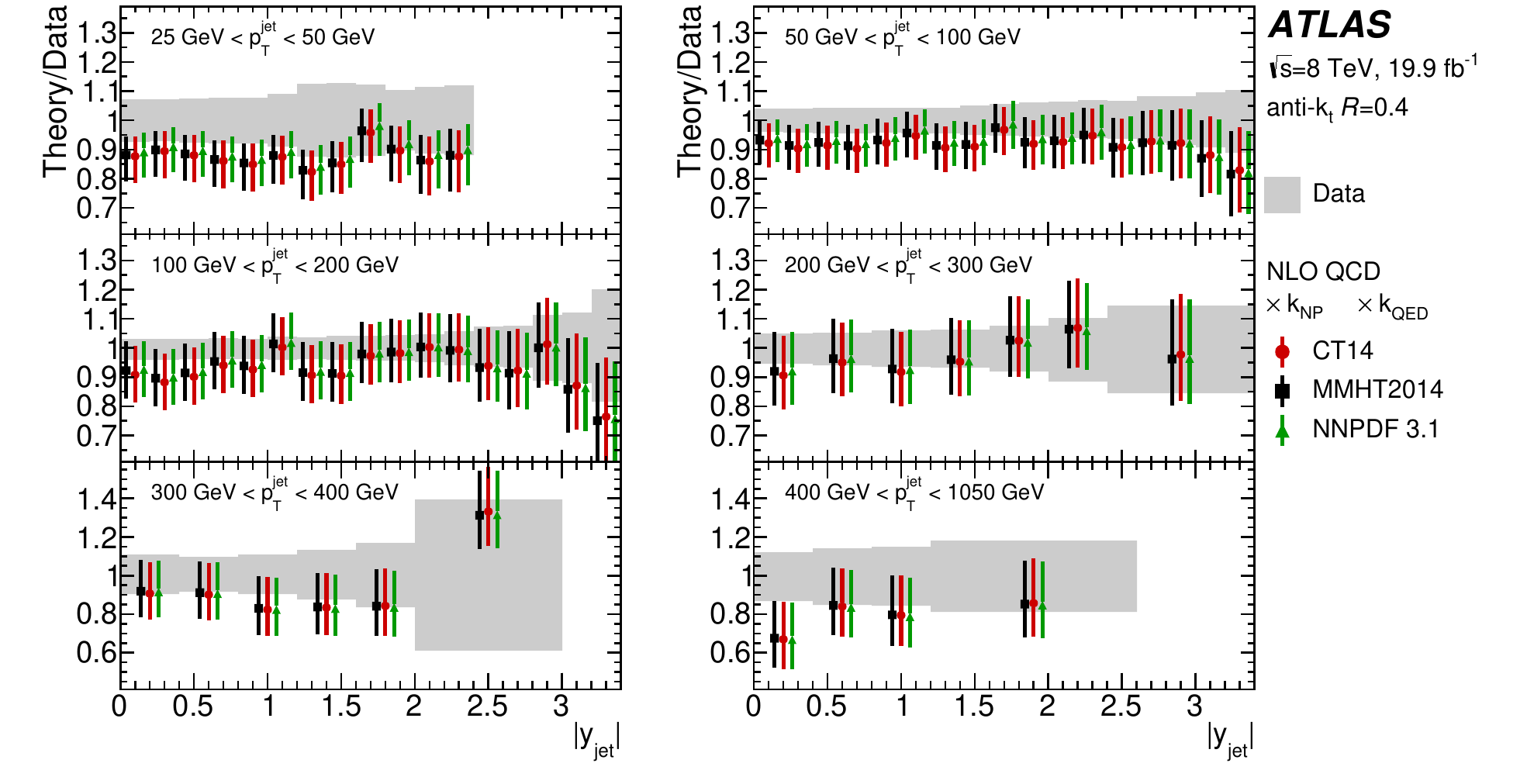}
\caption[CS]{
Ratio of the measured \Zjets{} production \xs{} and the NLO QCD predictions, obtained using MCFM,
corrected for the non-perturbative and QED radiation effects as a function of \absyj{} and \ptj{} bins.
Theoretical predictions are calculated using various PDF sets.
The coloured error bars represent the sum in quadrature of the effects of the PDF, scale, and \alphas{} uncertainties, and the uncertainties from the non-perturbative and QED radiation corrections.
The grey band shows the sum in quadrature of the statistical and systematic uncertainties in the measurement except for the luminosity uncertainty of 1.9\%.
}
\label{fig:data_pdf_comp_1}
\end{figure}

\begin{figure}
\centering
\includegraphics[width=1.0\textwidth]{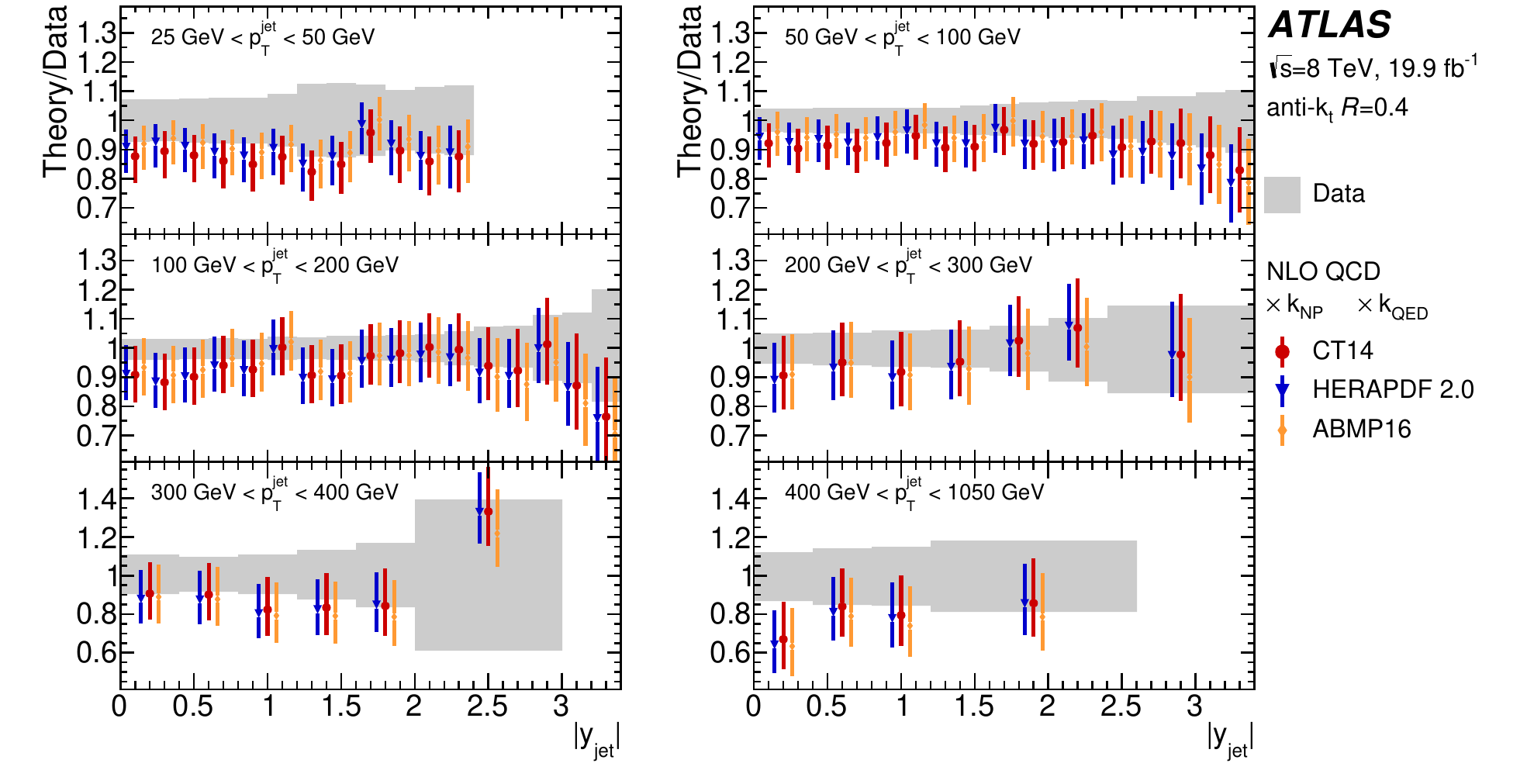}
\caption[CS]{
Ratio of the measured \Zjets{} production \xs{} and the NLO QCD predictions, obtained using MCFM,
corrected for the non-perturbative and QED radiation effects as a function of \absyj{} and \ptj{} bins.
Theoretical predictions are calculated using various PDF sets.
The coloured error bars represent the sum in quadrature of the effects of the PDF, scale, and \alphas{} uncertainties, and the uncertainties from the non-perturbative and QED radiation corrections.
The grey band shows the sum in quadrature of the statistical and systematic uncertainties in the measurement except for the luminosity uncertainty of 1.9\%.
}
\label{fig:data_pdf_comp_2}
\end{figure}

\begin{figure}
\centering
\includegraphics[width=1.0\textwidth]{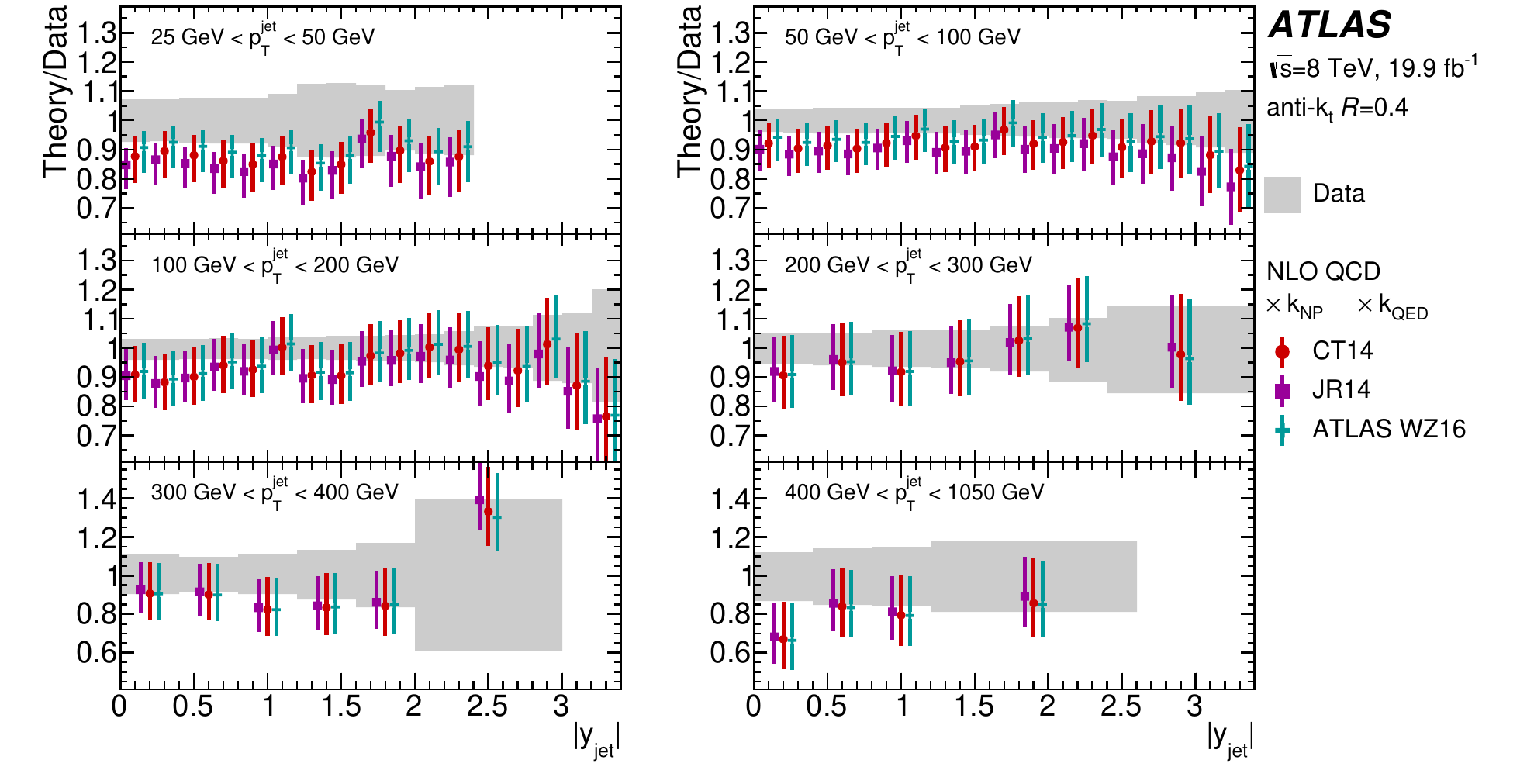}
\caption[CS]{
Ratio of the measured \Zjets{} production \xs{} and the NLO QCD predictions, obtained using MCFM,
corrected for the non-perturbative and QED radiation effects as a function of \absyj{} and \ptj{} bins.
Theoretical predictions are calculated using various PDF sets.
The coloured error bars represent the sum in quadrature of the effects of the PDF, scale, and \alphas{} uncertainties, and the uncertainties from the non-perturbative and QED radiation corrections.
The grey band shows the sum in quadrature of the statistical and systematic uncertainties in the measurement except for the luminosity uncertainty of 1.9\%.
}
\label{fig:data_pdf_comp_3}
\end{figure}

\begin{figure}
\centering
\includegraphics[width=1.0\textwidth]{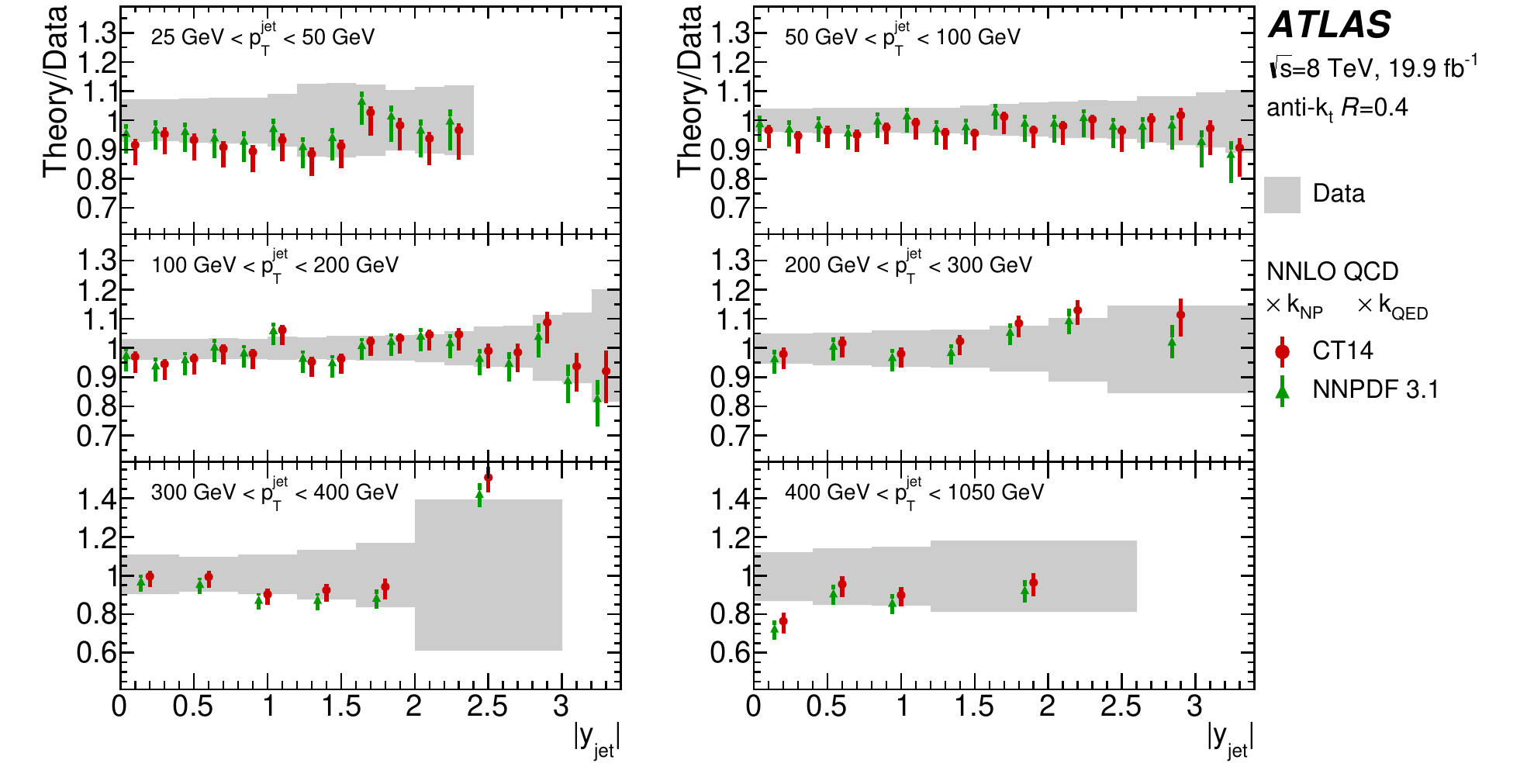}
\caption[CS]{
Ratio of the measured \Zjets{} production \xs{} and the NNLO QCD predictions, obtained using NNLOJET,
corrected for the non-perturbative and QED radiation effects as a function of \absyj{} and \ptj{} bins.
Theoretical predictions are calculated using various PDF sets.
The coloured error bars represent the sum in quadrature of the effects of the PDF, scale, and \alphas{} uncertainties, and the uncertainties from the non-perturbative and QED radiation corrections.
The grey band shows the sum in quadrature of the statistical and systematic uncertainties in the measurement except for the luminosity uncertainty of 1.9\%.
}
\label{fig:data_pdf_comp_6}
\end{figure}

The ratios between the measured \Zjets{} production \xss{} and the NLO predictions, calculated with various PDF sets,
are shown in~\Cref{fig:data_pdf_comp_1,fig:data_pdf_comp_2,fig:data_pdf_comp_3}.
The calculations with MMHT2014 and NNPDF3.1 predict 1\%--2\% larger \xss{} compared to those using the  CT14 PDF.
The \xss{} calculated with ATLAS-epWZ16 PDF are larger by 2\%--3\%.
The ABMP16 and HERAPDF2.0 \xs{} predictions in the $\absyj<2.0$ and $\ptj<100\GeV$ regions are 3\%--5\% larger than those from the CT14 PDF,
while in other bins of the measurement their predictions are up to 5\% lower than those obtained with the CT14 PDF.
The JR14 PDF predictions are 2\%--5\% lower than those from the CT14 PDF in the $25\GeV < \ptj < 200\GeV$ region and higher by 2\% in the $\ptj > 200\GeV$ region.
The differences between the \xss{} calculated at NLO accuracy with various PDF sets are covered by the theoretical uncertainties.
In the NNLO calculations, the difference between CT14 PDF and NNPDF3.1 predictions is 2\%--5\%, which is comparable to the size of the theoretical uncertainties, as shown in~\Cref{fig:data_pdf_comp_6}.

\FloatBarrier

\section{Quantitative data and  theory comparison\label{sec:chi2}}
\newcommand{\pstt}{\ensuremath{<10^{-3}}}
\newcommand{\psch}{\ensuremath{\chi^2}}
\newcommand{\pschc}{\ensuremath{\chi^2_\textrm{corr}}}
\newcommand{\pschu}{\ensuremath{\chi^2_\textrm{uncorr}}}
\newcommand{\pscht}{\ensuremath{\chi^2_\textrm{total}}}
 
The fixed-order pQCD predictions at NNLO accuracy, corrected for electroweak and non-perturbative effects, are quantitatively compared with the measured cross-section using a \psch{} function that accounts for both the experimental and theoretical uncertainties
 
\begin{equation}\label{eqn:chi2}
\psch \left(\mathbf{\beta^\textrm{data}},\mathbf{\beta^\textrm{theory}}\right)= \sum\limits_{i=1}^{N_\textrm{bins}} \frac{\left( \sigma^\textrm{data}_i + \sum_{\mu}\Gamma_{i;\mu}^\textrm{data}\beta_\mu^\textrm{data} - \sigma^\textrm{theory}_i - \sum_{\nu}\Gamma_{i;\nu}^\textrm{theory}\beta_\nu^\textrm{theory} \right)^2}{\Delta_i^2}
+ \sum\limits_{\mu} \left(\beta_\mu^\textrm{data}\right)^2
+  \sum\limits_{\nu} \left(\beta_\nu^\textrm{theory}\right)^2,
\end{equation}
 
where experimental~(theoretical) uncertainties are included in the calculation using the nuisance parameter vectors   $\mathbf{\beta^\textrm{data}}\left(\mathbf{\beta^\textrm{theory}}\right)$ and their influence on the data and the theory predictions is described by the respective $\Gamma_{l;\rho}$ matrices.
The Latin indices run over bins of measurements  and the Greek indices render sources of uncertainties.
The measured cross-sections  and their theory predictions in each bin are represented by
$\sigma^\textrm{data}_i$ and  $\sigma^\textrm{theory}_i$, respectively.
Uncorrelated uncertainties in data are denoted by $\Delta_i$.
The theoretical uncertainties include those arising from renormalisation and factorisation scales variations, PDF uncertainties, uncertainties in calculations of non-perturbative and electroweak effects as well as from the
$\alphas(m_Z)$ uncertainty.  All experimental and theoretical systematic uncertainties are assumed to be independent of each other, and fully correlated across the bins of the measurement.
The negligible correlations of statistical uncertainties are not included in the \psch{} tests presented here.
 
The minimisation of  Eq.~\eqref{eqn:chi2}, for the case of symmetric systematic uncertainties, results in
a system of linear equations for the shifts of systematic uncertainties, $\beta_\rho$.
The asymmetries in systematic uncertainties are accounted for using an iterative procedure.
Here, the influences $\Gamma_{l;\rho} $ are recalculated  as
 
\begin{equation*}
\Gamma_{l;\rho} \to  {\Gamma}_{l;\rho}	+   {{\Omega}}_{l;\rho}  \beta_{\rho},
\end{equation*}
 
where $\Gamma_{l;\rho}  = \frac{1}{2}   \left(   \Gamma_{l;\rho}^+  - \Gamma_{l;\rho}^-\right) $
and $ \Omega_{l;\rho} = \frac{1}{2} \left( \Gamma_{l;\rho}^+  + \Gamma_{l;\rho}^-\right)$,
after each iteration using the shifts $\beta_\rho$ from the previous iteration.
The $\Gamma_{l;\rho}^+$ and $\Gamma_{l;\rho}^-$ are positive and negative components of systematic uncertainties, respectively.
The \psch{} values at the minimum provide a measure of the probability of compatibility between the  measurements and the predictions.

Table \ref{tab:chi2bin} shows a summary of the calculated \pschu{}, the first term in Eq.~\eqref{eqn:chi2}, together with \pschc, the sum of squared shifts of nuisance parameters, for each \ptj{} bin separately.
A good agreement between measurements and theory is seen for the fits in individual \ptj{} bins in the $\ptj>50$~\GeV{} range, with
not so good agreement in the  $25<\ptj<50$~\GeV{} range. The level of agreement between data and predictions is very similar for different PDF sets.
 
\begin{table}
\centering
\caption{\label{tab:chi2bin} Values of \pschu{} and \pschc{} evaluated for theory predictions corrected for non-perturbative and electroweak effects and measured \Zjets{} cross-sections.
The total $\chi^2$ is equal to the sum of \pschu{} and \pschc.  The fits are performed individually in each \ptj{} bin.
The predictions are calculated using several  NNLO QCD PDF sets and one NLO QCD PDF set, CT14nlo.}
\small
\begin{tabular}{l|r|rr|rr|rr|rr|rr}
\hline
\hline
\multicolumn{2}{c|}{}                 &   \multicolumn{2}{c|}{CT14nlo} &   \multicolumn{2}{c|}{CT14} & \multicolumn{2}{c|}{NNPDF3.1} & \multicolumn{2}{c|}{MMHT2014} & \multicolumn{2}{c}{ABMP16}  \\ \hline
\multicolumn{1}{c|}{\ptj{} range [\GeV]}   &  \multicolumn{1}{c|}{$n_\text{bins}$} & \multicolumn{1}{c}{\pschu}  &  \multicolumn{1}{c|}{\pschc} & \multicolumn{1}{c}{\pschu}  & \multicolumn{1}{c|}{\pschc} & \multicolumn{1}{c}{\pschu} & \multicolumn{1}{c|}{\pschc} & \multicolumn{1}{c}{\pschu} & \multicolumn{1}{c|}{\pschc} & \multicolumn{1}{c}{\pschu} & \multicolumn{1}{c}{\pschc} \\
\hline
\hline
$25<\ptj<50$      & 12   & 31.5  & 14.7  & 32.2  & 15.6  & 33.6  & 15.7  & 32.7  & 15.9  & 31.8  & 13.8 \\
$50<\ptj<100$     & 17   & 23.6  & 2.6 & 24.2  & 2.3 & 27.1  & 2.3 & 26.3  & 2.1 & 24.9  & 2.5 \\
$100<\ptj<200$    & 17   & 24.9  & 3.6 & 24.8  & 2.5 & 26.1  & 1.8 & 27.2  & 2.8 & 22.6  & 1.5 \\
$200<\ptj<300$    & 7    & 3.1 & 0.9 & 2.9 & 0.7 & 3.6 & 0.1 & 4.5 & 0.5 & 2.7 & 0.2 \\
$300<\ptj<400$    & 6    & 2.7 & 0.1 & 2.7 & 0.1 & 2.9 & 0.1 & 3.2 & 0.0 & 2.5 & 0.3 \\
$400<\ptj<1050$   & 4    & 1.9 & 0.4 & 1.9 & 0.4 & 1.9 & 0.5 & 2.0 & 0.3 & 1.7 & 0.8 \\
\hline
\hline
\end{tabular}
\end{table}

In addition to fits of the predictions  to measured cross-sections in the individual \ptj{} bins, all  measured data points  are  fitted 	simultaneously.  Several ranges of \ptj{} are considered. The results of the global fits are presented in  Table \ref{tab:chi2global}.
Very good agreement between measurement and calculation is observed when using the $\ptj>50$~\GeV\ bins, while not so good agreement is observed when the $25<\ptj<50$~\GeV{} bin is included in the global fit.
 
The results of the \psch{} tests strongly depend on what is assumed about the correlation of systematic uncertainties. In general, the correlations of uncertainties related to detector measurements  are carefully studied and well known \cite{PERF-2012-01,PERF-2016-04}.  In contrast, the assumption of 100\% correlations of uncertainties resulting from simple comparisons of two (or more) different MC simulations (two-point systematic uncertainties) are less justified.  In order to investigate the impact of these assumptions on the results of \psch{} tests performed in this section, the uncertainties that are derived from comparisons of two different MC models, namely uncertainties in the jet flavour composition and jet flavour response,  were split into two subcomponents \cite{STDM-2015-01,STDM-2016-03}. The first subcomponent is derived by multiplying the original nuisance parameter with a linear  function of \ptj{} and jet absolute rapidity  and the second subcomponent is constructed such that the sum in quadrature of both subcomponents is equal to the original nuisance parameter.  These decorrelations did not result in a large improvement in the $\chi^2$ values.
 
\begin{table}
\centering
\caption{\label{tab:chi2global} Values of \psch{} evaluated from the comparison of theory predictions corrected for non-perturbative and electroweak effects with the measured \Zjets{} cross-sections. The fits are performed globally in all bins of the measurement within several \ptj{} ranges. The predictions are calculated using several NNLO~QCD PDF sets and one NLO~QCD PDF set, CT14nlo.}
\small
\begin{tabular}{l|l|r|r|r|r|r}
\hline
\hline
& \ptj{} range [\GeV]  & CT14nlo   &   CT14 &  NNPDF3.1  &  MMHT2014 & ABMP16  \\
\hline
\hline
\multicolumn{6}{c}{$\ptj>25$~\GeV}\\
\hline
\multirow{6}{*}{\pschu}
& $25<\ptj<50$    & 38.9  & 40.5  & 42.3  & 41.3   &  38.7 \\
& $50<\ptj<100$   & 32.1  & 33.0  & 37.5  & 39.2   &  31.6  \\
& $100<\ptj<200$  & 26.4  & 27.8  & 31.0  & 31.7   &  27.8  \\
& $200<\ptj<300$  & 6.3   & 6.3   & 5.1   & 5.6    &  4.1 \\
& $300<\ptj<400$  & 2.9   & 3.0   & 2.9   & 3.1    &  2.5 \\
& $400<\ptj<1050$ & 2.2   & 2.4   & 2.2   & 2.3    &  1.7 \\
\hline
\pschc  &         & 21.2  & 19.8  & 19.3  & 18.7   &  17.8  \\
\hline
$\chi^2/n_\text{bins}$  & &129.9/63   &132.6/63 & 140.0/63 & 141.9/63 & 124.3/63 \\
\hline
\multicolumn{6}{c}{$\ptj>50$~\GeV}\\
\hline
\multirow{5}{*}{\pschu}
& $50<\ptj<100$    & 24.4    &   24.8  &  26.9   & 27.1  & 24.8  \\
& $100<\ptj<200$   & 24.4    &   24.6  &  26.6   & 27.7  & 22.7   \\
& $200<\ptj<300$   & 4.4     &  4.2    &  4.4    & 4.7   & 3.4   \\
& $300<\ptj<400$   & 2.7     &  2.8    &  3.0    & 3.1   & 2.5   \\
& $400<\ptj<1050$  & 3.6     &  4.0    &  3.8    & 3.9   & 2.9   \\
\hline
\pschc  &          & 6.5     & 4.7     &  4.3    & 5.1   & 4.1   \\
\hline
$\chi^2/n_\text{bins}$  &  & 66.1/51   & 65.2/51 & 69.0/51 & 71.6/51 & 60.4/51 \\
\hline
\multicolumn{6}{c}{$\ptj>100$~\GeV}\\
\hline
\multirow{4}{*}{\pschu}
& $100<\ptj<200$   & 24.8      & 25.0  & 25.9  & 26.6  & 22.4  \\
& $200<\ptj<300$   & 3.2       & 3.3   & 4.1   & 4.4   & 3.3  \\
& $300<\ptj<400$   & 2.7       & 2.8   & 3.0   & 3.1   & 2.6  \\
& $400<\ptj<1050$  & 3.4       & 3.8   & 3.6   & 3.6   & 3.3  \\
\hline
\pschc &           &  4.9      & 3.7   & 2.7   & 4.1   & 2.3  \\
\hline
$\chi^2/n_\text{bins}$  &    & 39.0/34  & 38.5/34  & 39.3/34 & 41.8/34 & 33.8/34 \\
\hline
\hline
\end{tabular}
\end{table}
 
\FloatBarrier
 
\section{Conclusions\label{sec:conclusions}}

The double-differential \Zjets{} \xs{}, with the \ZBoson{} decaying into an electron--positron pair,
is measured using proton--proton collision data with an integrated luminosity \LUMI{} collected by the \ATLAS{} experiment at the LHC in 2012 at \COME{} \CoM{} energy.
The measurement is performed as a function of the absolute jet rapidity and the jet transverse momentum.
 
The measured \xs{} is corrected for detector effects and the results are provided at the particle level.
The measurements are compared with theory predictions, calculated using the multi-leg matrix element MC generators \sherpa{} and \alpgenpythia{},
supplemented with parton shower simulations.
\sherpav{} and \alpgenpythia{} describe well the shape of the \Zjets{} distribution as a function of \absyj{},
but not so well as a function of \ptj{}.
\sherpattv{} is in good agreement with data in all bins of the measurement.

The parton-level fixed-order \Zjets{} predictions, corrected for hadronisation, underlying-event and QED radiation effects, agree with the data within the uncertainties.

The uncertainties in the measured \xss{} are about half of the theoretical uncertainties in the NLO calculations in most bins of the measurement
and are approximately similar to the uncertainties in the NNLO calculations.
 
The measured double-differential \Zjets{} \xs{} provides a precision input to constrain the parton distribution functions.

\FloatBarrier

\section*{Acknowledgements}
 
 
We thank CERN for the very successful operation of the LHC, as well as the
support staff from our institutions without whom ATLAS could not be
operated efficiently.
 
We acknowledge the support of ANPCyT, Argentina; YerPhI, Armenia; ARC, Australia; BMWFW and FWF, Austria; ANAS, Azerbaijan; SSTC, Belarus; CNPq and FAPESP, Brazil; NSERC, NRC and CFI, Canada; CERN; CONICYT, Chile; CAS, MOST and NSFC, China; COLCIENCIAS, Colombia; MSMT CR, MPO CR and VSC CR, Czech Republic; DNRF and DNSRC, Denmark; IN2P3-CNRS, CEA-DRF/IRFU, France; SRNSFG, Georgia; BMBF, HGF, and MPG, Germany; GSRT, Greece; RGC, Hong Kong SAR, China; ISF and Benoziyo Center, Israel; INFN, Italy; MEXT and JSPS, Japan; CNRST, Morocco; NWO, Netherlands; RCN, Norway; MNiSW and NCN, Poland; FCT, Portugal; MNE/IFA, Romania; MES of Russia and NRC KI, Russian Federation; JINR; MESTD, Serbia; MSSR, Slovakia; ARRS and MIZ\v{S}, Slovenia; DST/NRF, South Africa; MINECO, Spain; SRC and Wallenberg Foundation, Sweden; SERI, SNSF and Cantons of Bern and Geneva, Switzerland; MOST, Taiwan; TAEK, Turkey; STFC, United Kingdom; DOE and NSF, United States of America. In addition, individual groups and members have received support from BCKDF, CANARIE, CRC and Compute Canada, Canada; COST, ERC, ERDF, Horizon 2020, and Marie Sk{\l}odowska-Curie Actions, European Union; Investissements d' Avenir Labex and Idex, ANR, France; DFG and AvH Foundation, Germany; Herakleitos, Thales and Aristeia programmes co-financed by EU-ESF and the Greek NSRF, Greece; BSF-NSF and GIF, Israel; CERCA Programme Generalitat de Catalunya, Spain; The Royal Society and Leverhulme Trust, United Kingdom.
 
The crucial computing support from all WLCG partners is acknowledged gratefully, in particular from CERN, the ATLAS Tier-1 facilities at TRIUMF (Canada), NDGF (Denmark, Norway, Sweden), CC-IN2P3 (France), KIT/GridKA (Germany), INFN-CNAF (Italy), NL-T1 (Netherlands), PIC (Spain), ASGC (Taiwan), RAL (UK) and BNL (USA), the Tier-2 facilities worldwide and large non-WLCG resource providers. Major contributors of computing resources are listed in Ref.~\cite{ATL-GEN-PUB-2016-002}.
 

\clearpage
\appendix
\part*{Appendix}
\addcontentsline{toc}{part}{Appendix}
 
\section{Tables of measured \xss{}\label{sec:tables}}
 
 
\newcolumntype{R}[1]{>{\raggedleft}m{#1}}
\begin{landscape}
\begin{table}
\centering
\caption[]{The measured double-differential \Zjets{} production \xss{} as a function of \absyj{} in the $25\GeV < \ptj < 50\GeV$ range.
$\delta_{\text{data}}^{\text{stat}}$ and $\delta_{\text{MC}}^{\text{stat}}$ are the statistical uncertainties in data and MC simulation, respectively.
$\delta_{\text{tot}}^{\text{sys}}$ is the total systematic uncertainty and includes the following components:
uncertainties due to electron reconstruction ($\delta_{\text{rec}}^{\text{el}}$), identification ($\delta_{\text{ID}}^{\text{el}}$) and trigger ($\delta_{\text{trig}}^{\text{el}}$) efficiencies;
electron energy scale ($\delta_{\text{scale}}^{\text{el}}$) and energy resolution ($\delta_{\text{res}}^{\text{el}}$) uncertainties;
sum in quadrature of the uncertainties from JES in situ methods ($\delta_{\text{in situ}}^{\text{JES}}$);
sum in quadrature of the uncertainties from JES $\eta$-intercalibration methods ($\delta_{\eta\text{-int}}^{\text{JES}}$);
an uncertainty of the measured single-hadron response ($\delta_{\text{hadron}}^{\text{JES}}$);
MC non-closure uncertainty ($\delta_{\text{closure}}^{\text{JES}}$);
sum in quadrature of the uncertainties due to pile-up corrections of the jet kinematics ($\delta_{\text{pile-up}}^{\text{JES}}$);
sum in quadrature of the flavour-based uncertainties ($\delta_{\text{flavour}}^{\text{JES}}$);
punch-through uncertainty ($\delta_{\text{pthrough}}^{\text{JES}}$); JER uncertainty ($\delta^{\text{JER}}$);
JVF uncertainty ($\delta^{\text{JVF}}$); sum in quadrature of the unfolding uncertainties ($\delta^{\text{unf}}$);
sum in quadrature of the uncertainties due to MC generated backgrounds normalisation ($\delta_{\text{MC}}^{\text{bg}}$);
sum in quadrature of the uncertainty due to combined multijet and \Wjets{} backgrounds ($\delta_{\text{mult}}^{\text{bg}}$);
uncertainty due to jet quality selection ($\delta^{\text{qual}}$).
All uncertainties are given in \%. The luminosity uncertainty of 1.9\% is not shown and not included in the total uncertainty and its components.}
\tiny
\setlength\tabcolsep{2pt}
\begin{tabular}{c S[table-format=-4.3] S[table-format=-2.2] S[table-format=-2.2] R{0.8cm} | R{0.8cm}  R{0.8cm}  R{0.8cm}  R{0.8cm}  R{0.8cm}  R{0.8cm}  R{0.8cm}  R{0.8cm}  R{0.8cm}  R{0.8cm}  R{0.8cm}  R{0.8cm}  R{0.8cm}  R{0.8cm}  R{0.8cm}  R{0.8cm}  R{0.8cm}  R{0.8cm}  }
\hline
\rule{0pt}{10pt}
\absyj{} & \multicolumn{1}{c}{$\frac{\text{d}^2\sigma}{\text{d}\absyj \text{d}\ptj}$} & \multicolumn{1}{c}{$\delta_{\text{data}}^{\text{stat}}$} &  \multicolumn{1}{c}{$\delta_{\text{MC}}^{\text{stat}}$} &  \multicolumn{1}{c|}{$\delta_{\text{tot}}^{\text{sys}}$}  & \multicolumn{1}{c}{$\delta_{\text{rec}}^{\text{el}}$}  & \multicolumn{1}{c}{$\delta_{\text{ID}}^{\text{el}}$}  & \multicolumn{1}{c}{$\delta_{\text{trig}}^{\text{el}}$}  & \multicolumn{1}{c}{$\delta_{\text{scale}}^{\text{el}}$}  & \multicolumn{1}{c}{$\delta_{\text{res}}^{\text{el}}$}  & \multicolumn{1}{c}{$\delta_{\text{in situ}}^{\text{JES}}$}  & \multicolumn{1}{c}{$\delta_{\eta\text{-int}}^{\text{JES}}$}  & \multicolumn{1}{c}{$\delta_{\text{hadron}}^{\text{JES}}$}  & \multicolumn{1}{c}{$\delta_{\text{closure}}^{\text{JES}}$}  & \multicolumn{1}{c}{$\delta_{\text{pile-up}}^{\text{JES}}$}  & \multicolumn{1}{c}{$\delta_{\text{flavour}}^{\text{JES}}$}  & \multicolumn{1}{c}{$\delta_{\text{pthrough}}^{\text{JES}}$}  & \multicolumn{1}{c}{$\delta^{\text{JER}}$}  & \multicolumn{1}{c}{$\delta^{\text{JVF}}$}  & \multicolumn{1}{c}{$\delta^{\text{unf}}$}  & \multicolumn{1}{c}{$\delta_{\text{MC}}^{\text{bg}}$}  & \multicolumn{1}{c}{$\delta_{\text{mult}}^{\text{bg}}$}  & \multicolumn{1}{c}{$\delta^{\text{qual}}$}  \tabularnewline[8pt]
& \multicolumn{1}{c}{[fb/\GeV]} & \multicolumn{1}{c}{[\%]} & \multicolumn{1}{c}{[\%]} & \multicolumn{1}{c|}{[\%]} & \multicolumn{1}{c}{[\%]}  & \multicolumn{1}{c}{[\%]}  & \multicolumn{1}{c}{[\%]}  & \multicolumn{1}{c}{[\%]}  & \multicolumn{1}{c}{[\%]}  & \multicolumn{1}{c}{[\%]}  & \multicolumn{1}{c}{[\%]}  & \multicolumn{1}{c}{[\%]}  & \multicolumn{1}{c}{[\%]}  & \multicolumn{1}{c}{[\%]}  & \multicolumn{1}{c}{[\%]}  & \multicolumn{1}{c}{[\%]}  & \multicolumn{1}{c}{[\%]}  & \multicolumn{1}{c}{[\%]}  & \multicolumn{1}{c}{[\%]}  & \multicolumn{1}{c}{[\%]}  & \multicolumn{1}{c}{[\%]}  & \multicolumn{1}{c}{[\%]} \tabularnewline[2pt]
\hline
\rule{0pt}{10pt}
0.0--0.2 & 1643.603 & 0.42 & 0.51	 & \begin{tabular}{r} $+$7.28 \\ $-$7.37 \end{tabular}	 & \begin{tabular}{r} $+$0.08 \\ $-$0.08 \end{tabular}	 & \begin{tabular}{r} $+$0.23 \\ $-$0.27 \end{tabular}	 & \begin{tabular}{r} $+$0.31 \\ $-$0.44 \end{tabular}	 & \begin{tabular}{r} $-$0.16 \\ $+$0.14 \end{tabular}	 & \begin{tabular}{r} $+$0.01 \\ $-$0.01 \end{tabular}	 & \begin{tabular}{r} $+$3.04 \\ $-$3.04 \end{tabular}	 & \begin{tabular}{r} $+$0.30 \\ $-$0.49 \end{tabular}	 & \begin{tabular}{r} $+$0.00 \\ $-$0.01 \end{tabular}	 & \begin{tabular}{r} $+$0.01 \\ $-$0.01 \end{tabular}	 & \begin{tabular}{r} $+$1.63 \\ $-$1.68 \end{tabular}	 & \begin{tabular}{r} $+$2.95 \\ $-$3.04 \end{tabular}	 & \begin{tabular}{r} $+$0.00 \\ $-$0.02 \end{tabular}	 & \begin{tabular}{r} $-$3.83 \\ $+$3.83 \end{tabular}	 & \begin{tabular}{r} $+$0.47 \\ $-$0.72 \end{tabular}	 & \begin{tabular}{r} $+$2.84 \\ $-$2.84 \end{tabular}	 & \begin{tabular}{r} $+$0.06 \\ $-$0.05 \end{tabular}	 & \begin{tabular}{r} $+$0.14 \\ $-$0.38 \end{tabular}	 & \begin{tabular}{r} $+$1.00 \\ $-$1.00 \end{tabular}\tabularnewline[4pt]
0.2--0.4 & 1595.690 & 0.34 & 0.60	 & \begin{tabular}{r} $+$7.19 \\ $-$7.17 \end{tabular}	 & \begin{tabular}{r} $+$0.08 \\ $-$0.08 \end{tabular}	 & \begin{tabular}{r} $+$0.23 \\ $-$0.27 \end{tabular}	 & \begin{tabular}{r} $+$0.31 \\ $-$0.44 \end{tabular}	 & \begin{tabular}{r} $-$0.16 \\ $+$0.14 \end{tabular}	 & \begin{tabular}{r} $+$0.01 \\ $-$0.01 \end{tabular}	 & \begin{tabular}{r} $+$3.16 \\ $-$3.02 \end{tabular}	 & \begin{tabular}{r} $+$0.30 \\ $-$0.49 \end{tabular}	 & \begin{tabular}{r} $+$0.00 \\ $-$0.01 \end{tabular}	 & \begin{tabular}{r} $+$0.01 \\ $-$0.01 \end{tabular}	 & \begin{tabular}{r} $+$1.65 \\ $-$1.60 \end{tabular}	 & \begin{tabular}{r} $+$3.04 \\ $-$3.05 \end{tabular}	 & \begin{tabular}{r} $+$0.00 \\ $-$0.02 \end{tabular}	 & \begin{tabular}{r} $-$3.89 \\ $+$3.89 \end{tabular}	 & \begin{tabular}{r} $+$0.47 \\ $-$0.72 \end{tabular}	 & \begin{tabular}{r} $+$2.55 \\ $-$2.55 \end{tabular}	 & \begin{tabular}{r} $+$0.06 \\ $-$0.05 \end{tabular}	 & \begin{tabular}{r} $+$0.14 \\ $-$0.39 \end{tabular}	 & \begin{tabular}{r} $+$1.00 \\ $-$1.00 \end{tabular}\tabularnewline[4pt]
0.4--0.6 & 1587.440 & 0.37 & 0.60	 & \begin{tabular}{r} $+$7.57 \\ $-$7.69 \end{tabular}	 & \begin{tabular}{r} $+$0.08 \\ $-$0.08 \end{tabular}	 & \begin{tabular}{r} $+$0.23 \\ $-$0.27 \end{tabular}	 & \begin{tabular}{r} $+$0.31 \\ $-$0.44 \end{tabular}	 & \begin{tabular}{r} $-$0.16 \\ $+$0.14 \end{tabular}	 & \begin{tabular}{r} $+$0.01 \\ $-$0.01 \end{tabular}	 & \begin{tabular}{r} $+$3.17 \\ $-$3.14 \end{tabular}	 & \begin{tabular}{r} $+$0.30 \\ $-$0.49 \end{tabular}	 & \begin{tabular}{r} $+$0.00 \\ $-$0.01 \end{tabular}	 & \begin{tabular}{r} $+$0.01 \\ $-$0.01 \end{tabular}	 & \begin{tabular}{r} $+$1.59 \\ $-$1.72 \end{tabular}	 & \begin{tabular}{r} $+$3.08 \\ $-$3.23 \end{tabular}	 & \begin{tabular}{r} $+$0.00 \\ $-$0.02 \end{tabular}	 & \begin{tabular}{r} $-$4.17 \\ $+$4.17 \end{tabular}	 & \begin{tabular}{r} $+$0.46 \\ $-$0.67 \end{tabular}	 & \begin{tabular}{r} $+$2.86 \\ $-$2.86 \end{tabular}	 & \begin{tabular}{r} $+$0.06 \\ $-$0.05 \end{tabular}	 & \begin{tabular}{r} $+$0.15 \\ $-$0.39 \end{tabular}	 & \begin{tabular}{r} $+$1.00 \\ $-$1.00 \end{tabular}\tabularnewline[4pt]
0.6--0.8 & 1569.884 & 0.38 & 0.60	 & \begin{tabular}{r} $+$7.72 \\ $-$7.92 \end{tabular}	 & \begin{tabular}{r} $+$0.08 \\ $-$0.08 \end{tabular}	 & \begin{tabular}{r} $+$0.23 \\ $-$0.27 \end{tabular}	 & \begin{tabular}{r} $+$0.31 \\ $-$0.44 \end{tabular}	 & \begin{tabular}{r} $-$0.16 \\ $+$0.14 \end{tabular}	 & \begin{tabular}{r} $+$0.01 \\ $-$0.01 \end{tabular}	 & \begin{tabular}{r} $+$3.19 \\ $-$3.25 \end{tabular}	 & \begin{tabular}{r} $+$0.30 \\ $-$0.49 \end{tabular}	 & \begin{tabular}{r} $+$0.00 \\ $-$0.01 \end{tabular}	 & \begin{tabular}{r} $+$0.01 \\ $-$0.01 \end{tabular}	 & \begin{tabular}{r} $+$1.62 \\ $-$1.82 \end{tabular}	 & \begin{tabular}{r} $+$3.26 \\ $-$3.46 \end{tabular}	 & \begin{tabular}{r} $+$0.00 \\ $-$0.02 \end{tabular}	 & \begin{tabular}{r} $-$3.74 \\ $+$3.74 \end{tabular}	 & \begin{tabular}{r} $+$0.46 \\ $-$0.67 \end{tabular}	 & \begin{tabular}{r} $+$3.22 \\ $-$3.22 \end{tabular}	 & \begin{tabular}{r} $+$0.05 \\ $-$0.05 \end{tabular}	 & \begin{tabular}{r} $+$0.15 \\ $-$0.40 \end{tabular}	 & \begin{tabular}{r} $+$1.00 \\ $-$1.00 \end{tabular}\tabularnewline[4pt]
0.8--1.0 & 1520.883 & 0.36 & 0.59	 & \begin{tabular}{r} $+$7.80 \\ $-$7.85 \end{tabular}	 & \begin{tabular}{r} $+$0.08 \\ $-$0.08 \end{tabular}	 & \begin{tabular}{r} $+$0.23 \\ $-$0.27 \end{tabular}	 & \begin{tabular}{r} $+$0.31 \\ $-$0.33 \end{tabular}	 & \begin{tabular}{r} $-$0.16 \\ $+$0.14 \end{tabular}	 & \begin{tabular}{r} $+$0.01 \\ $-$0.01 \end{tabular}	 & \begin{tabular}{r} $+$3.33 \\ $-$3.25 \end{tabular}	 & \begin{tabular}{r} $+$0.30 \\ $-$0.49 \end{tabular}	 & \begin{tabular}{r} $+$0.00 \\ $-$0.01 \end{tabular}	 & \begin{tabular}{r} $+$0.01 \\ $-$0.01 \end{tabular}	 & \begin{tabular}{r} $+$1.80 \\ $-$1.88 \end{tabular}	 & \begin{tabular}{r} $+$3.54 \\ $-$3.61 \end{tabular}	 & \begin{tabular}{r} $+$0.00 \\ $-$0.02 \end{tabular}	 & \begin{tabular}{r} $-$2.88 \\ $+$2.88 \end{tabular}	 & \begin{tabular}{r} $+$0.46 \\ $-$0.56 \end{tabular}	 & \begin{tabular}{r} $+$3.48 \\ $-$3.48 \end{tabular}	 & \begin{tabular}{r} $+$0.05 \\ $-$0.05 \end{tabular}	 & \begin{tabular}{r} $+$0.15 \\ $-$0.39 \end{tabular}	 & \begin{tabular}{r} $+$1.00 \\ $-$1.00 \end{tabular}\tabularnewline[4pt]
1.0--1.2 & 1393.139 & 0.38 & 0.64	 & \begin{tabular}{r} $+$9.12 \\ $-$8.96 \end{tabular}	 & \begin{tabular}{r} $+$0.08 \\ $-$0.08 \end{tabular}	 & \begin{tabular}{r} $+$0.23 \\ $-$0.27 \end{tabular}	 & \begin{tabular}{r} $+$0.31 \\ $-$0.33 \end{tabular}	 & \begin{tabular}{r} $-$0.16 \\ $+$0.14 \end{tabular}	 & \begin{tabular}{r} $+$0.01 \\ $-$0.01 \end{tabular}	 & \begin{tabular}{r} $+$3.60 \\ $-$3.35 \end{tabular}	 & \begin{tabular}{r} $+$0.76 \\ $-$0.49 \end{tabular}	 & \begin{tabular}{r} $+$0.00 \\ $-$0.01 \end{tabular}	 & \begin{tabular}{r} $+$0.01 \\ $-$0.01 \end{tabular}	 & \begin{tabular}{r} $+$1.97 \\ $-$1.81 \end{tabular}	 & \begin{tabular}{r} $+$3.93 \\ $-$3.87 \end{tabular}	 & \begin{tabular}{r} $+$0.00 \\ $-$0.02 \end{tabular}	 & \begin{tabular}{r} $-$5.18 \\ $+$5.18 \end{tabular}	 & \begin{tabular}{r} $+$0.46 \\ $-$0.56 \end{tabular}	 & \begin{tabular}{r} $+$3.32 \\ $-$3.32 \end{tabular}	 & \begin{tabular}{r} $+$0.05 \\ $-$0.05 \end{tabular}	 & \begin{tabular}{r} $+$0.16 \\ $-$0.41 \end{tabular}	 & \begin{tabular}{r} $+$1.00 \\ $-$1.00 \end{tabular}\tabularnewline[4pt]
1.2--1.4 & 1377.328 & 0.47 & 0.57	 & \begin{tabular}{r} $+$12.60 \\ $-$12.33 \end{tabular}	 & \begin{tabular}{r} $+$0.08 \\ $-$0.08 \end{tabular}	 & \begin{tabular}{r} $+$0.23 \\ $-$0.27 \end{tabular}	 & \begin{tabular}{r} $+$0.31 \\ $-$0.33 \end{tabular}	 & \begin{tabular}{r} $-$0.16 \\ $+$0.14 \end{tabular}	 & \begin{tabular}{r} $+$0.01 \\ $-$0.01 \end{tabular}	 & \begin{tabular}{r} $+$3.69 \\ $-$3.29 \end{tabular}	 & \begin{tabular}{r} $+$0.76 \\ $-$0.69 \end{tabular}	 & \begin{tabular}{r} $+$0.00 \\ $-$0.01 \end{tabular}	 & \begin{tabular}{r} $+$0.01 \\ $-$0.01 \end{tabular}	 & \begin{tabular}{r} $+$2.04 \\ $-$1.78 \end{tabular}	 & \begin{tabular}{r} $+$4.49 \\ $-$4.14 \end{tabular}	 & \begin{tabular}{r} $+$0.00 \\ $-$0.02 \end{tabular}	 & \begin{tabular}{r} $-$8.88 \\ $+$8.88 \end{tabular}	 & \begin{tabular}{r} $+$0.46 \\ $-$0.56 \end{tabular}	 & \begin{tabular}{r} $+$4.47 \\ $-$4.47 \end{tabular}	 & \begin{tabular}{r} $+$0.05 \\ $-$0.05 \end{tabular}	 & \begin{tabular}{r} $+$0.17 \\ $-$0.41 \end{tabular}	 & \begin{tabular}{r} $+$1.00 \\ $-$1.00 \end{tabular}\tabularnewline[4pt]
1.4--1.6 & 1228.213 & 0.42 & 0.60	 & \begin{tabular}{r} $+$12.88 \\ $-$12.59 \end{tabular}	 & \begin{tabular}{r} $+$0.08 \\ $-$0.08 \end{tabular}	 & \begin{tabular}{r} $+$0.23 \\ $-$0.27 \end{tabular}	 & \begin{tabular}{r} $+$0.31 \\ $-$0.33 \end{tabular}	 & \begin{tabular}{r} $-$0.16 \\ $+$0.14 \end{tabular}	 & \begin{tabular}{r} $+$0.01 \\ $-$0.01 \end{tabular}	 & \begin{tabular}{r} $+$3.65 \\ $-$3.22 \end{tabular}	 & \begin{tabular}{r} $+$1.44 \\ $-$1.14 \end{tabular}	 & \begin{tabular}{r} $+$0.00 \\ $-$0.01 \end{tabular}	 & \begin{tabular}{r} $+$0.01 \\ $-$0.01 \end{tabular}	 & \begin{tabular}{r} $+$2.00 \\ $-$1.74 \end{tabular}	 & \begin{tabular}{r} $+$4.37 \\ $-$4.02 \end{tabular}	 & \begin{tabular}{r} $+$0.00 \\ $-$0.02 \end{tabular}	 & \begin{tabular}{r} $-$9.07 \\ $+$9.07 \end{tabular}	 & \begin{tabular}{r} $+$0.71 \\ $-$0.69 \end{tabular}	 & \begin{tabular}{r} $+$4.67 \\ $-$4.67 \end{tabular}	 & \begin{tabular}{r} $+$0.05 \\ $-$0.05 \end{tabular}	 & \begin{tabular}{r} $+$0.18 \\ $-$0.42 \end{tabular}	 & \begin{tabular}{r} $+$1.00 \\ $-$1.00 \end{tabular}\tabularnewline[4pt]
1.6--1.8 & 987.654 & 0.48 & 0.64	 & \begin{tabular}{r} $+$12.31 \\ $-$12.17 \end{tabular}	 & \begin{tabular}{r} $+$0.08 \\ $-$0.08 \end{tabular}	 & \begin{tabular}{r} $+$0.23 \\ $-$0.27 \end{tabular}	 & \begin{tabular}{r} $+$0.31 \\ $-$0.33 \end{tabular}	 & \begin{tabular}{r} $-$0.16 \\ $+$0.14 \end{tabular}	 & \begin{tabular}{r} $+$0.01 \\ $-$0.01 \end{tabular}	 & \begin{tabular}{r} $+$3.37 \\ $-$3.10 \end{tabular}	 & \begin{tabular}{r} $+$1.42 \\ $-$1.25 \end{tabular}	 & \begin{tabular}{r} $+$0.00 \\ $-$0.01 \end{tabular}	 & \begin{tabular}{r} $+$0.01 \\ $-$0.01 \end{tabular}	 & \begin{tabular}{r} $+$1.83 \\ $-$1.62 \end{tabular}	 & \begin{tabular}{r} $+$3.60 \\ $-$3.51 \end{tabular}	 & \begin{tabular}{r} $+$0.00 \\ $-$0.02 \end{tabular}	 & \begin{tabular}{r} $-$10.46 \\ $+$10.46 \end{tabular}	 & \begin{tabular}{r} $+$0.71 \\ $-$0.69 \end{tabular}	 & \begin{tabular}{r} $+$2.32 \\ $-$2.32 \end{tabular}	 & \begin{tabular}{r} $+$0.06 \\ $-$0.05 \end{tabular}	 & \begin{tabular}{r} $+$0.17 \\ $-$0.43 \end{tabular}	 & \begin{tabular}{r} $+$1.00 \\ $-$1.00 \end{tabular}\tabularnewline[4pt]
1.8--2.0 & 944.560 & 0.45 & 0.65	 & \begin{tabular}{r} $+$10.50 \\ $-$10.28 \end{tabular}	 & \begin{tabular}{r} $+$0.08 \\ $-$0.08 \end{tabular}	 & \begin{tabular}{r} $+$0.23 \\ $-$0.27 \end{tabular}	 & \begin{tabular}{r} $+$0.40 \\ $-$0.33 \end{tabular}	 & \begin{tabular}{r} $-$0.16 \\ $+$0.14 \end{tabular}	 & \begin{tabular}{r} $+$0.01 \\ $-$0.01 \end{tabular}	 & \begin{tabular}{r} $+$3.38 \\ $-$3.09 \end{tabular}	 & \begin{tabular}{r} $+$1.58 \\ $-$1.25 \end{tabular}	 & \begin{tabular}{r} $+$0.00 \\ $-$0.01 \end{tabular}	 & \begin{tabular}{r} $+$0.01 \\ $-$0.01 \end{tabular}	 & \begin{tabular}{r} $+$1.90 \\ $-$1.66 \end{tabular}	 & \begin{tabular}{r} $+$3.38 \\ $-$3.23 \end{tabular}	 & \begin{tabular}{r} $+$0.00 \\ $-$0.02 \end{tabular}	 & \begin{tabular}{r} $-$8.31 \\ $+$8.31 \end{tabular}	 & \begin{tabular}{r} $+$0.55 \\ $-$0.55 \end{tabular}	 & \begin{tabular}{r} $+$2.32 \\ $-$2.32 \end{tabular}	 & \begin{tabular}{r} $+$0.05 \\ $-$0.05 \end{tabular}	 & \begin{tabular}{r} $+$0.16 \\ $-$0.43 \end{tabular}	 & \begin{tabular}{r} $+$1.00 \\ $-$1.00 \end{tabular}\tabularnewline[4pt]
2.0--2.2 & 871.035 & 0.49 & 0.85	 & \begin{tabular}{r} $+$11.46 \\ $-$11.35 \end{tabular}	 & \begin{tabular}{r} $+$0.08 \\ $-$0.08 \end{tabular}	 & \begin{tabular}{r} $+$0.23 \\ $-$0.27 \end{tabular}	 & \begin{tabular}{r} $+$0.40 \\ $-$0.33 \end{tabular}	 & \begin{tabular}{r} $-$0.16 \\ $+$0.14 \end{tabular}	 & \begin{tabular}{r} $+$0.01 \\ $-$0.01 \end{tabular}	 & \begin{tabular}{r} $+$3.65 \\ $-$3.41 \end{tabular}	 & \begin{tabular}{r} $+$1.96 \\ $-$1.74 \end{tabular}	 & \begin{tabular}{r} $+$0.00 \\ $-$0.01 \end{tabular}	 & \begin{tabular}{r} $+$0.01 \\ $-$0.01 \end{tabular}	 & \begin{tabular}{r} $+$2.05 \\ $-$1.94 \end{tabular}	 & \begin{tabular}{r} $+$3.56 \\ $-$3.60 \end{tabular}	 & \begin{tabular}{r} $+$0.00 \\ $-$0.02 \end{tabular}	 & \begin{tabular}{r} $-$6.74 \\ $+$6.74 \end{tabular}	 & \begin{tabular}{r} $+$0.55 \\ $-$0.55 \end{tabular}	 & \begin{tabular}{r} $+$5.01 \\ $-$5.01 \end{tabular}	 & \begin{tabular}{r} $+$0.05 \\ $-$0.05 \end{tabular}	 & \begin{tabular}{r} $+$0.17 \\ $-$0.43 \end{tabular}	 & \begin{tabular}{r} $+$1.00 \\ $-$1.00 \end{tabular}\tabularnewline[4pt]
2.2--2.4 & 749.498 & 0.54 & 0.80	 & \begin{tabular}{r} $+$11.98 \\ $-$11.78 \end{tabular}	 & \begin{tabular}{r} $+$0.08 \\ $-$0.08 \end{tabular}	 & \begin{tabular}{r} $+$0.23 \\ $-$0.27 \end{tabular}	 & \begin{tabular}{r} $+$0.40 \\ $-$0.33 \end{tabular}	 & \begin{tabular}{r} $-$0.16 \\ $+$0.14 \end{tabular}	 & \begin{tabular}{r} $+$0.01 \\ $-$0.01 \end{tabular}	 & \begin{tabular}{r} $+$4.23 \\ $-$3.95 \end{tabular}	 & \begin{tabular}{r} $+$2.57 \\ $-$2.33 \end{tabular}	 & \begin{tabular}{r} $+$0.00 \\ $-$0.01 \end{tabular}	 & \begin{tabular}{r} $+$0.01 \\ $-$0.01 \end{tabular}	 & \begin{tabular}{r} $+$2.70 \\ $-$2.50 \end{tabular}	 & \begin{tabular}{r} $+$3.91 \\ $-$3.85 \end{tabular}	 & \begin{tabular}{r} $+$0.00 \\ $-$0.02 \end{tabular}	 & \begin{tabular}{r} $-$7.48 \\ $+$7.48 \end{tabular}	 & \begin{tabular}{r} $+$0.55 \\ $-$0.55 \end{tabular}	 & \begin{tabular}{r} $+$4.42 \\ $-$4.42 \end{tabular}	 & \begin{tabular}{r} $+$0.05 \\ $-$0.05 \end{tabular}	 & \begin{tabular}{r} $+$0.18 \\ $-$0.42 \end{tabular}	 & \begin{tabular}{r} $+$1.00 \\ $-$1.00 \end{tabular}\tabularnewline[4pt]
\hline
\end{tabular}
\end{table}
\end{landscape}

\newcolumntype{R}[1]{>{\raggedleft}m{#1}}
\begin{landscape}
\begin{table}
\centering
\caption[]{The measured double-differential \Zjets{} production \xss{} as a function of \absyj{} in the $50\GeV < \ptj < 100\GeV$ range.
$\delta_{\text{data}}^{\text{stat}}$ and $\delta_{\text{MC}}^{\text{stat}}$ are the statistical uncertainties in data and MC simulation, respectively.
$\delta_{\text{tot}}^{\text{sys}}$ is the total systematic uncertainty and includes the following components:
uncertainties due to electron reconstruction ($\delta_{\text{rec}}^{\text{el}}$), identification ($\delta_{\text{ID}}^{\text{el}}$) and trigger ($\delta_{\text{trig}}^{\text{el}}$) efficiencies;
electron energy scale ($\delta_{\text{scale}}^{\text{el}}$) and energy resolution ($\delta_{\text{res}}^{\text{el}}$) uncertainties;
sum in quadrature of the uncertainties from JES in situ methods ($\delta_{\text{in situ}}^{\text{JES}}$);
sum in quadrature of the uncertainties from JES $\eta$-intercalibration methods ($\delta_{\eta\text{-int}}^{\text{JES}}$);
an uncertainty of the measured single-hadron response ($\delta_{\text{hadron}}^{\text{JES}}$);
MC non-closure uncertainty ($\delta_{\text{closure}}^{\text{JES}}$);
sum in quadrature of the uncertainties due to pile-up corrections of the jet kinematics ($\delta_{\text{pile-up}}^{\text{JES}}$);
sum in quadrature of the flavour-based uncertainties ($\delta_{\text{flavour}}^{\text{JES}}$);
punch-through uncertainty ($\delta_{\text{pthrough}}^{\text{JES}}$); JER uncertainty ($\delta^{\text{JER}}$);
JVF uncertainty ($\delta^{\text{JVF}}$); sum in quadrature of the unfolding uncertainties ($\delta^{\text{unf}}$);
sum in quadrature of the uncertainties due to MC generated backgrounds normalisation ($\delta_{\text{MC}}^{\text{bg}}$);
sum in quadrature of the uncertainty due to combined multijet and \Wjets{} backgrounds ($\delta_{\text{mult}}^{\text{bg}}$);
uncertainty due to jet quality selection ($\delta^{\text{qual}}$).
All uncertainties are given in \%. The luminosity uncertainty of 1.9\% is not shown and not included in the total uncertainty and its components.}
\tiny
\setlength\tabcolsep{2pt}
\begin{tabular}{c S[table-format=-4.3] S[table-format=-2.2] S[table-format=-2.2] R{0.8cm} | R{0.8cm}  R{0.8cm}  R{0.8cm}  R{0.8cm}  R{0.8cm}  R{0.8cm}  R{0.8cm}  R{0.8cm}  R{0.8cm}  R{0.8cm}  R{0.8cm}  R{0.8cm}  R{0.8cm}  R{0.8cm}  R{0.8cm}  R{0.8cm}  R{0.8cm}  R{0.8cm}  }
\hline
\rule{0pt}{10pt}
\absyj{} & \multicolumn{1}{c}{$\frac{\text{d}^2\sigma}{\text{d}\absyj \text{d}\ptj}$} & \multicolumn{1}{c}{$\delta_{\text{data}}^{\text{stat}}$} &  \multicolumn{1}{c}{$\delta_{\text{MC}}^{\text{stat}}$} &  \multicolumn{1}{c|}{$\delta_{\text{tot}}^{\text{sys}}$}  & \multicolumn{1}{c}{$\delta_{\text{rec}}^{\text{el}}$}  & \multicolumn{1}{c}{$\delta_{\text{ID}}^{\text{el}}$}  & \multicolumn{1}{c}{$\delta_{\text{trig}}^{\text{el}}$}  & \multicolumn{1}{c}{$\delta_{\text{scale}}^{\text{el}}$}  & \multicolumn{1}{c}{$\delta_{\text{res}}^{\text{el}}$}  & \multicolumn{1}{c}{$\delta_{\text{in situ}}^{\text{JES}}$}  & \multicolumn{1}{c}{$\delta_{\eta\text{-int}}^{\text{JES}}$}  & \multicolumn{1}{c}{$\delta_{\text{hadron}}^{\text{JES}}$}  & \multicolumn{1}{c}{$\delta_{\text{closure}}^{\text{JES}}$}  & \multicolumn{1}{c}{$\delta_{\text{pile-up}}^{\text{JES}}$}  & \multicolumn{1}{c}{$\delta_{\text{flavour}}^{\text{JES}}$}  & \multicolumn{1}{c}{$\delta_{\text{pthrough}}^{\text{JES}}$}  & \multicolumn{1}{c}{$\delta^{\text{JER}}$}  & \multicolumn{1}{c}{$\delta^{\text{JVF}}$}  & \multicolumn{1}{c}{$\delta^{\text{unf}}$}  & \multicolumn{1}{c}{$\delta_{\text{MC}}^{\text{bg}}$}  & \multicolumn{1}{c}{$\delta_{\text{mult}}^{\text{bg}}$}  & \multicolumn{1}{c}{$\delta^{\text{qual}}$}  \tabularnewline[8pt]
& \multicolumn{1}{c}{[fb/\GeV]} & \multicolumn{1}{c}{[\%]} & \multicolumn{1}{c}{[\%]} & \multicolumn{1}{c|}{[\%]} & \multicolumn{1}{c}{[\%]}  & \multicolumn{1}{c}{[\%]}  & \multicolumn{1}{c}{[\%]}  & \multicolumn{1}{c}{[\%]}  & \multicolumn{1}{c}{[\%]}  & \multicolumn{1}{c}{[\%]}  & \multicolumn{1}{c}{[\%]}  & \multicolumn{1}{c}{[\%]}  & \multicolumn{1}{c}{[\%]}  & \multicolumn{1}{c}{[\%]}  & \multicolumn{1}{c}{[\%]}  & \multicolumn{1}{c}{[\%]}  & \multicolumn{1}{c}{[\%]}  & \multicolumn{1}{c}{[\%]}  & \multicolumn{1}{c}{[\%]}  & \multicolumn{1}{c}{[\%]}  & \multicolumn{1}{c}{[\%]}  & \multicolumn{1}{c}{[\%]} \tabularnewline[2pt]
\hline
\rule{0pt}{10pt}
0.0--0.2 & 349.964 & 0.56 & 0.80	 & \begin{tabular}{r} $+$3.97 \\ $-$3.92 \end{tabular}	 & \begin{tabular}{r} $+$0.03 \\ $-$0.08 \end{tabular}	 & \begin{tabular}{r} $+$0.15 \\ $-$0.21 \end{tabular}	 & \begin{tabular}{r} $+$0.24 \\ $-$0.29 \end{tabular}	 & \begin{tabular}{r} $-$0.25 \\ $+$0.19 \end{tabular}	 & \begin{tabular}{r} $+$0.00 \\ $-$0.05 \end{tabular}	 & \begin{tabular}{r} $+$2.67 \\ $-$2.61 \end{tabular}	 & \begin{tabular}{r} $+$0.31 \\ $-$0.26 \end{tabular}	 & \begin{tabular}{r} $+$0.00 \\ $-$0.03 \end{tabular}	 & \begin{tabular}{r} $+$0.00 \\ $-$0.02 \end{tabular}	 & \begin{tabular}{r} $+$1.02 \\ $-$0.92 \end{tabular}	 & \begin{tabular}{r} $+$0.81 \\ $-$0.77 \end{tabular}	 & \begin{tabular}{r} $+$0.00 \\ $-$0.02 \end{tabular}	 & \begin{tabular}{r} $-$1.47 \\ $+$1.47 \end{tabular}	 & \begin{tabular}{r} $-$0.46 \\ $+$0.49 \end{tabular}	 & \begin{tabular}{r} $+$1.28 \\ $-$1.28 \end{tabular}	 & \begin{tabular}{r} $+$0.17 \\ $-$0.15 \end{tabular}	 & \begin{tabular}{r} $+$0.15 \\ $-$0.42 \end{tabular}	 & \begin{tabular}{r} $+$1.00 \\ $-$1.00 \end{tabular}\tabularnewline[4pt]
0.2--0.4 & 352.217 & 0.71 & 0.80	 & \begin{tabular}{r} $+$3.90 \\ $-$3.98 \end{tabular}	 & \begin{tabular}{r} $+$0.03 \\ $-$0.08 \end{tabular}	 & \begin{tabular}{r} $+$0.15 \\ $-$0.21 \end{tabular}	 & \begin{tabular}{r} $+$0.24 \\ $-$0.29 \end{tabular}	 & \begin{tabular}{r} $-$0.25 \\ $+$0.19 \end{tabular}	 & \begin{tabular}{r} $+$0.00 \\ $-$0.05 \end{tabular}	 & \begin{tabular}{r} $+$2.56 \\ $-$2.70 \end{tabular}	 & \begin{tabular}{r} $+$0.31 \\ $-$0.26 \end{tabular}	 & \begin{tabular}{r} $+$0.00 \\ $-$0.03 \end{tabular}	 & \begin{tabular}{r} $+$0.00 \\ $-$0.02 \end{tabular}	 & \begin{tabular}{r} $+$1.02 \\ $-$0.92 \end{tabular}	 & \begin{tabular}{r} $+$0.81 \\ $-$0.77 \end{tabular}	 & \begin{tabular}{r} $+$0.00 \\ $-$0.02 \end{tabular}	 & \begin{tabular}{r} $-$1.47 \\ $+$1.47 \end{tabular}	 & \begin{tabular}{r} $-$0.46 \\ $+$0.49 \end{tabular}	 & \begin{tabular}{r} $+$1.28 \\ $-$1.28 \end{tabular}	 & \begin{tabular}{r} $+$0.16 \\ $-$0.15 \end{tabular}	 & \begin{tabular}{r} $+$0.15 \\ $-$0.42 \end{tabular}	 & \begin{tabular}{r} $+$1.00 \\ $-$1.00 \end{tabular}\tabularnewline[4pt]
0.4--0.6 & 338.924 & 0.74 & 0.81	 & \begin{tabular}{r} $+$4.04 \\ $-$4.24 \end{tabular}	 & \begin{tabular}{r} $+$0.03 \\ $-$0.08 \end{tabular}	 & \begin{tabular}{r} $+$0.15 \\ $-$0.21 \end{tabular}	 & \begin{tabular}{r} $+$0.24 \\ $-$0.29 \end{tabular}	 & \begin{tabular}{r} $-$0.25 \\ $+$0.19 \end{tabular}	 & \begin{tabular}{r} $+$0.00 \\ $-$0.05 \end{tabular}	 & \begin{tabular}{r} $+$2.52 \\ $-$2.86 \end{tabular}	 & \begin{tabular}{r} $+$0.31 \\ $-$0.26 \end{tabular}	 & \begin{tabular}{r} $+$0.00 \\ $-$0.03 \end{tabular}	 & \begin{tabular}{r} $+$0.00 \\ $-$0.02 \end{tabular}	 & \begin{tabular}{r} $+$1.07 \\ $-$0.92 \end{tabular}	 & \begin{tabular}{r} $+$0.81 \\ $-$0.77 \end{tabular}	 & \begin{tabular}{r} $+$0.00 \\ $-$0.02 \end{tabular}	 & \begin{tabular}{r} $-$1.60 \\ $+$1.60 \end{tabular}	 & \begin{tabular}{r} $-$0.46 \\ $+$0.49 \end{tabular}	 & \begin{tabular}{r} $+$1.43 \\ $-$1.43 \end{tabular}	 & \begin{tabular}{r} $+$0.16 \\ $-$0.15 \end{tabular}	 & \begin{tabular}{r} $+$0.15 \\ $-$0.43 \end{tabular}	 & \begin{tabular}{r} $+$1.00 \\ $-$1.00 \end{tabular}\tabularnewline[4pt]
0.6--0.8 & 328.606 & 0.72 & 0.93	 & \begin{tabular}{r} $+$4.23 \\ $-$4.38 \end{tabular}	 & \begin{tabular}{r} $+$0.03 \\ $-$0.08 \end{tabular}	 & \begin{tabular}{r} $+$0.15 \\ $-$0.21 \end{tabular}	 & \begin{tabular}{r} $+$0.24 \\ $-$0.29 \end{tabular}	 & \begin{tabular}{r} $-$0.25 \\ $+$0.19 \end{tabular}	 & \begin{tabular}{r} $+$0.00 \\ $-$0.05 \end{tabular}	 & \begin{tabular}{r} $+$2.75 \\ $-$2.86 \end{tabular}	 & \begin{tabular}{r} $+$0.31 \\ $-$0.26 \end{tabular}	 & \begin{tabular}{r} $+$0.00 \\ $-$0.03 \end{tabular}	 & \begin{tabular}{r} $+$0.00 \\ $-$0.02 \end{tabular}	 & \begin{tabular}{r} $+$1.07 \\ $-$1.08 \end{tabular}	 & \begin{tabular}{r} $+$1.28 \\ $-$1.44 \end{tabular}	 & \begin{tabular}{r} $+$0.00 \\ $-$0.02 \end{tabular}	 & \begin{tabular}{r} $-$1.47 \\ $+$1.47 \end{tabular}	 & \begin{tabular}{r} $-$0.46 \\ $+$0.49 \end{tabular}	 & \begin{tabular}{r} $+$1.41 \\ $-$1.41 \end{tabular}	 & \begin{tabular}{r} $+$0.16 \\ $-$0.14 \end{tabular}	 & \begin{tabular}{r} $+$0.16 \\ $-$0.43 \end{tabular}	 & \begin{tabular}{r} $+$1.00 \\ $-$1.00 \end{tabular}\tabularnewline[4pt]
0.8--1.0 & 303.475 & 0.69 & 0.87	 & \begin{tabular}{r} $+$4.19 \\ $-$4.03 \end{tabular}	 & \begin{tabular}{r} $+$0.03 \\ $-$0.08 \end{tabular}	 & \begin{tabular}{r} $+$0.15 \\ $-$0.21 \end{tabular}	 & \begin{tabular}{r} $+$0.24 \\ $-$0.29 \end{tabular}	 & \begin{tabular}{r} $-$0.25 \\ $+$0.19 \end{tabular}	 & \begin{tabular}{r} $+$0.00 \\ $-$0.05 \end{tabular}	 & \begin{tabular}{r} $+$2.98 \\ $-$2.76 \end{tabular}	 & \begin{tabular}{r} $+$0.31 \\ $-$0.26 \end{tabular}	 & \begin{tabular}{r} $+$0.00 \\ $-$0.03 \end{tabular}	 & \begin{tabular}{r} $+$0.00 \\ $-$0.02 \end{tabular}	 & \begin{tabular}{r} $+$1.07 \\ $-$1.08 \end{tabular}	 & \begin{tabular}{r} $+$1.50 \\ $-$1.44 \end{tabular}	 & \begin{tabular}{r} $+$0.00 \\ $-$0.02 \end{tabular}	 & \begin{tabular}{r} $-$1.21 \\ $+$1.21 \end{tabular}	 & \begin{tabular}{r} $-$0.46 \\ $+$0.49 \end{tabular}	 & \begin{tabular}{r} $+$1.07 \\ $-$1.07 \end{tabular}	 & \begin{tabular}{r} $+$0.15 \\ $-$0.14 \end{tabular}	 & \begin{tabular}{r} $+$0.17 \\ $-$0.45 \end{tabular}	 & \begin{tabular}{r} $+$1.00 \\ $-$1.00 \end{tabular}\tabularnewline[4pt]
1.0--1.2 & 274.407 & 0.71 & 1.05	 & \begin{tabular}{r} $+$4.00 \\ $-$4.22 \end{tabular}	 & \begin{tabular}{r} $+$0.03 \\ $-$0.08 \end{tabular}	 & \begin{tabular}{r} $+$0.15 \\ $-$0.21 \end{tabular}	 & \begin{tabular}{r} $+$0.24 \\ $-$0.29 \end{tabular}	 & \begin{tabular}{r} $-$0.25 \\ $+$0.19 \end{tabular}	 & \begin{tabular}{r} $+$0.00 \\ $-$0.05 \end{tabular}	 & \begin{tabular}{r} $+$2.74 \\ $-$2.83 \end{tabular}	 & \begin{tabular}{r} $+$0.31 \\ $-$0.96 \end{tabular}	 & \begin{tabular}{r} $+$0.00 \\ $-$0.03 \end{tabular}	 & \begin{tabular}{r} $+$0.00 \\ $-$0.02 \end{tabular}	 & \begin{tabular}{r} $+$1.01 \\ $-$1.19 \end{tabular}	 & \begin{tabular}{r} $+$1.49 \\ $-$1.44 \end{tabular}	 & \begin{tabular}{r} $+$0.00 \\ $-$0.02 \end{tabular}	 & \begin{tabular}{r} $-$1.21 \\ $+$1.21 \end{tabular}	 & \begin{tabular}{r} $-$0.46 \\ $+$0.49 \end{tabular}	 & \begin{tabular}{r} $+$1.07 \\ $-$1.07 \end{tabular}	 & \begin{tabular}{r} $+$0.15 \\ $-$0.14 \end{tabular}	 & \begin{tabular}{r} $+$0.19 \\ $-$0.46 \end{tabular}	 & \begin{tabular}{r} $+$1.00 \\ $-$1.00 \end{tabular}\tabularnewline[4pt]
1.2--1.4 & 261.553 & 0.81 & 0.84	 & \begin{tabular}{r} $+$4.28 \\ $-$4.44 \end{tabular}	 & \begin{tabular}{r} $+$0.03 \\ $-$0.08 \end{tabular}	 & \begin{tabular}{r} $+$0.15 \\ $-$0.21 \end{tabular}	 & \begin{tabular}{r} $+$0.24 \\ $-$0.29 \end{tabular}	 & \begin{tabular}{r} $-$0.25 \\ $+$0.19 \end{tabular}	 & \begin{tabular}{r} $+$0.00 \\ $-$0.05 \end{tabular}	 & \begin{tabular}{r} $+$2.86 \\ $-$2.86 \end{tabular}	 & \begin{tabular}{r} $+$1.25 \\ $-$0.96 \end{tabular}	 & \begin{tabular}{r} $+$0.00 \\ $-$0.03 \end{tabular}	 & \begin{tabular}{r} $+$0.00 \\ $-$0.02 \end{tabular}	 & \begin{tabular}{r} $+$1.01 \\ $-$1.19 \end{tabular}	 & \begin{tabular}{r} $+$1.54 \\ $-$1.94 \end{tabular}	 & \begin{tabular}{r} $+$0.00 \\ $-$0.02 \end{tabular}	 & \begin{tabular}{r} $-$1.21 \\ $+$1.21 \end{tabular}	 & \begin{tabular}{r} $-$0.46 \\ $+$0.49 \end{tabular}	 & \begin{tabular}{r} $+$1.07 \\ $-$1.07 \end{tabular}	 & \begin{tabular}{r} $+$0.14 \\ $-$0.13 \end{tabular}	 & \begin{tabular}{r} $+$0.21 \\ $-$0.49 \end{tabular}	 & \begin{tabular}{r} $+$1.00 \\ $-$1.00 \end{tabular}\tabularnewline[4pt]
1.4--1.6 & 233.170 & 0.75 & 1.02	 & \begin{tabular}{r} $+$4.81 \\ $-$4.71 \end{tabular}	 & \begin{tabular}{r} $+$0.03 \\ $-$0.08 \end{tabular}	 & \begin{tabular}{r} $+$0.15 \\ $-$0.21 \end{tabular}	 & \begin{tabular}{r} $+$0.24 \\ $-$0.29 \end{tabular}	 & \begin{tabular}{r} $-$0.25 \\ $+$0.19 \end{tabular}	 & \begin{tabular}{r} $+$0.00 \\ $-$0.05 \end{tabular}	 & \begin{tabular}{r} $+$2.82 \\ $-$2.74 \end{tabular}	 & \begin{tabular}{r} $+$1.25 \\ $-$0.96 \end{tabular}	 & \begin{tabular}{r} $+$0.00 \\ $-$0.03 \end{tabular}	 & \begin{tabular}{r} $+$0.00 \\ $-$0.02 \end{tabular}	 & \begin{tabular}{r} $+$1.01 \\ $-$1.05 \end{tabular}	 & \begin{tabular}{r} $+$1.78 \\ $-$1.69 \end{tabular}	 & \begin{tabular}{r} $+$0.00 \\ $-$0.02 \end{tabular}	 & \begin{tabular}{r} $-$1.42 \\ $+$1.42 \end{tabular}	 & \begin{tabular}{r} $-$0.46 \\ $+$0.40 \end{tabular}	 & \begin{tabular}{r} $+$1.73 \\ $-$1.73 \end{tabular}	 & \begin{tabular}{r} $+$0.13 \\ $-$0.12 \end{tabular}	 & \begin{tabular}{r} $+$0.37 \\ $-$0.60 \end{tabular}	 & \begin{tabular}{r} $+$1.00 \\ $-$1.00 \end{tabular}\tabularnewline[4pt]
1.6--1.8 & 192.405 & 0.92 & 1.16	 & \begin{tabular}{r} $+$5.40 \\ $-$5.11 \end{tabular}	 & \begin{tabular}{r} $+$0.03 \\ $-$0.08 \end{tabular}	 & \begin{tabular}{r} $+$0.15 \\ $-$0.21 \end{tabular}	 & \begin{tabular}{r} $+$0.24 \\ $-$0.29 \end{tabular}	 & \begin{tabular}{r} $-$0.25 \\ $+$0.19 \end{tabular}	 & \begin{tabular}{r} $+$0.00 \\ $-$0.05 \end{tabular}	 & \begin{tabular}{r} $+$3.09 \\ $-$2.92 \end{tabular}	 & \begin{tabular}{r} $+$2.10 \\ $-$1.97 \end{tabular}	 & \begin{tabular}{r} $+$0.00 \\ $-$0.03 \end{tabular}	 & \begin{tabular}{r} $+$0.00 \\ $-$0.02 \end{tabular}	 & \begin{tabular}{r} $+$1.34 \\ $-$1.05 \end{tabular}	 & \begin{tabular}{r} $+$2.01 \\ $-$1.69 \end{tabular}	 & \begin{tabular}{r} $+$0.00 \\ $-$0.02 \end{tabular}	 & \begin{tabular}{r} $-$1.42 \\ $+$1.42 \end{tabular}	 & \begin{tabular}{r} $-$0.46 \\ $+$0.40 \end{tabular}	 & \begin{tabular}{r} $+$1.73 \\ $-$1.73 \end{tabular}	 & \begin{tabular}{r} $+$0.13 \\ $-$0.12 \end{tabular}	 & \begin{tabular}{r} $+$0.19 \\ $-$0.51 \end{tabular}	 & \begin{tabular}{r} $+$1.00 \\ $-$1.00 \end{tabular}\tabularnewline[4pt]
1.8--2.0 & 174.081 & 0.90 & 1.18	 & \begin{tabular}{r} $+$6.01 \\ $-$5.43 \end{tabular}	 & \begin{tabular}{r} $+$0.03 \\ $-$0.08 \end{tabular}	 & \begin{tabular}{r} $+$0.15 \\ $-$0.21 \end{tabular}	 & \begin{tabular}{r} $+$0.24 \\ $-$0.29 \end{tabular}	 & \begin{tabular}{r} $-$0.25 \\ $+$0.19 \end{tabular}	 & \begin{tabular}{r} $+$0.00 \\ $-$0.05 \end{tabular}	 & \begin{tabular}{r} $+$3.38 \\ $-$3.05 \end{tabular}	 & \begin{tabular}{r} $+$2.61 \\ $-$2.27 \end{tabular}	 & \begin{tabular}{r} $+$0.00 \\ $-$0.03 \end{tabular}	 & \begin{tabular}{r} $+$0.00 \\ $-$0.02 \end{tabular}	 & \begin{tabular}{r} $+$1.34 \\ $-$1.05 \end{tabular}	 & \begin{tabular}{r} $+$2.25 \\ $-$1.66 \end{tabular}	 & \begin{tabular}{r} $+$0.00 \\ $-$0.02 \end{tabular}	 & \begin{tabular}{r} $-$1.89 \\ $+$1.89 \end{tabular}	 & \begin{tabular}{r} $-$0.34 \\ $+$0.40 \end{tabular}	 & \begin{tabular}{r} $+$1.73 \\ $-$1.73 \end{tabular}	 & \begin{tabular}{r} $+$0.12 \\ $-$0.11 \end{tabular}	 & \begin{tabular}{r} $+$0.20 \\ $-$0.55 \end{tabular}	 & \begin{tabular}{r} $+$1.00 \\ $-$1.00 \end{tabular}\tabularnewline[4pt]
2.0--2.2 & 145.578 & 0.94 & 1.11	 & \begin{tabular}{r} $+$6.30 \\ $-$5.73 \end{tabular}	 & \begin{tabular}{r} $+$0.03 \\ $-$0.08 \end{tabular}	 & \begin{tabular}{r} $+$0.15 \\ $-$0.21 \end{tabular}	 & \begin{tabular}{r} $+$0.24 \\ $-$0.29 \end{tabular}	 & \begin{tabular}{r} $-$0.25 \\ $+$0.19 \end{tabular}	 & \begin{tabular}{r} $+$0.00 \\ $-$0.05 \end{tabular}	 & \begin{tabular}{r} $+$3.48 \\ $-$3.02 \end{tabular}	 & \begin{tabular}{r} $+$2.80 \\ $-$2.36 \end{tabular}	 & \begin{tabular}{r} $+$0.00 \\ $-$0.03 \end{tabular}	 & \begin{tabular}{r} $+$0.00 \\ $-$0.02 \end{tabular}	 & \begin{tabular}{r} $+$1.40 \\ $-$0.92 \end{tabular}	 & \begin{tabular}{r} $+$1.91 \\ $-$1.66 \end{tabular}	 & \begin{tabular}{r} $+$0.00 \\ $-$0.02 \end{tabular}	 & \begin{tabular}{r} $-$1.89 \\ $+$1.89 \end{tabular}	 & \begin{tabular}{r} $-$0.34 \\ $+$0.40 \end{tabular}	 & \begin{tabular}{r} $+$2.16 \\ $-$2.16 \end{tabular}	 & \begin{tabular}{r} $+$0.11 \\ $-$0.10 \end{tabular}	 & \begin{tabular}{r} $+$0.20 \\ $-$0.56 \end{tabular}	 & \begin{tabular}{r} $+$1.00 \\ $-$1.00 \end{tabular}\tabularnewline[4pt]
2.2--2.4 & 117.333 & 1.08 & 1.37	 & \begin{tabular}{r} $+$6.63 \\ $-$5.91 \end{tabular}	 & \begin{tabular}{r} $+$0.03 \\ $-$0.08 \end{tabular}	 & \begin{tabular}{r} $+$0.15 \\ $-$0.21 \end{tabular}	 & \begin{tabular}{r} $+$0.24 \\ $-$0.29 \end{tabular}	 & \begin{tabular}{r} $-$0.25 \\ $+$0.19 \end{tabular}	 & \begin{tabular}{r} $+$0.00 \\ $-$0.05 \end{tabular}	 & \begin{tabular}{r} $+$3.56 \\ $-$2.83 \end{tabular}	 & \begin{tabular}{r} $+$2.78 \\ $-$2.23 \end{tabular}	 & \begin{tabular}{r} $+$0.00 \\ $-$0.03 \end{tabular}	 & \begin{tabular}{r} $+$0.00 \\ $-$0.02 \end{tabular}	 & \begin{tabular}{r} $+$1.40 \\ $-$0.92 \end{tabular}	 & \begin{tabular}{r} $+$1.91 \\ $-$1.66 \end{tabular}	 & \begin{tabular}{r} $+$0.00 \\ $-$0.02 \end{tabular}	 & \begin{tabular}{r} $-$0.76 \\ $+$0.76 \end{tabular}	 & \begin{tabular}{r} $-$0.34 \\ $+$0.40 \end{tabular}	 & \begin{tabular}{r} $+$2.83 \\ $-$2.83 \end{tabular}	 & \begin{tabular}{r} $+$0.11 \\ $-$0.10 \end{tabular}	 & \begin{tabular}{r} $+$0.21 \\ $-$0.58 \end{tabular}	 & \begin{tabular}{r} $+$1.00 \\ $-$1.00 \end{tabular}\tabularnewline[4pt]
2.4--2.6 & 98.813 & 1.31 & 1.42	 & \begin{tabular}{r} $+$6.32 \\ $-$5.97 \end{tabular}	 & \begin{tabular}{r} $+$0.03 \\ $-$0.08 \end{tabular}	 & \begin{tabular}{r} $+$0.15 \\ $-$0.21 \end{tabular}	 & \begin{tabular}{r} $+$0.24 \\ $-$0.29 \end{tabular}	 & \begin{tabular}{r} $-$0.25 \\ $+$0.19 \end{tabular}	 & \begin{tabular}{r} $+$0.00 \\ $-$0.05 \end{tabular}	 & \begin{tabular}{r} $+$3.32 \\ $-$2.87 \end{tabular}	 & \begin{tabular}{r} $+$2.56 \\ $-$2.20 \end{tabular}	 & \begin{tabular}{r} $+$0.00 \\ $-$0.03 \end{tabular}	 & \begin{tabular}{r} $+$0.00 \\ $-$0.02 \end{tabular}	 & \begin{tabular}{r} $+$0.87 \\ $-$0.92 \end{tabular}	 & \begin{tabular}{r} $+$1.91 \\ $-$1.82 \end{tabular}	 & \begin{tabular}{r} $+$0.00 \\ $-$0.02 \end{tabular}	 & \begin{tabular}{r} $-$0.76 \\ $+$0.76 \end{tabular}	 & \begin{tabular}{r} $-$0.34 \\ $+$0.40 \end{tabular}	 & \begin{tabular}{r} $+$2.83 \\ $-$2.83 \end{tabular}	 & \begin{tabular}{r} $+$0.11 \\ $-$0.10 \end{tabular}	 & \begin{tabular}{r} $+$0.23 \\ $-$0.64 \end{tabular}	 & \begin{tabular}{r} $+$1.00 \\ $-$1.00 \end{tabular}\tabularnewline[4pt]
2.6--2.8 & 75.900 & 1.47 & 1.67	 & \begin{tabular}{r} $+$7.88 \\ $-$7.43 \end{tabular}	 & \begin{tabular}{r} $+$0.03 \\ $-$0.08 \end{tabular}	 & \begin{tabular}{r} $+$0.15 \\ $-$0.21 \end{tabular}	 & \begin{tabular}{r} $+$0.24 \\ $-$0.29 \end{tabular}	 & \begin{tabular}{r} $-$0.25 \\ $+$0.19 \end{tabular}	 & \begin{tabular}{r} $+$0.00 \\ $-$0.05 \end{tabular}	 & \begin{tabular}{r} $+$3.21 \\ $-$2.85 \end{tabular}	 & \begin{tabular}{r} $+$2.88 \\ $-$2.20 \end{tabular}	 & \begin{tabular}{r} $+$0.00 \\ $-$0.03 \end{tabular}	 & \begin{tabular}{r} $+$0.00 \\ $-$0.02 \end{tabular}	 & \begin{tabular}{r} $+$0.87 \\ $-$0.92 \end{tabular}	 & \begin{tabular}{r} $+$2.25 \\ $-$1.82 \end{tabular}	 & \begin{tabular}{r} $+$0.00 \\ $-$0.02 \end{tabular}	 & \begin{tabular}{r} $-$0.76 \\ $+$0.76 \end{tabular}	 & \begin{tabular}{r} $-$0.34 \\ $+$0.40 \end{tabular}	 & \begin{tabular}{r} $+$4.23 \\ $-$4.23 \end{tabular}	 & \begin{tabular}{r} $+$0.10 \\ $-$0.10 \end{tabular}	 & \begin{tabular}{r} $+$0.24 \\ $-$0.67 \end{tabular}	 & \begin{tabular}{r} $+$1.00 \\ $-$1.00 \end{tabular}\tabularnewline[4pt]
2.8--3.0 & 58.038 & 1.59 & 2.21	 & \begin{tabular}{r} $+$7.67 \\ $-$7.89 \end{tabular}	 & \begin{tabular}{r} $+$0.03 \\ $-$0.08 \end{tabular}	 & \begin{tabular}{r} $+$0.15 \\ $-$0.21 \end{tabular}	 & \begin{tabular}{r} $+$0.24 \\ $-$0.29 \end{tabular}	 & \begin{tabular}{r} $-$0.25 \\ $+$0.19 \end{tabular}	 & \begin{tabular}{r} $+$0.00 \\ $-$0.05 \end{tabular}	 & \begin{tabular}{r} $+$3.08 \\ $-$2.85 \end{tabular}	 & \begin{tabular}{r} $+$2.88 \\ $-$3.79 \end{tabular}	 & \begin{tabular}{r} $+$0.00 \\ $-$0.03 \end{tabular}	 & \begin{tabular}{r} $+$0.00 \\ $-$0.02 \end{tabular}	 & \begin{tabular}{r} $+$0.87 \\ $-$0.92 \end{tabular}	 & \begin{tabular}{r} $+$2.25 \\ $-$1.82 \end{tabular}	 & \begin{tabular}{r} $+$0.00 \\ $-$0.02 \end{tabular}	 & \begin{tabular}{r} $-$0.76 \\ $+$0.76 \end{tabular}	 & \begin{tabular}{r} $-$0.34 \\ $+$0.40 \end{tabular}	 & \begin{tabular}{r} $+$4.08 \\ $-$4.08 \end{tabular}	 & \begin{tabular}{r} $+$0.10 \\ $-$0.09 \end{tabular}	 & \begin{tabular}{r} $+$0.25 \\ $-$0.69 \end{tabular}	 & \begin{tabular}{r} $+$1.00 \\ $-$1.00 \end{tabular}\tabularnewline[4pt]
3.0--3.2 & 44.324 & 1.58 & 2.56	 & \begin{tabular}{r} $+$9.22 \\ $-$8.78 \end{tabular}	 & \begin{tabular}{r} $+$0.03 \\ $-$0.08 \end{tabular}	 & \begin{tabular}{r} $+$0.15 \\ $-$0.21 \end{tabular}	 & \begin{tabular}{r} $+$0.24 \\ $-$0.29 \end{tabular}	 & \begin{tabular}{r} $-$0.25 \\ $+$0.19 \end{tabular}	 & \begin{tabular}{r} $+$0.00 \\ $-$0.05 \end{tabular}	 & \begin{tabular}{r} $+$3.08 \\ $-$2.72 \end{tabular}	 & \begin{tabular}{r} $+$4.36 \\ $-$3.79 \end{tabular}	 & \begin{tabular}{r} $+$0.00 \\ $-$0.03 \end{tabular}	 & \begin{tabular}{r} $+$0.00 \\ $-$0.02 \end{tabular}	 & \begin{tabular}{r} $+$0.87 \\ $-$0.92 \end{tabular}	 & \begin{tabular}{r} $+$2.25 \\ $-$1.82 \end{tabular}	 & \begin{tabular}{r} $+$0.00 \\ $-$0.02 \end{tabular}	 & \begin{tabular}{r} $-$0.76 \\ $+$0.76 \end{tabular}	 & \begin{tabular}{r} $-$0.34 \\ $+$0.40 \end{tabular}	 & \begin{tabular}{r} $+$4.94 \\ $-$4.94 \end{tabular}	 & \begin{tabular}{r} $+$0.09 \\ $-$0.09 \end{tabular}	 & \begin{tabular}{r} $+$0.26 \\ $-$0.70 \end{tabular}	 & \begin{tabular}{r} $+$1.00 \\ $-$1.00 \end{tabular}\tabularnewline[4pt]
3.2--3.4 & 32.909 & 2.09 & 2.91	 & \begin{tabular}{r} $+$9.89 \\ $-$10.65 \end{tabular}	 & \begin{tabular}{r} $+$0.03 \\ $-$0.08 \end{tabular}	 & \begin{tabular}{r} $+$0.15 \\ $-$0.21 \end{tabular}	 & \begin{tabular}{r} $+$0.24 \\ $-$0.29 \end{tabular}	 & \begin{tabular}{r} $-$0.25 \\ $+$0.19 \end{tabular}	 & \begin{tabular}{r} $+$0.00 \\ $-$0.05 \end{tabular}	 & \begin{tabular}{r} $+$3.08 \\ $-$2.72 \end{tabular}	 & \begin{tabular}{r} $+$4.91 \\ $-$6.55 \end{tabular}	 & \begin{tabular}{r} $+$0.00 \\ $-$0.03 \end{tabular}	 & \begin{tabular}{r} $+$0.00 \\ $-$0.02 \end{tabular}	 & \begin{tabular}{r} $+$0.87 \\ $-$0.92 \end{tabular}	 & \begin{tabular}{r} $+$2.25 \\ $-$1.82 \end{tabular}	 & \begin{tabular}{r} $+$0.00 \\ $-$0.02 \end{tabular}	 & \begin{tabular}{r} $-$0.76 \\ $+$0.76 \end{tabular}	 & \begin{tabular}{r} $-$0.34 \\ $+$0.40 \end{tabular}	 & \begin{tabular}{r} $+$5.32 \\ $-$5.32 \end{tabular}	 & \begin{tabular}{r} $+$0.09 \\ $-$0.09 \end{tabular}	 & \begin{tabular}{r} $+$0.25 \\ $-$0.70 \end{tabular}	 & \begin{tabular}{r} $+$1.00 \\ $-$1.00 \end{tabular}\tabularnewline[4pt]
\hline
\end{tabular}
\end{table}
\end{landscape}

\newcolumntype{R}[1]{>{\raggedleft}m{#1}}
\begin{landscape}
\begin{table}
\centering
\caption[]{The measured double-differential \Zjets{} production \xss{} as a function of \absyj{} in the $100\GeV < \ptj < 200\GeV$ range.
$\delta_{\text{data}}^{\text{stat}}$ and $\delta_{\text{MC}}^{\text{stat}}$ are the statistical uncertainties in data and MC simulation, respectively.
$\delta_{\text{tot}}^{\text{sys}}$ is the total systematic uncertainty and includes the following components:
uncertainties due to electron reconstruction ($\delta_{\text{rec}}^{\text{el}}$), identification ($\delta_{\text{ID}}^{\text{el}}$) and trigger ($\delta_{\text{trig}}^{\text{el}}$) efficiencies;
electron energy scale ($\delta_{\text{scale}}^{\text{el}}$) and energy resolution ($\delta_{\text{res}}^{\text{el}}$) uncertainties;
sum in quadrature of the uncertainties from JES in situ methods ($\delta_{\text{in situ}}^{\text{JES}}$);
sum in quadrature of the uncertainties from JES $\eta$-intercalibration methods ($\delta_{\eta\text{-int}}^{\text{JES}}$);
an uncertainty of the measured single-hadron response ($\delta_{\text{hadron}}^{\text{JES}}$);
MC non-closure uncertainty ($\delta_{\text{closure}}^{\text{JES}}$);
sum in quadrature of the uncertainties due to pile-up corrections of the jet kinematics ($\delta_{\text{pile-up}}^{\text{JES}}$);
sum in quadrature of the flavour-based uncertainties ($\delta_{\text{flavour}}^{\text{JES}}$);
punch-through uncertainty ($\delta_{\text{pthrough}}^{\text{JES}}$); JER uncertainty ($\delta^{\text{JER}}$);
JVF uncertainty ($\delta^{\text{JVF}}$); sum in quadrature of the unfolding uncertainties ($\delta^{\text{unf}}$);
sum in quadrature of the uncertainties due to MC generated backgrounds normalisation ($\delta_{\text{MC}}^{\text{bg}}$);
sum in quadrature of the uncertainty due to combined multijet and \Wjets{} backgrounds ($\delta_{\text{mult}}^{\text{bg}}$);
uncertainty due to jet quality selection ($\delta^{\text{qual}}$).
All uncertainties are given in \%. The luminosity uncertainty of 1.9\% is not shown and not included in the total uncertainty and its components.}
\tiny
\setlength\tabcolsep{2pt}
\begin{tabular}{c S[table-format=-4.3] S[table-format=-2.2] S[table-format=-2.2] R{0.8cm} | R{0.8cm}  R{0.8cm}  R{0.8cm}  R{0.8cm}  R{0.8cm}  R{0.8cm}  R{0.8cm}  R{0.8cm}  R{0.8cm}  R{0.8cm}  R{0.8cm}  R{0.8cm}  R{0.8cm}  R{0.8cm}  R{0.8cm}  R{0.8cm}  R{0.8cm}  R{0.8cm}  }
\hline
\rule{0pt}{10pt}
\absyj{} & \multicolumn{1}{c}{$\frac{\text{d}^2\sigma}{\text{d}\absyj \text{d}\ptj}$} & \multicolumn{1}{c}{$\delta_{\text{data}}^{\text{stat}}$} &  \multicolumn{1}{c}{$\delta_{\text{MC}}^{\text{stat}}$} &  \multicolumn{1}{c|}{$\delta_{\text{tot}}^{\text{sys}}$}  & \multicolumn{1}{c}{$\delta_{\text{rec}}^{\text{el}}$}  & \multicolumn{1}{c}{$\delta_{\text{ID}}^{\text{el}}$}  & \multicolumn{1}{c}{$\delta_{\text{trig}}^{\text{el}}$}  & \multicolumn{1}{c}{$\delta_{\text{scale}}^{\text{el}}$}  & \multicolumn{1}{c}{$\delta_{\text{res}}^{\text{el}}$}  & \multicolumn{1}{c}{$\delta_{\text{in situ}}^{\text{JES}}$}  & \multicolumn{1}{c}{$\delta_{\eta\text{-int}}^{\text{JES}}$}  & \multicolumn{1}{c}{$\delta_{\text{hadron}}^{\text{JES}}$}  & \multicolumn{1}{c}{$\delta_{\text{closure}}^{\text{JES}}$}  & \multicolumn{1}{c}{$\delta_{\text{pile-up}}^{\text{JES}}$}  & \multicolumn{1}{c}{$\delta_{\text{flavour}}^{\text{JES}}$}  & \multicolumn{1}{c}{$\delta_{\text{pthrough}}^{\text{JES}}$}  & \multicolumn{1}{c}{$\delta^{\text{JER}}$}  & \multicolumn{1}{c}{$\delta^{\text{JVF}}$}  & \multicolumn{1}{c}{$\delta^{\text{unf}}$}  & \multicolumn{1}{c}{$\delta_{\text{MC}}^{\text{bg}}$}  & \multicolumn{1}{c}{$\delta_{\text{mult}}^{\text{bg}}$}  & \multicolumn{1}{c}{$\delta^{\text{qual}}$}  \tabularnewline[8pt]
& \multicolumn{1}{c}{[fb/\GeV]} & \multicolumn{1}{c}{[\%]} & \multicolumn{1}{c}{[\%]} & \multicolumn{1}{c|}{[\%]} & \multicolumn{1}{c}{[\%]}  & \multicolumn{1}{c}{[\%]}  & \multicolumn{1}{c}{[\%]}  & \multicolumn{1}{c}{[\%]}  & \multicolumn{1}{c}{[\%]}  & \multicolumn{1}{c}{[\%]}  & \multicolumn{1}{c}{[\%]}  & \multicolumn{1}{c}{[\%]}  & \multicolumn{1}{c}{[\%]}  & \multicolumn{1}{c}{[\%]}  & \multicolumn{1}{c}{[\%]}  & \multicolumn{1}{c}{[\%]}  & \multicolumn{1}{c}{[\%]}  & \multicolumn{1}{c}{[\%]}  & \multicolumn{1}{c}{[\%]}  & \multicolumn{1}{c}{[\%]}  & \multicolumn{1}{c}{[\%]}  & \multicolumn{1}{c}{[\%]} \tabularnewline[2pt]
\hline
\rule{0pt}{10pt}
0.0--0.2 & 45.769 & 1.28 & 1.29	 & \begin{tabular}{r} $+$2.59 \\ $-$3.54 \end{tabular}	 & \begin{tabular}{r} $-$0.04 \\ $+$0.00 \end{tabular}	 & \begin{tabular}{r} $+$0.02 \\ $-$0.08 \end{tabular}	 & \begin{tabular}{r} $+$0.11 \\ $-$0.15 \end{tabular}	 & \begin{tabular}{r} $-$0.29 \\ $+$0.38 \end{tabular}	 & \begin{tabular}{r} $+$0.06 \\ $+$0.00 \end{tabular}	 & \begin{tabular}{r} $+$1.44 \\ $-$2.11 \end{tabular}	 & \begin{tabular}{r} $+$0.50 \\ $-$0.67 \end{tabular}	 & \begin{tabular}{r} $+$0.00 \\ $-$0.06 \end{tabular}	 & \begin{tabular}{r} $+$0.00 \\ $-$0.03 \end{tabular}	 & \begin{tabular}{r} $+$0.83 \\ $-$1.13 \end{tabular}	 & \begin{tabular}{r} $+$1.51 \\ $-$2.20 \end{tabular}	 & \begin{tabular}{r} $+$0.00 \\ $-$0.06 \end{tabular}	 & \begin{tabular}{r} $-$0.36 \\ $+$0.36 \end{tabular}	 & \begin{tabular}{r} $-$0.04 \\ $+$0.00 \end{tabular}	 & \begin{tabular}{r} $+$0.14 \\ $-$0.14 \end{tabular}	 & \begin{tabular}{r} $+$0.19 \\ $-$0.18 \end{tabular}	 & \begin{tabular}{r} $+$0.18 \\ $-$0.43 \end{tabular}	 & \begin{tabular}{r} $+$1.00 \\ $-$1.00 \end{tabular}\tabularnewline[4pt]
0.2--0.4 & 46.342 & 1.22 & 1.39	 & \begin{tabular}{r} $+$2.59 \\ $-$3.54 \end{tabular}	 & \begin{tabular}{r} $-$0.04 \\ $+$0.00 \end{tabular}	 & \begin{tabular}{r} $+$0.02 \\ $-$0.08 \end{tabular}	 & \begin{tabular}{r} $+$0.11 \\ $-$0.15 \end{tabular}	 & \begin{tabular}{r} $-$0.29 \\ $+$0.38 \end{tabular}	 & \begin{tabular}{r} $+$0.06 \\ $+$0.00 \end{tabular}	 & \begin{tabular}{r} $+$1.44 \\ $-$2.11 \end{tabular}	 & \begin{tabular}{r} $+$0.50 \\ $-$0.67 \end{tabular}	 & \begin{tabular}{r} $+$0.00 \\ $-$0.06 \end{tabular}	 & \begin{tabular}{r} $+$0.00 \\ $-$0.03 \end{tabular}	 & \begin{tabular}{r} $+$0.83 \\ $-$1.13 \end{tabular}	 & \begin{tabular}{r} $+$1.51 \\ $-$2.20 \end{tabular}	 & \begin{tabular}{r} $+$0.00 \\ $-$0.06 \end{tabular}	 & \begin{tabular}{r} $-$0.36 \\ $+$0.36 \end{tabular}	 & \begin{tabular}{r} $-$0.04 \\ $+$0.00 \end{tabular}	 & \begin{tabular}{r} $+$0.14 \\ $-$0.14 \end{tabular}	 & \begin{tabular}{r} $+$0.18 \\ $-$0.17 \end{tabular}	 & \begin{tabular}{r} $+$0.18 \\ $-$0.44 \end{tabular}	 & \begin{tabular}{r} $+$1.00 \\ $-$1.00 \end{tabular}\tabularnewline[4pt]
0.4--0.6 & 43.964 & 1.25 & 1.47	 & \begin{tabular}{r} $+$2.59 \\ $-$3.21 \end{tabular}	 & \begin{tabular}{r} $-$0.04 \\ $+$0.00 \end{tabular}	 & \begin{tabular}{r} $+$0.02 \\ $-$0.08 \end{tabular}	 & \begin{tabular}{r} $+$0.11 \\ $-$0.15 \end{tabular}	 & \begin{tabular}{r} $-$0.29 \\ $+$0.38 \end{tabular}	 & \begin{tabular}{r} $+$0.06 \\ $+$0.00 \end{tabular}	 & \begin{tabular}{r} $+$1.44 \\ $-$1.93 \end{tabular}	 & \begin{tabular}{r} $+$0.50 \\ $-$0.67 \end{tabular}	 & \begin{tabular}{r} $+$0.00 \\ $-$0.06 \end{tabular}	 & \begin{tabular}{r} $+$0.00 \\ $-$0.03 \end{tabular}	 & \begin{tabular}{r} $+$0.83 \\ $-$1.13 \end{tabular}	 & \begin{tabular}{r} $+$1.51 \\ $-$1.84 \end{tabular}	 & \begin{tabular}{r} $+$0.00 \\ $-$0.06 \end{tabular}	 & \begin{tabular}{r} $-$0.36 \\ $+$0.36 \end{tabular}	 & \begin{tabular}{r} $-$0.04 \\ $+$0.00 \end{tabular}	 & \begin{tabular}{r} $+$0.14 \\ $-$0.14 \end{tabular}	 & \begin{tabular}{r} $+$0.19 \\ $-$0.17 \end{tabular}	 & \begin{tabular}{r} $+$0.17 \\ $-$0.43 \end{tabular}	 & \begin{tabular}{r} $+$1.00 \\ $-$1.00 \end{tabular}\tabularnewline[4pt]
0.6--0.8 & 40.076 & 1.40 & 1.67	 & \begin{tabular}{r} $+$2.59 \\ $-$3.15 \end{tabular}	 & \begin{tabular}{r} $-$0.04 \\ $+$0.00 \end{tabular}	 & \begin{tabular}{r} $+$0.02 \\ $-$0.08 \end{tabular}	 & \begin{tabular}{r} $+$0.11 \\ $-$0.15 \end{tabular}	 & \begin{tabular}{r} $-$0.29 \\ $+$0.38 \end{tabular}	 & \begin{tabular}{r} $+$0.06 \\ $+$0.00 \end{tabular}	 & \begin{tabular}{r} $+$1.44 \\ $-$1.93 \end{tabular}	 & \begin{tabular}{r} $+$0.50 \\ $-$0.67 \end{tabular}	 & \begin{tabular}{r} $+$0.00 \\ $-$0.06 \end{tabular}	 & \begin{tabular}{r} $+$0.00 \\ $-$0.03 \end{tabular}	 & \begin{tabular}{r} $+$0.83 \\ $-$1.13 \end{tabular}	 & \begin{tabular}{r} $+$1.51 \\ $-$1.73 \end{tabular}	 & \begin{tabular}{r} $+$0.00 \\ $-$0.06 \end{tabular}	 & \begin{tabular}{r} $-$0.36 \\ $+$0.36 \end{tabular}	 & \begin{tabular}{r} $-$0.04 \\ $+$0.00 \end{tabular}	 & \begin{tabular}{r} $+$0.14 \\ $-$0.14 \end{tabular}	 & \begin{tabular}{r} $+$0.18 \\ $-$0.17 \end{tabular}	 & \begin{tabular}{r} $+$0.20 \\ $-$0.47 \end{tabular}	 & \begin{tabular}{r} $+$1.00 \\ $-$1.00 \end{tabular}\tabularnewline[4pt]
0.8--1.0 & 37.981 & 1.40 & 1.38	 & \begin{tabular}{r} $+$2.50 \\ $-$3.15 \end{tabular}	 & \begin{tabular}{r} $-$0.04 \\ $+$0.00 \end{tabular}	 & \begin{tabular}{r} $+$0.02 \\ $-$0.08 \end{tabular}	 & \begin{tabular}{r} $+$0.11 \\ $-$0.15 \end{tabular}	 & \begin{tabular}{r} $-$0.29 \\ $+$0.38 \end{tabular}	 & \begin{tabular}{r} $+$0.06 \\ $+$0.00 \end{tabular}	 & \begin{tabular}{r} $+$1.44 \\ $-$1.93 \end{tabular}	 & \begin{tabular}{r} $+$0.50 \\ $-$0.67 \end{tabular}	 & \begin{tabular}{r} $+$0.00 \\ $-$0.06 \end{tabular}	 & \begin{tabular}{r} $+$0.00 \\ $-$0.03 \end{tabular}	 & \begin{tabular}{r} $+$0.83 \\ $-$1.13 \end{tabular}	 & \begin{tabular}{r} $+$1.37 \\ $-$1.73 \end{tabular}	 & \begin{tabular}{r} $+$0.00 \\ $-$0.06 \end{tabular}	 & \begin{tabular}{r} $-$0.36 \\ $+$0.36 \end{tabular}	 & \begin{tabular}{r} $-$0.04 \\ $+$0.00 \end{tabular}	 & \begin{tabular}{r} $+$0.14 \\ $-$0.14 \end{tabular}	 & \begin{tabular}{r} $+$0.17 \\ $-$0.16 \end{tabular}	 & \begin{tabular}{r} $+$0.17 \\ $-$0.46 \end{tabular}	 & \begin{tabular}{r} $+$1.00 \\ $-$1.00 \end{tabular}\tabularnewline[4pt]
1.0--1.2 & 32.122 & 1.63 & 1.68	 & \begin{tabular}{r} $+$2.96 \\ $-$3.28 \end{tabular}	 & \begin{tabular}{r} $-$0.04 \\ $+$0.00 \end{tabular}	 & \begin{tabular}{r} $+$0.02 \\ $-$0.08 \end{tabular}	 & \begin{tabular}{r} $+$0.11 \\ $-$0.15 \end{tabular}	 & \begin{tabular}{r} $-$0.29 \\ $+$0.38 \end{tabular}	 & \begin{tabular}{r} $+$0.06 \\ $+$0.00 \end{tabular}	 & \begin{tabular}{r} $+$1.44 \\ $-$1.82 \end{tabular}	 & \begin{tabular}{r} $+$0.50 \\ $-$0.67 \end{tabular}	 & \begin{tabular}{r} $+$0.00 \\ $-$0.06 \end{tabular}	 & \begin{tabular}{r} $+$0.00 \\ $-$0.03 \end{tabular}	 & \begin{tabular}{r} $+$0.83 \\ $-$0.92 \end{tabular}	 & \begin{tabular}{r} $+$1.37 \\ $-$1.45 \end{tabular}	 & \begin{tabular}{r} $+$0.00 \\ $-$0.06 \end{tabular}	 & \begin{tabular}{r} $-$0.36 \\ $+$0.36 \end{tabular}	 & \begin{tabular}{r} $-$0.04 \\ $+$0.00 \end{tabular}	 & \begin{tabular}{r} $+$1.13 \\ $-$1.13 \end{tabular}	 & \begin{tabular}{r} $+$0.18 \\ $-$0.16 \end{tabular}	 & \begin{tabular}{r} $+$0.18 \\ $-$0.48 \end{tabular}	 & \begin{tabular}{r} $+$1.00 \\ $-$1.00 \end{tabular}\tabularnewline[4pt]
1.2--1.4 & 31.772 & 1.33 & 1.53	 & \begin{tabular}{r} $+$3.20 \\ $-$3.29 \end{tabular}	 & \begin{tabular}{r} $-$0.04 \\ $+$0.00 \end{tabular}	 & \begin{tabular}{r} $+$0.02 \\ $-$0.08 \end{tabular}	 & \begin{tabular}{r} $+$0.11 \\ $-$0.15 \end{tabular}	 & \begin{tabular}{r} $-$0.29 \\ $+$0.38 \end{tabular}	 & \begin{tabular}{r} $+$0.06 \\ $+$0.00 \end{tabular}	 & \begin{tabular}{r} $+$1.87 \\ $-$1.82 \end{tabular}	 & \begin{tabular}{r} $+$0.50 \\ $-$0.67 \end{tabular}	 & \begin{tabular}{r} $+$0.00 \\ $-$0.06 \end{tabular}	 & \begin{tabular}{r} $+$0.00 \\ $-$0.03 \end{tabular}	 & \begin{tabular}{r} $+$0.83 \\ $-$0.92 \end{tabular}	 & \begin{tabular}{r} $+$1.37 \\ $-$1.45 \end{tabular}	 & \begin{tabular}{r} $+$0.00 \\ $-$0.06 \end{tabular}	 & \begin{tabular}{r} $-$0.36 \\ $+$0.36 \end{tabular}	 & \begin{tabular}{r} $-$0.04 \\ $+$0.00 \end{tabular}	 & \begin{tabular}{r} $+$1.13 \\ $-$1.13 \end{tabular}	 & \begin{tabular}{r} $+$0.17 \\ $-$0.15 \end{tabular}	 & \begin{tabular}{r} $+$0.19 \\ $-$0.53 \end{tabular}	 & \begin{tabular}{r} $+$1.00 \\ $-$1.00 \end{tabular}\tabularnewline[4pt]
1.4--1.6 & 27.737 & 1.34 & 1.85	 & \begin{tabular}{r} $+$3.39 \\ $-$3.34 \end{tabular}	 & \begin{tabular}{r} $-$0.04 \\ $+$0.00 \end{tabular}	 & \begin{tabular}{r} $+$0.02 \\ $-$0.08 \end{tabular}	 & \begin{tabular}{r} $+$0.11 \\ $-$0.15 \end{tabular}	 & \begin{tabular}{r} $-$0.29 \\ $+$0.38 \end{tabular}	 & \begin{tabular}{r} $+$0.06 \\ $+$0.00 \end{tabular}	 & \begin{tabular}{r} $+$1.87 \\ $-$1.82 \end{tabular}	 & \begin{tabular}{r} $+$0.50 \\ $-$0.67 \end{tabular}	 & \begin{tabular}{r} $+$0.00 \\ $-$0.06 \end{tabular}	 & \begin{tabular}{r} $+$0.00 \\ $-$0.03 \end{tabular}	 & \begin{tabular}{r} $+$0.83 \\ $-$0.92 \end{tabular}	 & \begin{tabular}{r} $+$1.70 \\ $-$1.45 \end{tabular}	 & \begin{tabular}{r} $+$0.00 \\ $-$0.06 \end{tabular}	 & \begin{tabular}{r} $-$0.36 \\ $+$0.36 \end{tabular}	 & \begin{tabular}{r} $-$0.04 \\ $+$0.00 \end{tabular}	 & \begin{tabular}{r} $+$1.13 \\ $-$1.13 \end{tabular}	 & \begin{tabular}{r} $+$0.17 \\ $-$0.16 \end{tabular}	 & \begin{tabular}{r} $+$0.52 \\ $-$0.79 \end{tabular}	 & \begin{tabular}{r} $+$1.00 \\ $-$1.00 \end{tabular}\tabularnewline[4pt]
1.6--1.8 & 21.873 & 1.85 & 2.01	 & \begin{tabular}{r} $+$3.36 \\ $-$3.30 \end{tabular}	 & \begin{tabular}{r} $-$0.04 \\ $+$0.00 \end{tabular}	 & \begin{tabular}{r} $+$0.02 \\ $-$0.08 \end{tabular}	 & \begin{tabular}{r} $+$0.11 \\ $-$0.15 \end{tabular}	 & \begin{tabular}{r} $-$0.29 \\ $+$0.38 \end{tabular}	 & \begin{tabular}{r} $+$0.06 \\ $+$0.00 \end{tabular}	 & \begin{tabular}{r} $+$1.87 \\ $-$1.82 \end{tabular}	 & \begin{tabular}{r} $+$0.50 \\ $-$0.67 \end{tabular}	 & \begin{tabular}{r} $+$0.00 \\ $-$0.06 \end{tabular}	 & \begin{tabular}{r} $+$0.00 \\ $-$0.03 \end{tabular}	 & \begin{tabular}{r} $+$0.83 \\ $-$0.92 \end{tabular}	 & \begin{tabular}{r} $+$1.70 \\ $-$1.45 \end{tabular}	 & \begin{tabular}{r} $+$0.00 \\ $-$0.06 \end{tabular}	 & \begin{tabular}{r} $-$0.36 \\ $+$0.36 \end{tabular}	 & \begin{tabular}{r} $-$0.04 \\ $+$0.00 \end{tabular}	 & \begin{tabular}{r} $+$1.13 \\ $-$1.13 \end{tabular}	 & \begin{tabular}{r} $+$0.16 \\ $-$0.15 \end{tabular}	 & \begin{tabular}{r} $+$0.24 \\ $-$0.60 \end{tabular}	 & \begin{tabular}{r} $+$1.00 \\ $-$1.00 \end{tabular}\tabularnewline[4pt]
1.8--2.0 & 17.806 & 1.88 & 2.00	 & \begin{tabular}{r} $+$3.52 \\ $-$3.48 \end{tabular}	 & \begin{tabular}{r} $-$0.04 \\ $+$0.00 \end{tabular}	 & \begin{tabular}{r} $+$0.02 \\ $-$0.08 \end{tabular}	 & \begin{tabular}{r} $+$0.11 \\ $-$0.15 \end{tabular}	 & \begin{tabular}{r} $-$0.29 \\ $+$0.38 \end{tabular}	 & \begin{tabular}{r} $+$0.06 \\ $+$0.00 \end{tabular}	 & \begin{tabular}{r} $+$1.98 \\ $-$1.95 \end{tabular}	 & \begin{tabular}{r} $+$0.50 \\ $-$0.67 \end{tabular}	 & \begin{tabular}{r} $+$0.00 \\ $-$0.06 \end{tabular}	 & \begin{tabular}{r} $+$0.00 \\ $-$0.03 \end{tabular}	 & \begin{tabular}{r} $+$1.17 \\ $-$0.92 \end{tabular}	 & \begin{tabular}{r} $+$1.70 \\ $-$1.66 \end{tabular}	 & \begin{tabular}{r} $+$0.00 \\ $-$0.06 \end{tabular}	 & \begin{tabular}{r} $-$0.36 \\ $+$0.36 \end{tabular}	 & \begin{tabular}{r} $-$0.04 \\ $+$0.00 \end{tabular}	 & \begin{tabular}{r} $+$1.13 \\ $-$1.13 \end{tabular}	 & \begin{tabular}{r} $+$0.15 \\ $-$0.14 \end{tabular}	 & \begin{tabular}{r} $+$0.26 \\ $-$0.68 \end{tabular}	 & \begin{tabular}{r} $+$1.00 \\ $-$1.00 \end{tabular}\tabularnewline[4pt]
2.0--2.2 & 13.820 & 2.26 & 2.26	 & \begin{tabular}{r} $+$3.52 \\ $-$3.48 \end{tabular}	 & \begin{tabular}{r} $-$0.04 \\ $+$0.00 \end{tabular}	 & \begin{tabular}{r} $+$0.02 \\ $-$0.08 \end{tabular}	 & \begin{tabular}{r} $+$0.11 \\ $-$0.15 \end{tabular}	 & \begin{tabular}{r} $-$0.29 \\ $+$0.38 \end{tabular}	 & \begin{tabular}{r} $+$0.06 \\ $+$0.00 \end{tabular}	 & \begin{tabular}{r} $+$1.98 \\ $-$1.95 \end{tabular}	 & \begin{tabular}{r} $+$0.50 \\ $-$0.67 \end{tabular}	 & \begin{tabular}{r} $+$0.00 \\ $-$0.06 \end{tabular}	 & \begin{tabular}{r} $+$0.00 \\ $-$0.03 \end{tabular}	 & \begin{tabular}{r} $+$1.17 \\ $-$0.92 \end{tabular}	 & \begin{tabular}{r} $+$1.70 \\ $-$1.66 \end{tabular}	 & \begin{tabular}{r} $+$0.00 \\ $-$0.06 \end{tabular}	 & \begin{tabular}{r} $-$0.36 \\ $+$0.36 \end{tabular}	 & \begin{tabular}{r} $-$0.04 \\ $+$0.00 \end{tabular}	 & \begin{tabular}{r} $+$1.13 \\ $-$1.13 \end{tabular}	 & \begin{tabular}{r} $+$0.15 \\ $-$0.14 \end{tabular}	 & \begin{tabular}{r} $+$0.26 \\ $-$0.70 \end{tabular}	 & \begin{tabular}{r} $+$1.00 \\ $-$1.00 \end{tabular}\tabularnewline[4pt]
2.2--2.4 & 10.613 & 2.55 & 2.81	 & \begin{tabular}{r} $+$4.14 \\ $-$4.52 \end{tabular}	 & \begin{tabular}{r} $-$0.04 \\ $+$0.00 \end{tabular}	 & \begin{tabular}{r} $+$0.02 \\ $-$0.08 \end{tabular}	 & \begin{tabular}{r} $+$0.11 \\ $-$0.15 \end{tabular}	 & \begin{tabular}{r} $-$0.29 \\ $+$0.38 \end{tabular}	 & \begin{tabular}{r} $+$0.06 \\ $+$0.00 \end{tabular}	 & \begin{tabular}{r} $+$2.23 \\ $-$1.95 \end{tabular}	 & \begin{tabular}{r} $+$0.50 \\ $-$2.96 \end{tabular}	 & \begin{tabular}{r} $+$0.00 \\ $-$0.06 \end{tabular}	 & \begin{tabular}{r} $+$0.00 \\ $-$0.03 \end{tabular}	 & \begin{tabular}{r} $+$1.17 \\ $-$0.92 \end{tabular}	 & \begin{tabular}{r} $+$2.57 \\ $-$1.66 \end{tabular}	 & \begin{tabular}{r} $+$0.00 \\ $-$0.06 \end{tabular}	 & \begin{tabular}{r} $-$0.36 \\ $+$0.36 \end{tabular}	 & \begin{tabular}{r} $-$0.04 \\ $+$0.00 \end{tabular}	 & \begin{tabular}{r} $+$1.13 \\ $-$1.13 \end{tabular}	 & \begin{tabular}{r} $+$0.15 \\ $-$0.14 \end{tabular}	 & \begin{tabular}{r} $+$0.25 \\ $-$0.70 \end{tabular}	 & \begin{tabular}{r} $+$1.00 \\ $-$1.00 \end{tabular}\tabularnewline[4pt]
2.4--2.6 & 8.152 & 3.12 & 2.94	 & \begin{tabular}{r} $+$5.83 \\ $-$4.55 \end{tabular}	 & \begin{tabular}{r} $-$0.04 \\ $+$0.00 \end{tabular}	 & \begin{tabular}{r} $+$0.02 \\ $-$0.08 \end{tabular}	 & \begin{tabular}{r} $+$0.11 \\ $-$0.15 \end{tabular}	 & \begin{tabular}{r} $-$0.29 \\ $+$0.38 \end{tabular}	 & \begin{tabular}{r} $+$0.06 \\ $+$0.00 \end{tabular}	 & \begin{tabular}{r} $+$2.23 \\ $-$1.95 \end{tabular}	 & \begin{tabular}{r} $+$4.14 \\ $-$2.96 \end{tabular}	 & \begin{tabular}{r} $+$0.00 \\ $-$0.06 \end{tabular}	 & \begin{tabular}{r} $+$0.00 \\ $-$0.03 \end{tabular}	 & \begin{tabular}{r} $+$1.17 \\ $-$0.92 \end{tabular}	 & \begin{tabular}{r} $+$2.57 \\ $-$1.66 \end{tabular}	 & \begin{tabular}{r} $+$0.00 \\ $-$0.06 \end{tabular}	 & \begin{tabular}{r} $-$0.36 \\ $+$0.36 \end{tabular}	 & \begin{tabular}{r} $-$0.04 \\ $+$0.00 \end{tabular}	 & \begin{tabular}{r} $+$1.13 \\ $-$1.13 \end{tabular}	 & \begin{tabular}{r} $+$0.15 \\ $-$0.14 \end{tabular}	 & \begin{tabular}{r} $+$0.32 \\ $-$0.87 \end{tabular}	 & \begin{tabular}{r} $+$1.00 \\ $-$1.00 \end{tabular}\tabularnewline[4pt]
2.6--2.8 & 5.663 & 3.22 & 3.91	 & \begin{tabular}{r} $+$5.84 \\ $-$4.59 \end{tabular}	 & \begin{tabular}{r} $-$0.04 \\ $+$0.00 \end{tabular}	 & \begin{tabular}{r} $+$0.02 \\ $-$0.08 \end{tabular}	 & \begin{tabular}{r} $+$0.11 \\ $-$0.15 \end{tabular}	 & \begin{tabular}{r} $-$0.29 \\ $+$0.38 \end{tabular}	 & \begin{tabular}{r} $+$0.06 \\ $+$0.00 \end{tabular}	 & \begin{tabular}{r} $+$2.23 \\ $-$1.95 \end{tabular}	 & \begin{tabular}{r} $+$4.14 \\ $-$2.96 \end{tabular}	 & \begin{tabular}{r} $+$0.00 \\ $-$0.06 \end{tabular}	 & \begin{tabular}{r} $+$0.00 \\ $-$0.03 \end{tabular}	 & \begin{tabular}{r} $+$1.17 \\ $-$0.92 \end{tabular}	 & \begin{tabular}{r} $+$2.57 \\ $-$1.66 \end{tabular}	 & \begin{tabular}{r} $+$0.00 \\ $-$0.06 \end{tabular}	 & \begin{tabular}{r} $-$0.36 \\ $+$0.36 \end{tabular}	 & \begin{tabular}{r} $-$0.04 \\ $+$0.00 \end{tabular}	 & \begin{tabular}{r} $+$1.13 \\ $-$1.13 \end{tabular}	 & \begin{tabular}{r} $+$0.17 \\ $-$0.16 \end{tabular}	 & \begin{tabular}{r} $+$0.44 \\ $-$1.04 \end{tabular}	 & \begin{tabular}{r} $+$1.00 \\ $-$1.00 \end{tabular}\tabularnewline[4pt]
2.8--3.0 & 3.248 & 3.91 & 4.78	 & \begin{tabular}{r} $+$9.49 \\ $-$9.52 \end{tabular}	 & \begin{tabular}{r} $-$0.04 \\ $+$0.00 \end{tabular}	 & \begin{tabular}{r} $+$0.02 \\ $-$0.08 \end{tabular}	 & \begin{tabular}{r} $+$0.11 \\ $-$0.15 \end{tabular}	 & \begin{tabular}{r} $-$0.29 \\ $+$0.38 \end{tabular}	 & \begin{tabular}{r} $+$0.06 \\ $+$0.00 \end{tabular}	 & \begin{tabular}{r} $+$2.23 \\ $-$1.95 \end{tabular}	 & \begin{tabular}{r} $+$8.54 \\ $-$7.71 \end{tabular}	 & \begin{tabular}{r} $+$0.00 \\ $-$0.06 \end{tabular}	 & \begin{tabular}{r} $+$0.00 \\ $-$0.03 \end{tabular}	 & \begin{tabular}{r} $+$1.17 \\ $-$0.92 \end{tabular}	 & \begin{tabular}{r} $+$2.57 \\ $-$4.65 \end{tabular}	 & \begin{tabular}{r} $+$0.00 \\ $-$0.06 \end{tabular}	 & \begin{tabular}{r} $-$0.36 \\ $+$0.36 \end{tabular}	 & \begin{tabular}{r} $-$0.04 \\ $+$0.00 \end{tabular}	 & \begin{tabular}{r} $+$1.13 \\ $-$1.13 \end{tabular}	 & \begin{tabular}{r} $+$0.18 \\ $-$0.17 \end{tabular}	 & \begin{tabular}{r} $+$0.40 \\ $-$1.07 \end{tabular}	 & \begin{tabular}{r} $+$1.00 \\ $-$1.00 \end{tabular}\tabularnewline[4pt]
3.0--3.2 & 2.169 & 5.43 & 5.73	 & \begin{tabular}{r} $+$9.55 \\ $-$9.53 \end{tabular}	 & \begin{tabular}{r} $-$0.04 \\ $+$0.00 \end{tabular}	 & \begin{tabular}{r} $+$0.02 \\ $-$0.08 \end{tabular}	 & \begin{tabular}{r} $+$0.11 \\ $-$0.15 \end{tabular}	 & \begin{tabular}{r} $-$0.29 \\ $+$0.38 \end{tabular}	 & \begin{tabular}{r} $+$0.06 \\ $+$0.00 \end{tabular}	 & \begin{tabular}{r} $+$2.47 \\ $-$1.95 \end{tabular}	 & \begin{tabular}{r} $+$8.54 \\ $-$7.71 \end{tabular}	 & \begin{tabular}{r} $+$0.00 \\ $-$0.06 \end{tabular}	 & \begin{tabular}{r} $+$0.00 \\ $-$0.03 \end{tabular}	 & \begin{tabular}{r} $+$1.17 \\ $-$0.92 \end{tabular}	 & \begin{tabular}{r} $+$2.57 \\ $-$4.65 \end{tabular}	 & \begin{tabular}{r} $+$0.00 \\ $-$0.06 \end{tabular}	 & \begin{tabular}{r} $-$0.36 \\ $+$0.36 \end{tabular}	 & \begin{tabular}{r} $-$0.04 \\ $+$0.00 \end{tabular}	 & \begin{tabular}{r} $+$1.13 \\ $-$1.13 \end{tabular}	 & \begin{tabular}{r} $+$0.18 \\ $-$0.17 \end{tabular}	 & \begin{tabular}{r} $+$0.46 \\ $-$1.09 \end{tabular}	 & \begin{tabular}{r} $+$1.00 \\ $-$1.00 \end{tabular}\tabularnewline[4pt]
3.2--3.4 & 1.234 & 7.36 & 9.27	 & \begin{tabular}{r} $+$15.91 \\ $-$13.62 \end{tabular}	 & \begin{tabular}{r} $-$0.04 \\ $+$0.00 \end{tabular}	 & \begin{tabular}{r} $+$0.02 \\ $-$0.08 \end{tabular}	 & \begin{tabular}{r} $+$0.11 \\ $-$0.15 \end{tabular}	 & \begin{tabular}{r} $-$0.29 \\ $+$0.38 \end{tabular}	 & \begin{tabular}{r} $+$0.06 \\ $+$0.00 \end{tabular}	 & \begin{tabular}{r} $+$2.47 \\ $-$1.95 \end{tabular}	 & \begin{tabular}{r} $+$15.33 \\ $-$12.42 \end{tabular}	 & \begin{tabular}{r} $+$0.00 \\ $-$0.06 \end{tabular}	 & \begin{tabular}{r} $+$0.00 \\ $-$0.03 \end{tabular}	 & \begin{tabular}{r} $+$1.17 \\ $-$0.92 \end{tabular}	 & \begin{tabular}{r} $+$2.57 \\ $-$4.65 \end{tabular}	 & \begin{tabular}{r} $+$0.00 \\ $-$0.06 \end{tabular}	 & \begin{tabular}{r} $-$0.36 \\ $+$0.36 \end{tabular}	 & \begin{tabular}{r} $-$0.04 \\ $+$0.00 \end{tabular}	 & \begin{tabular}{r} $+$1.13 \\ $-$1.13 \end{tabular}	 & \begin{tabular}{r} $+$0.15 \\ $-$0.13 \end{tabular}	 & \begin{tabular}{r} $+$0.41 \\ $-$1.07 \end{tabular}	 & \begin{tabular}{r} $+$1.00 \\ $-$1.00 \end{tabular}\tabularnewline[4pt]
\hline
\end{tabular}
\end{table}
\end{landscape}

\newcolumntype{R}[1]{>{\raggedleft}m{#1}}
\begin{landscape}
\begin{table}
\centering
\caption[]{The measured double-differential \Zjets{} production \xss{} as a function of \absyj{} in the $200\GeV < \ptj < 300\GeV$ range.
$\delta_{\text{data}}^{\text{stat}}$ and $\delta_{\text{MC}}^{\text{stat}}$ are the statistical uncertainties in data and MC simulation, respectively.
$\delta_{\text{tot}}^{\text{sys}}$ is the total systematic uncertainty and includes the following components:
uncertainties due to electron reconstruction ($\delta_{\text{rec}}^{\text{el}}$), identification ($\delta_{\text{ID}}^{\text{el}}$) and trigger ($\delta_{\text{trig}}^{\text{el}}$) efficiencies;
electron energy scale ($\delta_{\text{scale}}^{\text{el}}$) and energy resolution ($\delta_{\text{res}}^{\text{el}}$) uncertainties;
sum in quadrature of the uncertainties from JES in situ methods ($\delta_{\text{in situ}}^{\text{JES}}$);
sum in quadrature of the uncertainties from JES $\eta$-intercalibration methods ($\delta_{\eta\text{-int}}^{\text{JES}}$);
an uncertainty of the measured single-hadron response ($\delta_{\text{hadron}}^{\text{JES}}$);
MC non-closure uncertainty ($\delta_{\text{closure}}^{\text{JES}}$);
sum in quadrature of the uncertainties due to pile-up corrections of the jet kinematics ($\delta_{\text{pile-up}}^{\text{JES}}$);
sum in quadrature of the flavour-based uncertainties ($\delta_{\text{flavour}}^{\text{JES}}$);
punch-through uncertainty ($\delta_{\text{pthrough}}^{\text{JES}}$); JER uncertainty ($\delta^{\text{JER}}$);
JVF uncertainty ($\delta^{\text{JVF}}$); sum in quadrature of the unfolding uncertainties ($\delta^{\text{unf}}$);
sum in quadrature of the uncertainties due to MC generated backgrounds normalisation ($\delta_{\text{MC}}^{\text{bg}}$);
sum in quadrature of the uncertainty due to combined multijet and \Wjets{} backgrounds ($\delta_{\text{mult}}^{\text{bg}}$);
uncertainty due to jet quality selection ($\delta^{\text{qual}}$).
All uncertainties are given in \%. The luminosity uncertainty of 1.9\% is not shown and not included in the total uncertainty and its components.}
\tiny
\setlength\tabcolsep{2pt}
\begin{tabular}{c S[table-format=-4.3] S[table-format=-2.2] S[table-format=-2.2] R{0.8cm} | R{0.8cm}  R{0.8cm}  R{0.8cm}  R{0.8cm}  R{0.8cm}  R{0.8cm}  R{0.8cm}  R{0.8cm}  R{0.8cm}  R{0.8cm}  R{0.8cm}  R{0.8cm}  R{0.8cm}  R{0.8cm}  R{0.8cm}  R{0.8cm}  R{0.8cm}  R{0.8cm}  }
\hline
\rule{0pt}{10pt}
\absyj{} & \multicolumn{1}{c}{$\frac{\text{d}^2\sigma}{\text{d}\absyj \text{d}\ptj}$} & \multicolumn{1}{c}{$\delta_{\text{data}}^{\text{stat}}$} &  \multicolumn{1}{c}{$\delta_{\text{MC}}^{\text{stat}}$} &  \multicolumn{1}{c|}{$\delta_{\text{tot}}^{\text{sys}}$}  & \multicolumn{1}{c}{$\delta_{\text{rec}}^{\text{el}}$}  & \multicolumn{1}{c}{$\delta_{\text{ID}}^{\text{el}}$}  & \multicolumn{1}{c}{$\delta_{\text{trig}}^{\text{el}}$}  & \multicolumn{1}{c}{$\delta_{\text{scale}}^{\text{el}}$}  & \multicolumn{1}{c}{$\delta_{\text{res}}^{\text{el}}$}  & \multicolumn{1}{c}{$\delta_{\text{in situ}}^{\text{JES}}$}  & \multicolumn{1}{c}{$\delta_{\eta\text{-int}}^{\text{JES}}$}  & \multicolumn{1}{c}{$\delta_{\text{hadron}}^{\text{JES}}$}  & \multicolumn{1}{c}{$\delta_{\text{closure}}^{\text{JES}}$}  & \multicolumn{1}{c}{$\delta_{\text{pile-up}}^{\text{JES}}$}  & \multicolumn{1}{c}{$\delta_{\text{flavour}}^{\text{JES}}$}  & \multicolumn{1}{c}{$\delta_{\text{pthrough}}^{\text{JES}}$}  & \multicolumn{1}{c}{$\delta^{\text{JER}}$}  & \multicolumn{1}{c}{$\delta^{\text{JVF}}$}  & \multicolumn{1}{c}{$\delta^{\text{unf}}$}  & \multicolumn{1}{c}{$\delta_{\text{MC}}^{\text{bg}}$}  & \multicolumn{1}{c}{$\delta_{\text{mult}}^{\text{bg}}$}  & \multicolumn{1}{c}{$\delta^{\text{qual}}$}  \tabularnewline[8pt]
& \multicolumn{1}{c}{[fb/\GeV]} & \multicolumn{1}{c}{[\%]} & \multicolumn{1}{c}{[\%]} & \multicolumn{1}{c|}{[\%]} & \multicolumn{1}{c}{[\%]}  & \multicolumn{1}{c}{[\%]}  & \multicolumn{1}{c}{[\%]}  & \multicolumn{1}{c}{[\%]}  & \multicolumn{1}{c}{[\%]}  & \multicolumn{1}{c}{[\%]}  & \multicolumn{1}{c}{[\%]}  & \multicolumn{1}{c}{[\%]}  & \multicolumn{1}{c}{[\%]}  & \multicolumn{1}{c}{[\%]}  & \multicolumn{1}{c}{[\%]}  & \multicolumn{1}{c}{[\%]}  & \multicolumn{1}{c}{[\%]}  & \multicolumn{1}{c}{[\%]}  & \multicolumn{1}{c}{[\%]}  & \multicolumn{1}{c}{[\%]}  & \multicolumn{1}{c}{[\%]}  & \multicolumn{1}{c}{[\%]} \tabularnewline[2pt]
\hline
\rule{0pt}{10pt}
0.0--0.4 & 5.561 & 2.50 & 2.63	 & \begin{tabular}{r} $+$3.82 \\ $-$4.55 \end{tabular}	 & \begin{tabular}{r} $-$0.05 \\ $+$0.04 \end{tabular}	 & \begin{tabular}{r} $+$0.12 \\ $+$0.00 \end{tabular}	 & \begin{tabular}{r} $+$0.10 \\ $-$0.01 \end{tabular}	 & \begin{tabular}{r} $-$0.18 \\ $+$0.18 \end{tabular}	 & \begin{tabular}{r} $-$0.06 \\ $+$0.05 \end{tabular}	 & \begin{tabular}{r} $+$2.62 \\ $-$3.10 \end{tabular}	 & \begin{tabular}{r} $+$0.39 \\ $-$0.77 \end{tabular}	 & \begin{tabular}{r} $+$0.04 \\ $+$0.00 \end{tabular}	 & \begin{tabular}{r} $+$0.15 \\ $+$0.00 \end{tabular}	 & \begin{tabular}{r} $+$0.36 \\ $-$0.75 \end{tabular}	 & \begin{tabular}{r} $+$1.40 \\ $-$2.08 \end{tabular}	 & \begin{tabular}{r} $+$0.20 \\ $-$0.06 \end{tabular}	 & \begin{tabular}{r} $-$0.58 \\ $+$0.58 \end{tabular}	 & \begin{tabular}{r} $-$0.04 \\ $+$0.09 \end{tabular}	 & \begin{tabular}{r} $+$1.40 \\ $-$1.40 \end{tabular}	 & \begin{tabular}{r} $+$0.09 \\ $-$0.08 \end{tabular}	 & \begin{tabular}{r} $+$0.28 \\ $-$0.54 \end{tabular}	 & \begin{tabular}{r} $+$1.00 \\ $-$1.00 \end{tabular}\tabularnewline[4pt]
0.4--0.8 & 4.889 & 2.36 & 2.93	 & \begin{tabular}{r} $+$3.81 \\ $-$4.55 \end{tabular}	 & \begin{tabular}{r} $-$0.05 \\ $+$0.04 \end{tabular}	 & \begin{tabular}{r} $+$0.12 \\ $+$0.00 \end{tabular}	 & \begin{tabular}{r} $+$0.10 \\ $-$0.01 \end{tabular}	 & \begin{tabular}{r} $-$0.18 \\ $+$0.18 \end{tabular}	 & \begin{tabular}{r} $-$0.06 \\ $+$0.05 \end{tabular}	 & \begin{tabular}{r} $+$2.62 \\ $-$3.10 \end{tabular}	 & \begin{tabular}{r} $+$0.39 \\ $-$0.77 \end{tabular}	 & \begin{tabular}{r} $+$0.04 \\ $+$0.00 \end{tabular}	 & \begin{tabular}{r} $+$0.15 \\ $+$0.00 \end{tabular}	 & \begin{tabular}{r} $+$0.36 \\ $-$0.75 \end{tabular}	 & \begin{tabular}{r} $+$1.40 \\ $-$2.08 \end{tabular}	 & \begin{tabular}{r} $+$0.20 \\ $-$0.06 \end{tabular}	 & \begin{tabular}{r} $-$0.58 \\ $+$0.58 \end{tabular}	 & \begin{tabular}{r} $-$0.04 \\ $+$0.09 \end{tabular}	 & \begin{tabular}{r} $+$1.40 \\ $-$1.40 \end{tabular}	 & \begin{tabular}{r} $+$0.10 \\ $-$0.09 \end{tabular}	 & \begin{tabular}{r} $+$0.25 \\ $-$0.55 \end{tabular}	 & \begin{tabular}{r} $+$1.00 \\ $-$1.00 \end{tabular}\tabularnewline[4pt]
0.8--1.2 & 4.260 & 3.18 & 3.41	 & \begin{tabular}{r} $+$3.81 \\ $-$4.55 \end{tabular}	 & \begin{tabular}{r} $-$0.05 \\ $+$0.04 \end{tabular}	 & \begin{tabular}{r} $+$0.12 \\ $+$0.00 \end{tabular}	 & \begin{tabular}{r} $+$0.10 \\ $-$0.01 \end{tabular}	 & \begin{tabular}{r} $-$0.18 \\ $+$0.18 \end{tabular}	 & \begin{tabular}{r} $-$0.06 \\ $+$0.05 \end{tabular}	 & \begin{tabular}{r} $+$2.62 \\ $-$3.10 \end{tabular}	 & \begin{tabular}{r} $+$0.39 \\ $-$0.77 \end{tabular}	 & \begin{tabular}{r} $+$0.04 \\ $+$0.00 \end{tabular}	 & \begin{tabular}{r} $+$0.15 \\ $+$0.00 \end{tabular}	 & \begin{tabular}{r} $+$0.36 \\ $-$0.75 \end{tabular}	 & \begin{tabular}{r} $+$1.40 \\ $-$2.08 \end{tabular}	 & \begin{tabular}{r} $+$0.20 \\ $-$0.06 \end{tabular}	 & \begin{tabular}{r} $-$0.58 \\ $+$0.58 \end{tabular}	 & \begin{tabular}{r} $-$0.04 \\ $+$0.09 \end{tabular}	 & \begin{tabular}{r} $+$1.40 \\ $-$1.40 \end{tabular}	 & \begin{tabular}{r} $+$0.11 \\ $-$0.10 \end{tabular}	 & \begin{tabular}{r} $+$0.21 \\ $-$0.57 \end{tabular}	 & \begin{tabular}{r} $+$1.00 \\ $-$1.00 \end{tabular}\tabularnewline[4pt]
1.2--1.6 & 3.055 & 3.61 & 3.17	 & \begin{tabular}{r} $+$3.82 \\ $-$4.57 \end{tabular}	 & \begin{tabular}{r} $-$0.05 \\ $+$0.04 \end{tabular}	 & \begin{tabular}{r} $+$0.12 \\ $+$0.00 \end{tabular}	 & \begin{tabular}{r} $+$0.10 \\ $-$0.01 \end{tabular}	 & \begin{tabular}{r} $-$0.18 \\ $+$0.18 \end{tabular}	 & \begin{tabular}{r} $-$0.06 \\ $+$0.05 \end{tabular}	 & \begin{tabular}{r} $+$2.62 \\ $-$3.10 \end{tabular}	 & \begin{tabular}{r} $+$0.39 \\ $-$0.77 \end{tabular}	 & \begin{tabular}{r} $+$0.04 \\ $+$0.00 \end{tabular}	 & \begin{tabular}{r} $+$0.15 \\ $+$0.00 \end{tabular}	 & \begin{tabular}{r} $+$0.36 \\ $-$0.75 \end{tabular}	 & \begin{tabular}{r} $+$1.40 \\ $-$2.08 \end{tabular}	 & \begin{tabular}{r} $+$0.20 \\ $-$0.06 \end{tabular}	 & \begin{tabular}{r} $-$0.58 \\ $+$0.58 \end{tabular}	 & \begin{tabular}{r} $-$0.04 \\ $+$0.09 \end{tabular}	 & \begin{tabular}{r} $+$1.40 \\ $-$1.40 \end{tabular}	 & \begin{tabular}{r} $+$0.14 \\ $-$0.13 \end{tabular}	 & \begin{tabular}{r} $+$0.27 \\ $-$0.71 \end{tabular}	 & \begin{tabular}{r} $+$1.00 \\ $-$1.00 \end{tabular}\tabularnewline[4pt]
1.6--2.0 & 1.780 & 4.43 & 4.42	 & \begin{tabular}{r} $+$3.83 \\ $-$4.60 \end{tabular}	 & \begin{tabular}{r} $-$0.05 \\ $+$0.04 \end{tabular}	 & \begin{tabular}{r} $+$0.12 \\ $+$0.00 \end{tabular}	 & \begin{tabular}{r} $+$0.10 \\ $-$0.01 \end{tabular}	 & \begin{tabular}{r} $-$0.18 \\ $+$0.18 \end{tabular}	 & \begin{tabular}{r} $-$0.06 \\ $+$0.05 \end{tabular}	 & \begin{tabular}{r} $+$2.62 \\ $-$3.10 \end{tabular}	 & \begin{tabular}{r} $+$0.39 \\ $-$0.77 \end{tabular}	 & \begin{tabular}{r} $+$0.04 \\ $+$0.00 \end{tabular}	 & \begin{tabular}{r} $+$0.15 \\ $+$0.00 \end{tabular}	 & \begin{tabular}{r} $+$0.36 \\ $-$0.75 \end{tabular}	 & \begin{tabular}{r} $+$1.40 \\ $-$2.08 \end{tabular}	 & \begin{tabular}{r} $+$0.20 \\ $-$0.06 \end{tabular}	 & \begin{tabular}{r} $-$0.58 \\ $+$0.58 \end{tabular}	 & \begin{tabular}{r} $-$0.04 \\ $+$0.09 \end{tabular}	 & \begin{tabular}{r} $+$1.40 \\ $-$1.40 \end{tabular}	 & \begin{tabular}{r} $+$0.17 \\ $-$0.16 \end{tabular}	 & \begin{tabular}{r} $+$0.39 \\ $-$0.84 \end{tabular}	 & \begin{tabular}{r} $+$1.00 \\ $-$1.00 \end{tabular}\tabularnewline[4pt]
2.0--2.4 & 0.831 & 6.45 & 7.17	 & \begin{tabular}{r} $+$3.87 \\ $-$6.41 \end{tabular}	 & \begin{tabular}{r} $-$0.05 \\ $+$0.04 \end{tabular}	 & \begin{tabular}{r} $+$0.12 \\ $+$0.00 \end{tabular}	 & \begin{tabular}{r} $+$0.10 \\ $-$0.01 \end{tabular}	 & \begin{tabular}{r} $-$0.18 \\ $+$0.18 \end{tabular}	 & \begin{tabular}{r} $-$0.06 \\ $+$0.05 \end{tabular}	 & \begin{tabular}{r} $+$2.62 \\ $-$5.37 \end{tabular}	 & \begin{tabular}{r} $+$0.39 \\ $-$0.77 \end{tabular}	 & \begin{tabular}{r} $+$0.04 \\ $+$0.00 \end{tabular}	 & \begin{tabular}{r} $+$0.15 \\ $+$0.00 \end{tabular}	 & \begin{tabular}{r} $+$0.36 \\ $-$0.75 \end{tabular}	 & \begin{tabular}{r} $+$1.40 \\ $-$2.08 \end{tabular}	 & \begin{tabular}{r} $+$0.20 \\ $-$0.06 \end{tabular}	 & \begin{tabular}{r} $-$0.58 \\ $+$0.58 \end{tabular}	 & \begin{tabular}{r} $-$0.04 \\ $+$0.09 \end{tabular}	 & \begin{tabular}{r} $+$1.40 \\ $-$1.40 \end{tabular}	 & \begin{tabular}{r} $+$0.22 \\ $-$0.20 \end{tabular}	 & \begin{tabular}{r} $+$0.65 \\ $-$1.19 \end{tabular}	 & \begin{tabular}{r} $+$1.00 \\ $-$1.00 \end{tabular}\tabularnewline[4pt]
2.4--3.4 & 0.136 & 9.48 & 11.75	 & \begin{tabular}{r} $+$3.84 \\ $-$6.40 \end{tabular}	 & \begin{tabular}{r} $-$0.05 \\ $+$0.04 \end{tabular}	 & \begin{tabular}{r} $+$0.12 \\ $+$0.00 \end{tabular}	 & \begin{tabular}{r} $+$0.10 \\ $-$0.01 \end{tabular}	 & \begin{tabular}{r} $-$0.18 \\ $+$0.18 \end{tabular}	 & \begin{tabular}{r} $-$0.06 \\ $+$0.05 \end{tabular}	 & \begin{tabular}{r} $+$2.62 \\ $-$5.37 \end{tabular}	 & \begin{tabular}{r} $+$0.39 \\ $-$0.77 \end{tabular}	 & \begin{tabular}{r} $+$0.04 \\ $+$0.00 \end{tabular}	 & \begin{tabular}{r} $+$0.15 \\ $+$0.00 \end{tabular}	 & \begin{tabular}{r} $+$0.36 \\ $-$0.75 \end{tabular}	 & \begin{tabular}{r} $+$1.40 \\ $-$2.08 \end{tabular}	 & \begin{tabular}{r} $+$0.20 \\ $-$0.06 \end{tabular}	 & \begin{tabular}{r} $-$0.58 \\ $+$0.58 \end{tabular}	 & \begin{tabular}{r} $-$0.04 \\ $+$0.09 \end{tabular}	 & \begin{tabular}{r} $+$1.40 \\ $-$1.40 \end{tabular}	 & \begin{tabular}{r} $+$0.32 \\ $-$0.29 \end{tabular}	 & \begin{tabular}{r} $+$0.42 \\ $-$1.12 \end{tabular}	 & \begin{tabular}{r} $+$1.00 \\ $-$1.00 \end{tabular}\tabularnewline[4pt]
\hline
\end{tabular}
\end{table}
\end{landscape}

\newcolumntype{R}[1]{>{\raggedleft}m{#1}}
\begin{landscape}
\begin{table}
\centering
\caption[]{The measured double-differential \Zjets{} production \xss{} as a function of \absyj{} in the $300\GeV < \ptj < 400\GeV$ range.
$\delta_{\text{data}}^{\text{stat}}$ and $\delta_{\text{MC}}^{\text{stat}}$ are the statistical uncertainties in data and MC simulation, respectively.
$\delta_{\text{tot}}^{\text{sys}}$ is the total systematic uncertainty and includes the following components:
uncertainties due to electron reconstruction ($\delta_{\text{rec}}^{\text{el}}$), identification ($\delta_{\text{ID}}^{\text{el}}$) and trigger ($\delta_{\text{trig}}^{\text{el}}$) efficiencies;
electron energy scale ($\delta_{\text{scale}}^{\text{el}}$) and energy resolution ($\delta_{\text{res}}^{\text{el}}$) uncertainties;
sum in quadrature of the uncertainties from JES in situ methods ($\delta_{\text{in situ}}^{\text{JES}}$);
sum in quadrature of the uncertainties from JES $\eta$-intercalibration methods ($\delta_{\eta\text{-int}}^{\text{JES}}$);
an uncertainty of the measured single-hadron response ($\delta_{\text{hadron}}^{\text{JES}}$);
MC non-closure uncertainty ($\delta_{\text{closure}}^{\text{JES}}$);
sum in quadrature of the uncertainties due to pile-up corrections of the jet kinematics ($\delta_{\text{pile-up}}^{\text{JES}}$);
sum in quadrature of the flavour-based uncertainties ($\delta_{\text{flavour}}^{\text{JES}}$);
punch-through uncertainty ($\delta_{\text{pthrough}}^{\text{JES}}$); JER uncertainty ($\delta^{\text{JER}}$);
JVF uncertainty ($\delta^{\text{JVF}}$); sum in quadrature of the unfolding uncertainties ($\delta^{\text{unf}}$);
sum in quadrature of the uncertainties due to MC generated backgrounds normalisation ($\delta_{\text{MC}}^{\text{bg}}$);
sum in quadrature of the uncertainty due to combined multijet and \Wjets{} backgrounds ($\delta_{\text{mult}}^{\text{bg}}$);
uncertainty due to jet quality selection ($\delta^{\text{qual}}$).
All uncertainties are given in \%. The luminosity uncertainty of 1.9\% is not shown and not included in the total uncertainty and its components.}
\tiny
\setlength\tabcolsep{2pt}
\begin{tabular}{c S[table-format=-4.3] S[table-format=-2.2] S[table-format=-2.2] R{0.8cm} | R{0.8cm}  R{0.8cm}  R{0.8cm}  R{0.8cm}  R{0.8cm}  R{0.8cm}  R{0.8cm}  R{0.8cm}  R{0.8cm}  R{0.8cm}  R{0.8cm}  R{0.8cm}  R{0.8cm}  R{0.8cm}  R{0.8cm}  R{0.8cm}  R{0.8cm}  R{0.8cm}  }
\hline
\rule{0pt}{10pt}
\absyj{} & \multicolumn{1}{c}{$\frac{\text{d}^2\sigma}{\text{d}\absyj \text{d}\ptj}$} & \multicolumn{1}{c}{$\delta_{\text{data}}^{\text{stat}}$} &  \multicolumn{1}{c}{$\delta_{\text{MC}}^{\text{stat}}$} &  \multicolumn{1}{c|}{$\delta_{\text{tot}}^{\text{sys}}$}  & \multicolumn{1}{c}{$\delta_{\text{rec}}^{\text{el}}$}  & \multicolumn{1}{c}{$\delta_{\text{ID}}^{\text{el}}$}  & \multicolumn{1}{c}{$\delta_{\text{trig}}^{\text{el}}$}  & \multicolumn{1}{c}{$\delta_{\text{scale}}^{\text{el}}$}  & \multicolumn{1}{c}{$\delta_{\text{res}}^{\text{el}}$}  & \multicolumn{1}{c}{$\delta_{\text{in situ}}^{\text{JES}}$}  & \multicolumn{1}{c}{$\delta_{\eta\text{-int}}^{\text{JES}}$}  & \multicolumn{1}{c}{$\delta_{\text{hadron}}^{\text{JES}}$}  & \multicolumn{1}{c}{$\delta_{\text{closure}}^{\text{JES}}$}  & \multicolumn{1}{c}{$\delta_{\text{pile-up}}^{\text{JES}}$}  & \multicolumn{1}{c}{$\delta_{\text{flavour}}^{\text{JES}}$}  & \multicolumn{1}{c}{$\delta_{\text{pthrough}}^{\text{JES}}$}  & \multicolumn{1}{c}{$\delta^{\text{JER}}$}  & \multicolumn{1}{c}{$\delta^{\text{JVF}}$}  & \multicolumn{1}{c}{$\delta^{\text{unf}}$}  & \multicolumn{1}{c}{$\delta_{\text{MC}}^{\text{bg}}$}  & \multicolumn{1}{c}{$\delta_{\text{mult}}^{\text{bg}}$}  & \multicolumn{1}{c}{$\delta^{\text{qual}}$}  \tabularnewline[8pt]
& \multicolumn{1}{c}{[fb/\GeV]} & \multicolumn{1}{c}{[\%]} & \multicolumn{1}{c}{[\%]} & \multicolumn{1}{c|}{[\%]} & \multicolumn{1}{c}{[\%]}  & \multicolumn{1}{c}{[\%]}  & \multicolumn{1}{c}{[\%]}  & \multicolumn{1}{c}{[\%]}  & \multicolumn{1}{c}{[\%]}  & \multicolumn{1}{c}{[\%]}  & \multicolumn{1}{c}{[\%]}  & \multicolumn{1}{c}{[\%]}  & \multicolumn{1}{c}{[\%]}  & \multicolumn{1}{c}{[\%]}  & \multicolumn{1}{c}{[\%]}  & \multicolumn{1}{c}{[\%]}  & \multicolumn{1}{c}{[\%]}  & \multicolumn{1}{c}{[\%]}  & \multicolumn{1}{c}{[\%]}  & \multicolumn{1}{c}{[\%]}  & \multicolumn{1}{c}{[\%]}  & \multicolumn{1}{c}{[\%]} \tabularnewline[2pt]
\hline
\rule{0pt}{10pt}
0.0--0.4 & 1.190 & 5.83 & 6.75	 & \begin{tabular}{r} $+$5.95 \\ $-$2.33 \end{tabular}	 & \begin{tabular}{r} $+$0.74 \\ $+$0.00 \end{tabular}	 & \begin{tabular}{r} $+$0.56 \\ $+$0.00 \end{tabular}	 & \begin{tabular}{r} $+$0.00 \\ $+$0.51 \end{tabular}	 & \begin{tabular}{r} $+$0.00 \\ $+$0.72 \end{tabular}	 & \begin{tabular}{r} $+$0.76 \\ $+$0.00 \end{tabular}	 & \begin{tabular}{r} $+$4.03 \\ $-$1.63 \end{tabular}	 & \begin{tabular}{r} $+$1.41 \\ $+$0.00 \end{tabular}	 & \begin{tabular}{r} $+$0.68 \\ $+$0.00 \end{tabular}	 & \begin{tabular}{r} $+$0.46 \\ $+$0.00 \end{tabular}	 & \begin{tabular}{r} $+$2.42 \\ $-$0.13 \end{tabular}	 & \begin{tabular}{r} $+$2.24 \\ $+$0.00 \end{tabular}	 & \begin{tabular}{r} $+$0.54 \\ $+$0.00 \end{tabular}	 & \begin{tabular}{r} $-$1.16 \\ $+$1.16 \end{tabular}	 & \begin{tabular}{r} $+$0.79 \\ $+$0.00 \end{tabular}	 & \begin{tabular}{r} $+$0.08 \\ $-$0.08 \end{tabular}	 & \begin{tabular}{r} $+$0.10 \\ $-$0.09 \end{tabular}	 & \begin{tabular}{r} $+$0.30 \\ $-$0.65 \end{tabular}	 & \begin{tabular}{r} $+$1.00 \\ $-$1.00 \end{tabular}\tabularnewline[4pt]
0.4--0.8 & 1.083 & 5.52 & 5.50	 & \begin{tabular}{r} $+$5.94 \\ $-$2.31 \end{tabular}	 & \begin{tabular}{r} $+$0.74 \\ $+$0.00 \end{tabular}	 & \begin{tabular}{r} $+$0.56 \\ $+$0.00 \end{tabular}	 & \begin{tabular}{r} $+$0.00 \\ $+$0.51 \end{tabular}	 & \begin{tabular}{r} $+$0.00 \\ $+$0.72 \end{tabular}	 & \begin{tabular}{r} $+$0.76 \\ $+$0.00 \end{tabular}	 & \begin{tabular}{r} $+$4.03 \\ $-$1.63 \end{tabular}	 & \begin{tabular}{r} $+$1.41 \\ $+$0.00 \end{tabular}	 & \begin{tabular}{r} $+$0.68 \\ $+$0.00 \end{tabular}	 & \begin{tabular}{r} $+$0.46 \\ $+$0.00 \end{tabular}	 & \begin{tabular}{r} $+$2.42 \\ $-$0.13 \end{tabular}	 & \begin{tabular}{r} $+$2.24 \\ $+$0.00 \end{tabular}	 & \begin{tabular}{r} $+$0.54 \\ $+$0.00 \end{tabular}	 & \begin{tabular}{r} $-$1.16 \\ $+$1.16 \end{tabular}	 & \begin{tabular}{r} $+$0.79 \\ $+$0.00 \end{tabular}	 & \begin{tabular}{r} $+$0.08 \\ $-$0.08 \end{tabular}	 & \begin{tabular}{r} $+$0.11 \\ $-$0.10 \end{tabular}	 & \begin{tabular}{r} $+$0.23 \\ $-$0.55 \end{tabular}	 & \begin{tabular}{r} $+$1.00 \\ $-$1.00 \end{tabular}\tabularnewline[4pt]
0.8--1.2 & 0.946 & 6.68 & 6.87	 & \begin{tabular}{r} $+$5.95 \\ $-$2.34 \end{tabular}	 & \begin{tabular}{r} $+$0.74 \\ $+$0.00 \end{tabular}	 & \begin{tabular}{r} $+$0.56 \\ $+$0.00 \end{tabular}	 & \begin{tabular}{r} $+$0.00 \\ $+$0.51 \end{tabular}	 & \begin{tabular}{r} $+$0.00 \\ $+$0.72 \end{tabular}	 & \begin{tabular}{r} $+$0.76 \\ $+$0.00 \end{tabular}	 & \begin{tabular}{r} $+$4.03 \\ $-$1.63 \end{tabular}	 & \begin{tabular}{r} $+$1.41 \\ $+$0.00 \end{tabular}	 & \begin{tabular}{r} $+$0.68 \\ $+$0.00 \end{tabular}	 & \begin{tabular}{r} $+$0.46 \\ $+$0.00 \end{tabular}	 & \begin{tabular}{r} $+$2.42 \\ $-$0.13 \end{tabular}	 & \begin{tabular}{r} $+$2.24 \\ $+$0.00 \end{tabular}	 & \begin{tabular}{r} $+$0.54 \\ $+$0.00 \end{tabular}	 & \begin{tabular}{r} $-$1.16 \\ $+$1.16 \end{tabular}	 & \begin{tabular}{r} $+$0.79 \\ $+$0.00 \end{tabular}	 & \begin{tabular}{r} $+$0.08 \\ $-$0.08 \end{tabular}	 & \begin{tabular}{r} $+$0.12 \\ $-$0.11 \end{tabular}	 & \begin{tabular}{r} $+$0.30 \\ $-$0.68 \end{tabular}	 & \begin{tabular}{r} $+$1.00 \\ $-$1.00 \end{tabular}\tabularnewline[4pt]
1.2--1.6 & 0.628 & 8.15 & 8.34	 & \begin{tabular}{r} $+$5.96 \\ $-$2.44 \end{tabular}	 & \begin{tabular}{r} $+$0.74 \\ $+$0.00 \end{tabular}	 & \begin{tabular}{r} $+$0.56 \\ $+$0.00 \end{tabular}	 & \begin{tabular}{r} $+$0.00 \\ $+$0.51 \end{tabular}	 & \begin{tabular}{r} $+$0.00 \\ $+$0.72 \end{tabular}	 & \begin{tabular}{r} $+$0.76 \\ $+$0.00 \end{tabular}	 & \begin{tabular}{r} $+$4.03 \\ $-$1.63 \end{tabular}	 & \begin{tabular}{r} $+$1.41 \\ $+$0.00 \end{tabular}	 & \begin{tabular}{r} $+$0.68 \\ $+$0.00 \end{tabular}	 & \begin{tabular}{r} $+$0.46 \\ $+$0.00 \end{tabular}	 & \begin{tabular}{r} $+$2.42 \\ $-$0.13 \end{tabular}	 & \begin{tabular}{r} $+$2.24 \\ $+$0.00 \end{tabular}	 & \begin{tabular}{r} $+$0.54 \\ $+$0.00 \end{tabular}	 & \begin{tabular}{r} $-$1.16 \\ $+$1.16 \end{tabular}	 & \begin{tabular}{r} $+$0.79 \\ $+$0.00 \end{tabular}	 & \begin{tabular}{r} $+$0.08 \\ $-$0.08 \end{tabular}	 & \begin{tabular}{r} $+$0.19 \\ $-$0.17 \end{tabular}	 & \begin{tabular}{r} $+$0.45 \\ $-$0.95 \end{tabular}	 & \begin{tabular}{r} $+$1.00 \\ $-$1.00 \end{tabular}\tabularnewline[4pt]
1.6--2.0 & 0.322 & 11.56 & 11.42	 & \begin{tabular}{r} $+$5.97 \\ $-$2.46 \end{tabular}	 & \begin{tabular}{r} $+$0.74 \\ $+$0.00 \end{tabular}	 & \begin{tabular}{r} $+$0.56 \\ $+$0.00 \end{tabular}	 & \begin{tabular}{r} $+$0.00 \\ $+$0.51 \end{tabular}	 & \begin{tabular}{r} $+$0.00 \\ $+$0.72 \end{tabular}	 & \begin{tabular}{r} $+$0.76 \\ $+$0.00 \end{tabular}	 & \begin{tabular}{r} $+$4.03 \\ $-$1.63 \end{tabular}	 & \begin{tabular}{r} $+$1.41 \\ $+$0.00 \end{tabular}	 & \begin{tabular}{r} $+$0.68 \\ $+$0.00 \end{tabular}	 & \begin{tabular}{r} $+$0.46 \\ $+$0.00 \end{tabular}	 & \begin{tabular}{r} $+$2.42 \\ $-$0.13 \end{tabular}	 & \begin{tabular}{r} $+$2.24 \\ $+$0.00 \end{tabular}	 & \begin{tabular}{r} $+$0.54 \\ $+$0.00 \end{tabular}	 & \begin{tabular}{r} $-$1.16 \\ $+$1.16 \end{tabular}	 & \begin{tabular}{r} $+$0.79 \\ $+$0.00 \end{tabular}	 & \begin{tabular}{r} $+$0.08 \\ $-$0.08 \end{tabular}	 & \begin{tabular}{r} $+$0.26 \\ $-$0.24 \end{tabular}	 & \begin{tabular}{r} $+$0.51 \\ $-$1.00 \end{tabular}	 & \begin{tabular}{r} $+$1.00 \\ $-$1.00 \end{tabular}\tabularnewline[4pt]
2.0--3.0 & 0.032 & 26.98 & 24.63	 & \begin{tabular}{r} $+$6.01 \\ $-$2.64 \end{tabular}	 & \begin{tabular}{r} $+$0.74 \\ $+$0.00 \end{tabular}	 & \begin{tabular}{r} $+$0.56 \\ $+$0.00 \end{tabular}	 & \begin{tabular}{r} $+$0.00 \\ $+$0.51 \end{tabular}	 & \begin{tabular}{r} $+$0.00 \\ $+$0.72 \end{tabular}	 & \begin{tabular}{r} $+$0.76 \\ $+$0.00 \end{tabular}	 & \begin{tabular}{r} $+$4.03 \\ $-$1.63 \end{tabular}	 & \begin{tabular}{r} $+$1.41 \\ $+$0.00 \end{tabular}	 & \begin{tabular}{r} $+$0.68 \\ $+$0.00 \end{tabular}	 & \begin{tabular}{r} $+$0.46 \\ $+$0.00 \end{tabular}	 & \begin{tabular}{r} $+$2.42 \\ $-$0.13 \end{tabular}	 & \begin{tabular}{r} $+$2.24 \\ $+$0.00 \end{tabular}	 & \begin{tabular}{r} $+$0.54 \\ $+$0.00 \end{tabular}	 & \begin{tabular}{r} $-$1.16 \\ $+$1.16 \end{tabular}	 & \begin{tabular}{r} $+$0.79 \\ $+$0.00 \end{tabular}	 & \begin{tabular}{r} $+$0.08 \\ $-$0.08 \end{tabular}	 & \begin{tabular}{r} $+$0.69 \\ $-$0.63 \end{tabular}	 & \begin{tabular}{r} $+$0.63 \\ $-$1.25 \end{tabular}	 & \begin{tabular}{r} $+$1.00 \\ $-$1.00 \end{tabular}\tabularnewline[4pt]
\hline
\end{tabular}
\end{table}
\end{landscape}

\newcolumntype{R}[1]{>{\raggedleft}m{#1}}
\begin{landscape}
\begin{table}
\centering
\caption[]{The measured double-differential \Zjets{} production \xss{} as a function of \absyj{} in the $400\GeV< \ptj < 1050\GeV$ range.
$\delta_{\text{data}}^{\text{stat}}$ and $\delta_{\text{MC}}^{\text{stat}}$ are the statistical uncertainties in data and MC simulation, respectively.
$\delta_{\text{tot}}^{\text{sys}}$ is the total systematic uncertainty and includes the following components:
uncertainties due to electron reconstruction ($\delta_{\text{rec}}^{\text{el}}$), identification ($\delta_{\text{ID}}^{\text{el}}$) and trigger ($\delta_{\text{trig}}^{\text{el}}$) efficiencies;
electron energy scale ($\delta_{\text{scale}}^{\text{el}}$) and energy resolution ($\delta_{\text{res}}^{\text{el}}$) uncertainties;
sum in quadrature of the uncertainties from JES in situ methods ($\delta_{\text{in situ}}^{\text{JES}}$);
sum in quadrature of the uncertainties from JES $\eta$-intercalibration methods ($\delta_{\eta\text{-int}}^{\text{JES}}$);
an uncertainty of the measured single-hadron response ($\delta_{\text{hadron}}^{\text{JES}}$);
MC non-closure uncertainty ($\delta_{\text{closure}}^{\text{JES}}$);
sum in quadrature of the uncertainties due to pile-up corrections of the jet kinematics ($\delta_{\text{pile-up}}^{\text{JES}}$);
sum in quadrature of the flavour-based uncertainties ($\delta_{\text{flavour}}^{\text{JES}}$);
punch-through uncertainty ($\delta_{\text{pthrough}}^{\text{JES}}$); JER uncertainty ($\delta^{\text{JER}}$);
JVF uncertainty ($\delta^{\text{JVF}}$); sum in quadrature of the unfolding uncertainties ($\delta^{\text{unf}}$);
sum in quadrature of the uncertainties due to MC generated backgrounds normalisation ($\delta_{\text{MC}}^{\text{bg}}$);
sum in quadrature of the uncertainty due to combined multijet and \Wjets{} backgrounds ($\delta_{\text{mult}}^{\text{bg}}$);
uncertainty due to jet quality selection ($\delta^{\text{qual}}$).
All uncertainties are given in \%. The luminosity uncertainty of 1.9\% is not shown and not included in the total uncertainty and its components.}
\tiny
\setlength\tabcolsep{2pt}
\begin{tabular}{c S[table-format=-4.3] S[table-format=-2.2] S[table-format=-2.2] R{0.8cm} | R{0.8cm}  R{0.8cm}  R{0.8cm}  R{0.8cm}  R{0.8cm}  R{0.8cm}  R{0.8cm}  R{0.8cm}  R{0.8cm}  R{0.8cm}  R{0.8cm}  R{0.8cm}  R{0.8cm}  R{0.8cm}  R{0.8cm}  R{0.8cm}  R{0.8cm}  R{0.8cm}  }
\hline
\rule{0pt}{10pt}
\absyj{} & \multicolumn{1}{c}{$\frac{\text{d}^2\sigma}{\text{d}\absyj \text{d}\ptj}$} & \multicolumn{1}{c}{$\delta_{\text{data}}^{\text{stat}}$} &  \multicolumn{1}{c}{$\delta_{\text{MC}}^{\text{stat}}$} &  \multicolumn{1}{c|}{$\delta_{\text{tot}}^{\text{sys}}$}  & \multicolumn{1}{c}{$\delta_{\text{rec}}^{\text{el}}$}  & \multicolumn{1}{c}{$\delta_{\text{ID}}^{\text{el}}$}  & \multicolumn{1}{c}{$\delta_{\text{trig}}^{\text{el}}$}  & \multicolumn{1}{c}{$\delta_{\text{scale}}^{\text{el}}$}  & \multicolumn{1}{c}{$\delta_{\text{res}}^{\text{el}}$}  & \multicolumn{1}{c}{$\delta_{\text{in situ}}^{\text{JES}}$}  & \multicolumn{1}{c}{$\delta_{\eta\text{-int}}^{\text{JES}}$}  & \multicolumn{1}{c}{$\delta_{\text{hadron}}^{\text{JES}}$}  & \multicolumn{1}{c}{$\delta_{\text{closure}}^{\text{JES}}$}  & \multicolumn{1}{c}{$\delta_{\text{pile-up}}^{\text{JES}}$}  & \multicolumn{1}{c}{$\delta_{\text{flavour}}^{\text{JES}}$}  & \multicolumn{1}{c}{$\delta_{\text{pthrough}}^{\text{JES}}$}  & \multicolumn{1}{c}{$\delta^{\text{JER}}$}  & \multicolumn{1}{c}{$\delta^{\text{JVF}}$}  & \multicolumn{1}{c}{$\delta^{\text{unf}}$}  & \multicolumn{1}{c}{$\delta_{\text{MC}}^{\text{bg}}$}  & \multicolumn{1}{c}{$\delta_{\text{mult}}^{\text{bg}}$}  & \multicolumn{1}{c}{$\delta^{\text{qual}}$}  \tabularnewline[8pt]
& \multicolumn{1}{c}{[fb/\GeV]} & \multicolumn{1}{c}{[\%]} & \multicolumn{1}{c}{[\%]} & \multicolumn{1}{c|}{[\%]} & \multicolumn{1}{c}{[\%]}  & \multicolumn{1}{c}{[\%]}  & \multicolumn{1}{c}{[\%]}  & \multicolumn{1}{c}{[\%]}  & \multicolumn{1}{c}{[\%]}  & \multicolumn{1}{c}{[\%]}  & \multicolumn{1}{c}{[\%]}  & \multicolumn{1}{c}{[\%]}  & \multicolumn{1}{c}{[\%]}  & \multicolumn{1}{c}{[\%]}  & \multicolumn{1}{c}{[\%]}  & \multicolumn{1}{c}{[\%]}  & \multicolumn{1}{c}{[\%]}  & \multicolumn{1}{c}{[\%]}  & \multicolumn{1}{c}{[\%]}  & \multicolumn{1}{c}{[\%]}  & \multicolumn{1}{c}{[\%]}  & \multicolumn{1}{c}{[\%]} \tabularnewline[2pt]
\hline
\rule{0pt}{10pt}
0.0--0.4 & 0.110 & 6.84 & 8.88	 & \begin{tabular}{r} $+$3.15 \\ $-$6.01 \end{tabular}	 & \begin{tabular}{r} $-$0.34 \\ $+$0.00 \end{tabular}	 & \begin{tabular}{r} $+$0.00 \\ $-$0.26 \end{tabular}	 & \begin{tabular}{r} $+$0.00 \\ $-$0.43 \end{tabular}	 & \begin{tabular}{r} $-$0.82 \\ $+$0.00 \end{tabular}	 & \begin{tabular}{r} $-$0.41 \\ $+$0.00 \end{tabular}	 & \begin{tabular}{r} $+$2.01 \\ $-$4.74 \end{tabular}	 & \begin{tabular}{r} $+$0.22 \\ $-$0.50 \end{tabular}	 & \begin{tabular}{r} $+$0.00 \\ $-$0.36 \end{tabular}	 & \begin{tabular}{r} $+$0.00 \\ $-$0.33 \end{tabular}	 & \begin{tabular}{r} $+$0.10 \\ $-$0.91 \end{tabular}	 & \begin{tabular}{r} $+$0.33 \\ $-$2.13 \end{tabular}	 & \begin{tabular}{r} $+$0.13 \\ $-$0.58 \end{tabular}	 & \begin{tabular}{r} $-$0.14 \\ $+$0.14 \end{tabular}	 & \begin{tabular}{r} $+$0.01 \\ $-$0.32 \end{tabular}	 & \begin{tabular}{r} $+$1.47 \\ $-$1.47 \end{tabular}	 & \begin{tabular}{r} $+$0.15 \\ $-$0.14 \end{tabular}	 & \begin{tabular}{r} $+$0.59 \\ $-$0.84 \end{tabular}	 & \begin{tabular}{r} $+$1.00 \\ $-$1.00 \end{tabular}\tabularnewline[4pt]
0.4--0.8 & 0.076 & 9.45 & 9.66	 & \begin{tabular}{r} $+$3.13 \\ $-$5.99 \end{tabular}	 & \begin{tabular}{r} $-$0.34 \\ $+$0.00 \end{tabular}	 & \begin{tabular}{r} $+$0.00 \\ $-$0.26 \end{tabular}	 & \begin{tabular}{r} $+$0.00 \\ $-$0.43 \end{tabular}	 & \begin{tabular}{r} $-$0.82 \\ $+$0.00 \end{tabular}	 & \begin{tabular}{r} $-$0.41 \\ $+$0.00 \end{tabular}	 & \begin{tabular}{r} $+$2.01 \\ $-$4.74 \end{tabular}	 & \begin{tabular}{r} $+$0.22 \\ $-$0.50 \end{tabular}	 & \begin{tabular}{r} $+$0.00 \\ $-$0.36 \end{tabular}	 & \begin{tabular}{r} $+$0.00 \\ $-$0.33 \end{tabular}	 & \begin{tabular}{r} $+$0.10 \\ $-$0.91 \end{tabular}	 & \begin{tabular}{r} $+$0.33 \\ $-$2.13 \end{tabular}	 & \begin{tabular}{r} $+$0.13 \\ $-$0.58 \end{tabular}	 & \begin{tabular}{r} $-$0.14 \\ $+$0.14 \end{tabular}	 & \begin{tabular}{r} $+$0.01 \\ $-$0.32 \end{tabular}	 & \begin{tabular}{r} $+$1.47 \\ $-$1.47 \end{tabular}	 & \begin{tabular}{r} $+$0.18 \\ $-$0.16 \end{tabular}	 & \begin{tabular}{r} $+$0.42 \\ $-$0.71 \end{tabular}	 & \begin{tabular}{r} $+$1.00 \\ $-$1.00 \end{tabular}\tabularnewline[4pt]
0.8--1.2 & 0.058 & 11.67 & 11.68	 & \begin{tabular}{r} $+$3.15 \\ $-$6.01 \end{tabular}	 & \begin{tabular}{r} $-$0.34 \\ $+$0.00 \end{tabular}	 & \begin{tabular}{r} $+$0.00 \\ $-$0.26 \end{tabular}	 & \begin{tabular}{r} $+$0.00 \\ $-$0.43 \end{tabular}	 & \begin{tabular}{r} $-$0.82 \\ $+$0.00 \end{tabular}	 & \begin{tabular}{r} $-$0.41 \\ $+$0.00 \end{tabular}	 & \begin{tabular}{r} $+$2.01 \\ $-$4.74 \end{tabular}	 & \begin{tabular}{r} $+$0.22 \\ $-$0.50 \end{tabular}	 & \begin{tabular}{r} $+$0.00 \\ $-$0.36 \end{tabular}	 & \begin{tabular}{r} $+$0.00 \\ $-$0.33 \end{tabular}	 & \begin{tabular}{r} $+$0.10 \\ $-$0.91 \end{tabular}	 & \begin{tabular}{r} $+$0.33 \\ $-$2.13 \end{tabular}	 & \begin{tabular}{r} $+$0.13 \\ $-$0.58 \end{tabular}	 & \begin{tabular}{r} $-$0.14 \\ $+$0.14 \end{tabular}	 & \begin{tabular}{r} $+$0.01 \\ $-$0.32 \end{tabular}	 & \begin{tabular}{r} $+$1.47 \\ $-$1.47 \end{tabular}	 & \begin{tabular}{r} $+$0.21 \\ $-$0.20 \end{tabular}	 & \begin{tabular}{r} $+$0.54 \\ $-$0.86 \end{tabular}	 & \begin{tabular}{r} $+$1.00 \\ $-$1.00 \end{tabular}\tabularnewline[4pt]
1.2--2.6 & 0.012 & 12.84 & 13.36	 & \begin{tabular}{r} $+$3.20 \\ $-$6.08 \end{tabular}	 & \begin{tabular}{r} $-$0.34 \\ $+$0.00 \end{tabular}	 & \begin{tabular}{r} $+$0.00 \\ $-$0.26 \end{tabular}	 & \begin{tabular}{r} $+$0.00 \\ $-$0.43 \end{tabular}	 & \begin{tabular}{r} $-$0.82 \\ $+$0.00 \end{tabular}	 & \begin{tabular}{r} $-$0.41 \\ $+$0.00 \end{tabular}	 & \begin{tabular}{r} $+$2.01 \\ $-$4.74 \end{tabular}	 & \begin{tabular}{r} $+$0.22 \\ $-$0.50 \end{tabular}	 & \begin{tabular}{r} $+$0.00 \\ $-$0.36 \end{tabular}	 & \begin{tabular}{r} $+$0.00 \\ $-$0.33 \end{tabular}	 & \begin{tabular}{r} $+$0.10 \\ $-$0.91 \end{tabular}	 & \begin{tabular}{r} $+$0.33 \\ $-$2.13 \end{tabular}	 & \begin{tabular}{r} $+$0.13 \\ $-$0.58 \end{tabular}	 & \begin{tabular}{r} $-$0.14 \\ $+$0.14 \end{tabular}	 & \begin{tabular}{r} $+$0.01 \\ $-$0.32 \end{tabular}	 & \begin{tabular}{r} $+$1.47 \\ $-$1.47 \end{tabular}	 & \begin{tabular}{r} $+$0.37 \\ $-$0.34 \end{tabular}	 & \begin{tabular}{r} $+$0.73 \\ $-$1.24 \end{tabular}	 & \begin{tabular}{r} $+$1.00 \\ $-$1.00 \end{tabular}\tabularnewline[4pt]
\hline
\end{tabular}
\end{table}
\end{landscape}
 
\FloatBarrier
 
\printbibliography
\clearpage 
 
\begin{flushleft}
{\Large The ATLAS Collaboration}

\bigskip

G.~Aad$^\textrm{\scriptsize 101}$,    
B.~Abbott$^\textrm{\scriptsize 128}$,    
D.C.~Abbott$^\textrm{\scriptsize 102}$,    
O.~Abdinov$^\textrm{\scriptsize 13,*}$,    
A.~Abed~Abud$^\textrm{\scriptsize 70a,70b}$,    
K.~Abeling$^\textrm{\scriptsize 53}$,    
D.K.~Abhayasinghe$^\textrm{\scriptsize 93}$,    
S.H.~Abidi$^\textrm{\scriptsize 167}$,    
O.S.~AbouZeid$^\textrm{\scriptsize 40}$,    
N.L.~Abraham$^\textrm{\scriptsize 156}$
% ...
\bigskip
\\

% ...
$^{13}$Institute of Physics, Azerbaijan Academy of Sciences, Baku; Azerbaijan.\\
% ...
$^{40}$Niels Bohr Institute, University of Copenhagen, Copenhagen; Denmark.\\
% ...
$^{53}$II. Physikalisches Institut, Georg-August-Universit\"{a}t G\"ottingen, G\"ottingen; Germany.\\
% ...
$^{70}$$^{(a)}$INFN Sezione di Pavia;$^{(b)}$Dipartimento di Fisica, Universit\`a di Pavia, Pavia; Italy.\\
% ...
$^{93}$Department of Physics, Royal Holloway University of London, Egham; United Kingdom.\\
% ...
$^{101}$CPPM, Aix-Marseille Universit\'e, CNRS/IN2P3, Marseille; France.\\
$^{102}$Department of Physics, University of Massachusetts, Amherst MA; United States of America.\\
% ...
$^{128}$Homer L. Dodge Department of Physics and Astronomy, University of Oklahoma, Norman OK; United States of America.\\
% ...
$^{156}$Department of Physics and Astronomy, University of Sussex, Brighton; United Kingdom.\\
% ...
$^{167}$Department of Physics, University of Toronto, Toronto ON; Canada.\\
% ...
$^{a}$ Also at CERN, Geneva; Switzerland.\\
$^{b}$ Also at CPPM, Aix-Marseille Universit\'e, CNRS/IN2P3, Marseille; France.\\
% ...
$^{*}$ Deceased

\end{flushleft}


\end{document}